\documentclass{article}
\usepackage{amssymb}

\usepackage{graphicx}
\usepackage{amsmath}

\input{tcilatex}

\begin{document}

{\huge ''\textbf{New''Veneziano\ amplitudes\ from} }

\ \ \ \ \ \ \ \ \ \ {\huge ''\textbf{old}'' \textbf{Fermat (hyper)surfaces}}

\bigskip

\ \ \ \ \ \ \ \ \ \ \ \ \ \ \ \ \ \ \ \ \ \ \ \ \ \ \ \ \ \ \ \ \ \ \ \ \ \
A.L.Kholodenko\footnote{%
e-mail address: string@clemson.edu}

\bigskip

\emph{375 H.L.Hunter Laboratories, Clemson University, Clemson, SC} 29634,
USA

\ \ \ \ \ \ \ \ \ \ \ \ \ \ \ \ \ \ \ \ \ \ \ \ \ \ \ \ \bigskip\ \ \ \ \ \
\ \ \ \ \ \ \ \ \ 

\ \ \ \ \ \ \ \ \ \ \ \ \ \ \ \ \ \ \ \ \ \ \ \ \ \ \ \ \ \ \ \ \ \ \ \ \ \
\ \textbf{Abstract\ \ }\ 

The history of the discovery of bosonic string theory is well documented.
This theory evolved as an attempt to find a multidimensional analogue of
Euler's beta function to describe the multiparticle Veneziano amplitudes.
Such an analogue had in fact been known in mathematics at least in 1922. Its
mathematical meaning was studied subsequently from different angles by
mathematicians such as Selberg, Weil and Deligne among others. The
mathematical interpretation of this multidimensional beta function that was
developed subsequently is markedly different from that described in physics
literature. This work aims to bridge the gap between the mathematical and
physical treatments. Using some results of recent publications (e.g.
J.Geom.Phys.38 (2001) 81 ; ibid 43 (2002) 45) new topological,
algebro-geometric, number-theoretic and combinatorial treatment of the
multiparticle Veneziano amplitudes is developed. As a result, an entirely
new physical meaning of these amplitudes is emerging: they are periods of
differential forms associated with homology cycles on Fermat
(hyper)surfaces. Such (hyper)surfaces are considered as complex projective
varieties of Hodge type. Although the computational formalism developed in
this work resembles that used in mirror symmetry calculations, many
additional results from mathematics are used\ along with their suitable
physical interpretation. For instance, the Hodge spectrum of the Fermat
(hyper)surfaces is in one-to-one correspondence with the possible spectrum
of particle masses. The formalism also allows us to obtain correlation
functions of both conformal field theory and particle physics using the same
type of the Picard-Fuchs equations whose solutions are being interpreted in
terms of periods.

\bigskip


\bigskip

\emph{Subj.Class}: \ Particle physics, scattering theory, string theory,
conformal field theory, mathematical physics \ \ \ \ \ \ \ \ \ \ \ \ \ \ \ \
\ \ \ \ \ \ \ \ \ \ \ \ \ \ \ \ \ \ \ \ \ \ \ \ \ \ \ \ \ \ \ \ \ \ \ \ \ \
\ \ \ \ \ \ \ \ \ \ \ \ \ \ \ \ \ \ \ \ \ \ \ \ \ \ \ \ \ \ \ \ \ \ \ \ \ \
\ \ \ \ \ \ 

\bigskip


\pagebreak

\section{Introduction}

\subsection{Some properties of Veneziano and Virasoro amplitudes}

\bigskip

In 1968 Veneziano [1] had postulated 4-particle scattering amplitude $%
A(s,t,u)$ given (up to a common constant factor) by 
\begin{equation}
A(s,t,u)=V(s,t)+V(s,u)+V(t,u)  \tag{1.1}
\end{equation}
where 
\begin{equation}
V(s,t)=\int\limits_{0}^{1}x^{-\alpha (s)-1}(1-x)^{-\alpha (t)-1}dx\equiv
B(-\alpha (s),-\alpha (t))  \tag{1.2}
\end{equation}
is Euler's beta function and $\alpha (x)$ is the Regge trajectory usually
written as $\alpha (x)=\alpha (0)+\alpha ^{\prime }x$ with $\alpha (0)$ and $%
\alpha ^{\prime }$ being the Regge slope and the intercept, respectively. In
case of Lorentzian metric with signature $\{+,-,-,-\}$ the Mandelstam
variables $s$, $t$ and $u$ entering the Regge trajectory are given by [2] 
\begin{eqnarray}
s &=&(p_{1}+q_{1})^{2},  \TCItag{1.3} \\
t &=&(p_{1}-p_{2})^{2},  \notag \\
u &=&(p_{1}-q_{2})^{2}.  \notag
\end{eqnarray}
The 4-momenta $p_{i}$ and \ $q_{i}$ \ are constrained by the energy-momentum
conservation $p_{1}+q_{1}=p_{2}+q_{2}$ leading to relation between the
Mandelstam variables: 
\begin{equation}
s+t+u=\sum\limits_{i=1}^{4}m_{i}^{2}.  \tag{1.4}
\end{equation}
Already Veneziano [1] had noticed\footnote{%
To get our Eq.(1.5) from Eq.7 of Veneziano paper, it is sufficient to notice
that his $1-\alpha (s)$ corresponds to ours -$\alpha (s).$} that to fit the
experimental data the Regge trajectories should obey the constraint 
\begin{equation}
\alpha (s)+\alpha (t)+\alpha (u)=-1  \tag{1.5}
\end{equation}
consistent with Eq.(1.4). He also noticed that the amplitude $A(s,t,u)$ can
be equivalently rewritten with help of this constraint as follows 
\begin{equation}
A(s,t,u)=\Gamma (-\alpha (s))\Gamma (-\alpha (t))\Gamma (-\alpha (u))[\sin
\pi (-\alpha (s))+\sin \pi (-\alpha (t))+\sin \pi (-\alpha (u))].  \tag{1.6}
\end{equation}
This result looks strikingly similar to that suggested a bit later by
Virasoro [3]. Up to a constant it is given by 
\begin{equation}
\bar{A}(s,t,u)=\frac{\Gamma (a)\Gamma (b)\Gamma (c)}{\Gamma (a+b)\Gamma
(b+c)\Gamma (c+a)}  \tag{1.7}
\end{equation}
with $a=-\frac{1}{2}\alpha (s),$etc$.$ also subjected to the constraint: 
\begin{equation}
\frac{1}{2}\left( \alpha (s)+\alpha (t)+\alpha (u)\right) =-1.  \tag{1.8}
\end{equation}
Use of the formulas 
\begin{equation}
\Gamma (x)\Gamma (1-x)=\frac{\pi }{\sin \pi x}  \tag{1.9}
\end{equation}
and 
\begin{equation}
4\sin x\sin y\sin z=\sin (x+y-z)+\sin (y+z-x)+\sin (z+x-y)-\sin (x+y+z) 
\tag{1.10}
\end{equation}
permits us to rewrite Eq.(1.7) in the alternative form (up to unimportant
constant): 
\begin{eqnarray}
\bar{A}(s,t,u) &=&[\Gamma (-\frac{1}{2}\alpha (s))\Gamma (-\frac{1}{2}\alpha
(t))\Gamma (-\frac{1}{2}\alpha (u))]^{2}\times  \TCItag{1.11} \\
&&[\sin \pi (-\frac{1}{2}\alpha (s))+\sin \pi (-\frac{1}{2}\alpha (t))+\sin
\pi (-\frac{1}{2}\alpha (u))].  \notag
\end{eqnarray}
Although these two amplitudes look deceptively similar, mathematically, they
are markedly different. Indeed, by using Eq.(1.6) conveniently rewritten as 
\begin{equation}
A(a,b,c)=\Gamma (a)\Gamma (b)\Gamma (c)[\sin \pi a+\sin \pi b+\sin \pi c] 
\tag{1.12}
\end{equation}
and exploiting the identity 
\begin{equation}
\cos \frac{\pi z}{2}=\frac{\pi ^{z}}{2^{1-z}}\frac{1}{\Gamma (z)}\frac{\zeta
(1-z)}{\zeta (z)}  \tag{1.13}
\end{equation}
after some trigonometric calculations the following result is obtained: 
\begin{equation}
A(a,b,c)=\frac{\zeta (1-a)}{\zeta (a)}\frac{\zeta (1-b)}{\zeta (b)}\frac{%
\zeta (1-c)}{\zeta (c)},  \tag{1.14}
\end{equation}
provided that 
\begin{equation}
a+b+c=1.  \tag{1.15}
\end{equation}
For the Virasoro amplitude, apparently, no result like Eq.(1.14) can be
obtained. The differences between the Veneziano and the Virasoro amplitudes
are much more profound as the rest of this paper demonstrates.

The result, Eq.(1.14), is remarkable in the sense that it allows us to
interpret the Veneziano amplitude from the point of view of number theory,
the theory of dynamical systems, etc. Such interpretation is presented in
some detail in Section 3.1.while Section 2 provides the necessary
mathematical background.

Moreover, in our previous work, Ref. [4], the following partition function
describing (train track) dynamics of 2+1 gravity was obtained 
\begin{equation}
Z(\beta )=\frac{\zeta (\beta -1)}{\zeta (\beta )}.  \tag{1.16a}
\end{equation}
This function is meant to describe a dynamical transition from the pseudo
Anosov (Appendix B) to the Seifert fibered dynamical regime controlled by
the temperature-like parameter $\beta $. The exact physical nature of the
parameter $\beta $ was left unexplained. The train tracks ''live'' on the
surface of the punctured torus-the simplest surface of negative Euler
characteristic on which they can ''live'' [5]. Normally, torus without
puncture(s) can be foliated by vector fields without singularities. The
punctured torus (via Schottky double construction) can be converted into a
double torus, that is into the Riemann surface of genus 2. If one tries to
foliate such surface with some line field (\textbf{not} to be confused with
the vector field [6]), one can easily discover that singularities of such
field will occur inevitably. The number of singularities is controlled by
the Poincare-Hopf index theorem\footnote{%
It is possible to have different sets of singularities as long as their
total index remains the same [5,6]}. Foliations with singularities are known
as pseudo Anosov (Appendix B) type. For the punctured torus only 2
singularities are allowed and their scattering on each other is very much
like that which the Veneziano amplitude is describing. The Mandelstam
variables $s,t,u$ can be used in the present case as well thus replacing our
dynamical zeta function, Eq.(1.16a), with the product of zeta functions,
i.e. with the Veneziano amplitude, Eq.(1.14). Such replacement is not
complete however since our dynamical zeta function has ''wrong'' argument: $%
\beta -1,$ in the numerator of Eq.(1.16a), instead of $1-\beta .$
Fortunately, this deficiency is easy to correct if we recall the
presentation of the Riemann zeta function as infinite product over primes.
Then, instead of Eq.(1.16a), we obtain 
\begin{equation}
Z(\beta )=\prod\limits_{p}\frac{1-p^{-\beta }}{1-p^{1-\beta }},  \tag{1.16b}
\end{equation}
where the product is taken over all prime numbers. Such presentation invokes
immediately connections with the p-adic string theory to be described below.
For now, however, we still need to make several observations. To begin, let
us introduce the p-adic analogue of zeta function, in particular, 
\begin{equation*}
\zeta _{p}(\beta -1)=\frac{1}{1-p^{1-\beta }},
\end{equation*}
to be reobtained in Section 2, Eq.(2.15). We can equivalently present it as 
\begin{equation}
\zeta _{p}(\beta -1)=\frac{-1}{p^{1-\beta }}\frac{1}{1-p^{\beta -1}}%
=(-1)p^{\beta -1}\zeta _{p}(1-\beta ).  \tag{1.17}
\end{equation}
It should be clear from Eq.(1.17) that $\zeta _{p}(\beta -1)$ has the same
pole as $\zeta _{p}(1-\beta )$ and, hence, \emph{the Veneziano amplitude,
Eq.(1.14), is obtainable from dynamics of 2+1 gravity} \emph{and, therefore,
from the full Einsteinian gravity}. Such arguments, although plausible, not
quite rigorous yet. Indeed, in the case of the open string the world sheet
is the Poincare upper half plane (or, equivalently, the open disc). The
incoming/outgoing particles are represented as ''insertions on the
boundary'' in terminology of Ref.[7], page 46, or as cusps in mathematical
terminology. In the case of 4-particle Veneziano amplitude there are 4
''insertions on the boundary''. But, as Fig.3 of our work on dynamics of
train tracks, Ref.[4], indicates, in the case of the punctured torus there
are in the simplest case also 4 insertions at the boundary. Moreover, the
open disc Poincare model is not the model of the world sheet in the case of
2+1 gravity. It is used for the description of the dynamics of simplest
train tracks. This dynamics is taking place in the Teichm\"{u}ller space of
the punctured torus which is just the open Poincare disc [8]. It can be
shown [9], that the Teichm\"{u}ller space of the punctured torus is the same
as that of the 4 times punctured sphere (considered to be the world sheet
for the closed string theory, Ref.[7], page 35). Hence, already at the tree
level in string theory the description of physical processes actually takes
place not in real but in the Teichm\"{u}ller space. And, of course,
mathematically rigorous study of the dynamics of train tracks indicates
that, indeed, such dynamics is taking place in the Teichm\"{u}ller space.
The above picture, perhaps, can be generalized to surfaces of genus higher
than one\footnote{%
And, indeed, in a series of papers by Takhtajan et al (e.g see [10] for the
latest paper and references therein) a serious attempt has been made to
accomplish this task. To our knowledge, the multiparticle Veneziano
amplitudes have not been reobtained thus far in these papers.}. In this
paper, however, no attempts are made to do this in view of the fact that
there is much more mathematically elegant and efficient way to accomplish
the task. Before discussing details of this other way, we would like to make
several comments regarding connections of the result, Eq.(1.14), with the
p-adic string theory.

\subsection{\protect\bigskip Connections with the p-adic string theory}

Eq.(1.14) was used implicitly (but essentially) in the p-adic string theory
as can be seen from the review by Bekke and Freund, Ref.[11], and is
discussed below. Since Eq.(1.14) was also obtained in our earlier work,
Ref.[4], in connection with dynamics of train tracks representing dynamics
of 2+1 gravity, from the standpoint of an external observer, the evolution
of the train track patterns takes place in \ normal \ 2+1 or even 3+1
physical space-time\footnote{%
Moreover, such picture is not just imaginary. It can be made entirely real
if one looks at dynamics of disclinations in 2 dimensional liquid crystals.
Such liquid crystal dynamics is in one to one correspondence with the
dynamics of train tracks [5,6].}. Eq.(1.14), when it is obtained with help
of the p-adic bosonic string formalism, ''lives'' instead in 26 dimensional
space- time ( page 26 of Ref.[11]). Hence, it is worth discussing \ and
comparing in some detail both cases now.

Let us begin with the case of the p-adic string theory. It is commonly
believed that the only rationale of such theory lies in regularizing of the
string world sheet by placing such sheet on the Bruhat-Tits (known in
physics literature as Bethe lattice) tree (e.g see Appendix C) ''which can
be embedded in ordinary \textit{real} space-time'' (page 29 of Ref.[11]).
Consistency with known continuous results in bosonic string theory require,
however, that such real space-time is 26 dimensional. Such conclusion is
reached roughly speaking based on the following arguments. \ Consider the
identity 
\begin{equation}
\pi ^{-\frac{s}{2}}\Gamma (\frac{s}{2})\text{ }\zeta (s)=\pi ^{-\frac{1-s}{2}%
}\Gamma (\frac{1-s}{2})\text{ }\zeta (1-s).  \tag{1.18}
\end{equation}
Following Ref.[11] we define the adelic zeta function $\zeta _{\mathbf{A}%
}(s) $ by 
\begin{equation}
\zeta _{\mathbf{A}}(s)=\pi ^{-\frac{s}{2}}\Gamma (\frac{s}{2})\text{ }\zeta
(s)  \tag{1.19}
\end{equation}
then, Eq.(1.18) can be rewritten as follows 
\begin{equation}
\zeta _{\mathbf{A}}(s)=\zeta _{\mathbf{A}}(1-s).  \tag{1.20}
\end{equation}
Using the same reference, an adelic gamma function $\Gamma _{\mathbf{A}}(s)$
is then defined by 
\begin{equation}
\Gamma _{\mathbf{A}}(s)=\frac{\zeta _{\mathbf{A}}(s)}{\zeta _{\mathbf{A}%
}(1-s)}.  \tag{1.21}
\end{equation}
In view of Eq.(1.20), clearly, $\Gamma _{\mathbf{A}}(s)=1.$ This result is a
particular example of the so called ''product formula'' \ well known in the
algebraic number theory [12]. Following general rules of p-adic analysis
beautifully explained in the classical book by Artin [12], one can rewrite
the statement $\Gamma _{\mathbf{A}}(s)=1$ in a form of the product over both
finite and infinite places, i.e. 
\begin{equation}
\Gamma _{\mathbf{A}}(s)=\prod\limits_{p}\Gamma _{p}(s)=1,  \tag{1.22}
\end{equation}
where $p$ runs over all primes (just like in the case of Riemann zeta
function) and, in addition, over the ''prime at infinity''. In our case 
\begin{equation}
\Gamma _{p}(s)=\frac{1-p^{s-1}}{1-p^{-s}}\equiv \frac{\zeta _{p}(s)}{\zeta
_{p}(1-s)}  \tag{1.23}
\end{equation}
but 
\begin{equation}
\Gamma _{\infty }(s)=\frac{\zeta _{\infty }(s)}{\zeta _{\infty }(1-s)}=\pi
^{-s+\frac{1}{2}}\Gamma (\frac{s}{2})/\Gamma (\frac{1-s}{2}),  \tag{1.24}
\end{equation}
where $\zeta _{\infty }(s)=\pi ^{-\frac{s}{2}}\Gamma (\frac{s}{2})$ in view
of Eq.(1.18). Clearly, we get then 
\begin{equation}
\frac{\zeta _{\infty }(s)}{\zeta _{\infty }(1-s)}\prod\limits_{p\neq \infty
}^{{}}\Gamma _{p}(s)=1.  \tag{1.25}
\end{equation}
This leads to 
\begin{equation}
\prod\limits_{p\neq \infty }^{{}}\frac{\zeta _{p}(s)}{\zeta _{p}(1-s)}%
=\prod\limits_{p\neq \infty }^{{}}\Gamma _{p}(s)=\frac{\zeta _{\infty }(1-s)%
}{\zeta _{\infty }(s)}=\frac{\zeta (s)}{\zeta (1-s)}  \tag{1.26}
\end{equation}
in view of Eq.(1.18). But the l.h.s. of Eq.(1.26) is obtainable from the
p-adic string path integral [11,13]\footnote{%
Actually, the product of 3 factors corresponding respectively to s, t and u
channels is obtained.}. Hence, the p-adic open string scattering amplitude
is just \textbf{an inverse} of the standard Veneziano amplitude in view of
Eq.(1.14). Since the standard amplitude is obtainable via Polyakov path
integral in 26 dimensions, the p-adic string should also live in 26
dimensions in view of the product formula, Eq.(1.22). This is the conclusion
reached in Ref.[11]. Fortunately (or unfortunately!), this conclusion is
incorrect. Indeed, let us take another look at Eq-s (1.20)-(1.22). We have 
\begin{equation}
\prod\limits_{p}\zeta _{p}(s)=\prod\limits_{p}\zeta _{p}(1-s)  \tag{1.27}
\end{equation}
or, equivalently, 
\begin{equation}
\zeta _{\infty }(s)\prod\limits_{p\neq \infty }\zeta _{p}(s)=\zeta _{\infty
}(1-s)\prod\limits_{p\neq \infty }\zeta _{p}(1-s).  \tag{1.28}
\end{equation}
Evidently, it is only natural to expect that ''locally'' we should also have 
\begin{equation}
\zeta _{p}(s)=\zeta _{p}(1-s)  \tag{1.29}
\end{equation}
for all $p\prime s$, including $p$=$\infty .$ This is not the case, however,
in view of definitions of local zeta functions,e.g. see Eq.(1.23). This
negative observation cannot be simply repaired. Indeed, from p-adic analysis
[14] and, even more advanced theory (which includes p-adic analysis as its
part) of Drinfeld modules [15], it can be found that the existing in
mathematics (e.g.see Eq.(3.45) below) expressions for the p-adic gamma
functions (actually, there are several expressions [15]) differ markedly
from that given in Eq.(1.23). Moreover, the Bruhat-Tits tree used for
calculations of \ Veneziano and Virasoro amplitudes \ in Ref-s[11,13] \emph{%
is the discrete analogue} \emph{of the upper Poincar}$e^{\prime }$\emph{\
half plane} [ 16, 17] which is the universal covering space for any \textbf{%
hyperbolic} Riemann surface \emph{while the conventional string theory} 
\emph{formulations producing at the tree level} Veneziano and
Virasoro-Shapiro-like amplitudes \emph{involve flat metrics} \emph{which is
obviously} \emph{nonhyperbolic}. Hence, the whole chain of arguments leading
to the conclusion that p-adic strings are ''living'' in 26 dimensions is
apparently incorrect. Moreover, we argue in this work that use of the
Bruhat-Tits tree in the p-adic string theory is motivated \textbf{not} by
the necessity of regularization of the world sheet. It occurs rather
naturally \ and is very closely associated with complex multiplication
theory developed by Shimura and Taniyama [18]. It can be thought as some
analogue of the Fourier component in the Fourier expansion\footnote{%
More exactly, it occurs as an analogue of the individual term in Euler's
prime number product \ decomposition of zeta function.} of some continuous
function. Just like in the linear Fourier analysis our knowledge of one
Fourier component is practically sufficient for restoration of the whole
function, the knowledge of one p-adic component is sufficient for
restoration of the whole function as well. This statement is illustrated
below on examples of calculation of the p-adic analogue of the Veneziano
amplitude discussed in Sections 3.3.,4.3. and 5.3.1.

\subsection{\protect\bigskip Organization of the rest of this paper}

The history of development of the dual resonance models leading ultimately
to models of relativistic bosonic and fermionic strings is well documented
and can be found, for example, in the excellent collection of review
articles edited by Jacob, Ref.[19]. This development one way or another
stems from earlier efforts to develop the axiomatic S-matrix theory-the idea
which can be traced back to two papers by Heisenberg [20] of 1943.The
mathematical foundations of the S-matrix theory can be traced back to even
much earlier works of Kramers and Kr\"{o}nig dating back to years 1926 and
1927. The usefulness of their results (related originally to the light
scattering) to particle physics scattering is well documented in the
classical monograph by Bjorken and Drell, Ref.[21]. The contribution of
Kramers and Kr\"{o}nig to scattering problems lies in their observation of
the usefulness of the Cauchy integral formula to such problems. Surely, from
the modern perspective, one can talk equivalently about usefulness of the
Riemann-Hilbert problem [22] to scattering problems. And this observation
leads us ultimately to the Picard-Fuchs type equations considered in Section
5\footnote{%
The connection between the Picard-Fuchs equations and the Riemann-Hilbert
problem is discussed, for example, in Ref. [23]}. If one reformulates the
Kramers-Kronig results into the language of particle physics as it is done
by Bjorken and Drell [21], then in few words the essence of the scattering
problem can be formulated as the statement about the scattering amplitude 
\begin{equation}
\func{Im}f(\omega )=\frac{\omega }{4\pi }\sigma (\omega )  \tag{1.30}
\end{equation}
connecting the imaginary part of the forward scattering amplitude $\func{Im}%
f(\omega )$ with the total crossection $\sigma (\omega )$ for the
''frequency'' $\omega >0.$ Suppose, we know the amplitude $f$($\omega )$,
then, the familiar trick (which is just a corollary of the Cauchy theorem)
tells us that 
\begin{equation}
f(\omega +i\varepsilon )-f(\omega -i\varepsilon )=2\pi i\func{Im}f(\omega )%
\text{ , }\varepsilon \rightarrow 0^{+}  \tag{1.31}
\end{equation}
thus allowing us to determine the total crossection using Eq.(1.30).
Alternatively, if such a crossection can be obtained experimentally, one can
restore the real part of the amplitude through Kramers-Kr\"{o}nig relations
thus reconstructing the whole amplitude from the experimental data. Such
simple minded picture becomes very complex when, for example, one is trying
to analyze the analytic properties of the multiparticle vertex parts
(contributing to scattering amplitude) based on analysis of the appropriate
Feynman diagrams of quantum field theory. At the physical level of rigor
this problem was studied by Landau [24]. Subsequently, much more
sophisticated cohomological analysis of the whole problem was developed by
Pham [25] whose results were brought to perfection by Milnor[26] and
Brieskorn [27]. In physics literature the results of Pham were carefully
analyzed in two fundamental papers by Ponzano, Regge, Speer and Westwater
[28]. Unfortunately, subsequent development of particle physics went \ into
different direction(s). Nevertheless, the major problem of reconstruction of
the underlying particle physics model from scattering amplitudes with known
analytic properties had ultimately created what is known \ today as
string/membrane theory. As it was mentioned in Section 1.1., Veneziano had
not derived his 4-particle amplitude theoretically. This amplitude was
postulated and \ subsequently checked experimentally [7,19]. The attempts to
extend this amplitude to multiparticle scattering had lead ultimately to
formulation of string theory. Since mathematically, the simplest Veneziano
amplitude is just Euler's beta function, the extension of such amplitude to
the multiparticle case mathematically is reduced to the correct
multidimensional extension of Euler's beta function integral. Such an
extension can be found in the book by Edwards, e.g. see Ref [29], page 167,
published in1922.His result was rediscovered by many mathematicians. The
latest in this list of people, perhaps, Varchenko whose two papers [30] of
1989-90 contain the result of Edwards and, surely, much more. Ironically, in
1967-a year earlier than Veneziano paper was published, the paper by Chowla
and Selberg had appeared [31] which related Euler's beta function to the
periods of elliptic integrals. The usefulness of results of Chowla and
Selberg \ is discussed in Section 4 within the context of conformal and
particle physics theory. The result of Cowla and Selberg was generalized by
Andre Weil whose two influential papers on the same subject [32, 33] had
brought into picture periods of the Abelian varieties, Hodge rings, etc.
thus inspiring Benedict Gross to rederive Edwards result in 1978 [34]
without actually being aware of its existence. In the paper by Gross the
beta function appears as period of the differential form on the Jacobian of
the Fermat curve. His results as well as those of Rohlich (placed in the
appendix to Gross paper) have been subsequently documented in the book by
Lang [35]. These results are explained from physical standpoint in Section
3.2. where they are reinterpreted with help of Milnor's fundamental results
on singular points of complex hypersurfaces [26]. Milnor's work to a large
extent had shaped up the theory of singularities and, not surprisingly, all
the results of this paper could be considered as some practical applications
of the singularity theory as described, for example, in the monograph by
Arnol'd, Varchenko and Gussein-Zade [36].

Although in the paper by Gross the multidimensional extension of beta
function is considered, e.g. see page 207 of Gross paper [34], the details
were not provided however. The details are provided (to our knowledge) only
in this paper. They are based on ideas developed in lectures by Deligne [37]
delivered in 1978-1979. Deligne was seemingly unaware of both results of
Edwards and Gross. In developing of Hodge theory Deligne noticed that Hodge
theory requires some essential changes(e.g. introduction of mixed Hodge
structures,etc.) in the case if the Hodge-K\"{a}hler manifolds possess
singularities. We discuss these modifications briefly in Section 5 and in
Appendix D within the context of standard singularity theory. Although this
paper is an attempt to present a balanced treatment of both the
number-theoretic (including the p-adic ) and standard analytic aspects of
the problem of periods associated with Veneziano amplitudes, and in spite of
the fact that the paper came out rather long, in reality, it presents just a
gentle scratch on the surface of enormous amount of information. For
instance, the number-theoretic aspects of the mixed Hodge structures
involving theories of motives [37,38], of crystalline cohomology [37-39], of
Tannakian categories [37-38], etc. are left completely outside the scope of
this paper mainly because their physical meaning still remains to be
uncovered. Also, the symplectic aspects of our scattering problem [40]
needed for development of correct ''string''-theoretic model associated with
Veneziano amplitudes are left outside the scope of this paper as well. The
emerging opportunity to ''fill in the gaps'' and to demonstrate that the
correct formulation of ''string'' theory leads indeed to ''the theory of
everything'', hopefully, should provide an inspiration for our readers. In
Section 6 we provide some outline of what lies ahead. Very exciting
connections between many branches of so far ''pure'' mathematics and ''down
to earth'' and, hence,''dirty'' physics are yet to be unveiled.

\section{Number fields versus function fields}

\subsection{Valuations}

The existing physical literature uses real, complex and p-adic numbers. Use
of p-adic numbers is still rather exotic [11,13, 41] and to a large extent
artificial and, hence, unpopular. It is strange that function fields have
not been used in physics so far. In algebra [12,42] and number theory
[12,43] it was recognized long ago that function fields have practically the
same properties as number fields but in many cases are easier to handle. \
We would like to argue that the function fields are not only easier to
handle than the number fields but also that these fields are physically more
relevant. Not surprisingly, these fields found their widespread use in
discrete mathematics, computer science and coding theory [44.45].

Elementary number theory begins with the ring of integers \textbf{Z} . This
ring can be made into the field of fractions \textbf{Q} if the division
operation is included (provided that there is no divisions by zero). The
field of rational numbers \textbf{R} is obtained by the completion operation
involving the Cauchy sequences and the usual absolute values [46]. The field
of complex numbers \textbf{C} is obtained by algebraic extension of \textbf{Z%
}, \textbf{Q} or \textbf{R. }E.g.\textbf{\ }solution\textbf{\ }of\textbf{\ }%
equation\textbf{\ }$x^{2}+1=0$ is the simplest basic example leading to
complex numbers$.$ Evidently every number $n$ $\in \mathbf{Z}$ can be
decomposed into primes by the property of unique factorization, e.g. 
\begin{equation}
n=\pm p_{1}^{m_{1}}\cdot \cdot \cdot p_{r}^{m_{r}},  \tag{2.1}
\end{equation}
etc. At the same time, every polynomial $f(T)$ can also be uniquely
decomposed over the algebraically closed field \textbf{C} as follows: 
\begin{equation}
f(T)=c(T-\alpha _{1})^{m_{1}}\cdot \cdot \cdot (T-\alpha _{r})^{m_{r}}. 
\tag{2.2}
\end{equation}
Comparison between Eq.s (2.1) and (2.2) already suggests a close analogy. It
is not our purpose here to reproduce long excerpts from algebra related to
polynomial rings $\mathbf{F}[T]$ over the field $\mathbf{F}$ where $\mathbf{F%
}$ can be either one of the fields above or finite field of $q$ elements. In
the last case one usually writes $\mathbf{F}_{q}[T].$ Evidently, the
function ring $\mathbf{F}[T]$ is analogous to the number ring $\mathbf{Z}$
while the quotients of polynomials taken from the ring $\mathbf{F}[T]$ form
the field $\mathbf{F}(T)$ analogous to $\mathbf{Q}$\textbf{,}etc.
Computationally the function field has many advantages over the number
field. We shall discuss only those which are related to the goals we have in
mind. To this purpose, following Artin [12], we introduce the following

\bigskip

\textbf{Definition 2.1.} A valuation of a field \textbf{F} is a real-valued
function $\left| x\right| $,

defined for all $x\in \mathbf{F}$ satisfying the following requirements:

a) $\left| x\right| \geq 0;$ $\left| x\right| =0$ if and only if $x=0$;

b) $\left| xy\right| $=$\left| x\right| \left| y\right| ;$

c) $\left| x+y\right| \leq \left| x\right| +\left| y\right| $ .\ \ \ \ 

\bigskip

There are \textbf{many} ways to introduce valuations explicitly as explained
in Ref.[47] and we shall take advantage of this fact. Nevertheless, every
valuation is equivalent to a valuation for which the triangular inequality
holds. The triangular inequality can be sharpened for the \textit{%
non-archimedean} valuations. In this case instead of c) one has the
following inequality 
\begin{equation}
\left| x+y\right| \leq \max (\left| x\right| ,\left| y\right| ).  \tag{2.3}
\end{equation}
The valuation is called \textit{trivial} if $\left| x\right| =1$ $\forall
x\in \mathbf{F}^{\ast }$ (\textbf{F}$^{\ast }=\mathbf{F}\setminus 0).$In the
case of field \textbf{Q} it is convenient to choose some prime number $p$ so
that any number $a\in \mathbf{Q,}$ $a\neq 0,$can be presented in the form $%
a=p^{\nu }b$ with numerator and denominator of $b$ being prime to $p$.

\textbf{Definition 2.2}. The standard valuation at $p$ is defined according
to the rule: 
\begin{equation}
\left| a\right| _{p}=p^{-\nu },\text{ where }\nu =ord_{p}(a)=\text{\textbf{%
order} of }a\text{ at\textbf{\ }}p.  \tag{2.4}
\end{equation}
Clearly, with such definition the whole field $\mathbf{Q}$ is divided into
equivalence classes. Within such scheme the usual absolute value $\left|
a\right| \equiv \left| a\right| _{p_{\infty }}$. This rule serves as
definition of a fictitious ''prime at infinity''. The difference between the
prime at infinity and the rest of primes is profound and represents the
major difficulty in number theory as explained by Mazur, Ref. [48], page 15.

Fortunately, for the case of function fields this problem does not exist
altogether since the prime at infinity is treated as any other prime. Let $%
\mathbf{F}[T]$ be a polynomial ring and $\mathbf{F}(T)$ its quotient field
of rational functions. Let $p=p(T)$ be an irreducible polynomial (of degree $%
\geq 1)$ with leading coefficient 1 in $\mathbf{F}[T].$ For each rational
function $\varphi (t)\in \mathbf{F}(T)$ one can write 
\begin{equation}
\varphi (T)=\left( p(T)\right) ^{\nu }\frac{f(T)}{g(T)}  \tag{2.5}
\end{equation}
where $f(T)$ and $g(T)$ $\in \mathbf{F}[T]$ are polynomials and $p\nmid fg.$
By analogy with Eq.(2.4) the standard valuation can be defined now as 
\begin{equation}
\left| \varphi \right| _{p}=\exp \{-\nu \deg p\}.  \tag{2.6}
\end{equation}
In this case the analogue of the absolute value (prime at infinity) is just 
\begin{equation}
\left| \varphi \right| _{\infty }=\exp \{\deg \varphi \}.  \tag{2.7}
\end{equation}
These results can be made physically relevant if one considers the field of
meromorphic functions \ on the Riemann sphere, i.e. on $\mathbf{C+}\infty $.
Let $f(x)$ be such function defined on $\mathbf{C}$\textbf{. }Then, near
some point $x_{0}$ it can be presented as 
\begin{equation}
f(x)=(x-x_{0})^{ord_{x_{0}}f(x)}g(x),  \tag{2.8}
\end{equation}
provided that $g(x_{0})\neq 0$ or $\infty .$ Using these results one can
write 
\begin{equation}
\left| f(x)\right| _{x_{0}}=\exp \{-\nu \}\text{ where }\nu =ord_{x_{0}}f(x).
\tag{2.9}
\end{equation}
If $x_{0}$=$\infty $, one writes $f(x)=\left( \frac{1}{x}\right)
^{ord_{\infty }f(x)}g(\frac{1}{x}),$where $g(0)\neq 0$ or $\infty $, and \
using Eq.(2.7) one defines valuation at infinity as 
\begin{equation}
\left| f(x)\right| _{\infty }=\exp \{\deg f(x)\}  \tag{2.10}
\end{equation}
since in this case $ord_{\infty }f(x)=\deg f(x).$ For the meromorphic
functions 
\begin{equation}
\sum\limits_{x_{o_{i}}}ord_{x_{o_{i}}}^{{}}f(x)=0.  \tag{2.11}
\end{equation}
This result can be interpreted as some sort of conservation law for
Coulombic-like ''charges''. This analogy is not superficial. It will be
discussed in a separate publication.

Eq.(2.11) is valid for both number and function fields and is known
number-theoretically as the product formula 
\begin{equation}
\prod\nolimits_{p}\left| a\right| _{p}=1.  \tag{2.12}
\end{equation}
Taking log of both sides of Eq.(2.12) and using definitions of valuations
one reobtains Eq.(2.11) back as required. From the discussion we had so far
it is clear that the order and the degree have almost the same meaning. In
particular, every element $f(T)\in \mathbf{F}[T]$ has the form $f(T)$=$%
\alpha _{0}T^{n}+\alpha _{1}$\ $T^{n-1}$\ \ +$\cdot \cdot \cdot +\alpha _{n}$%
\ and if $\alpha _{0}\neq 0$ it can be said that $\deg (f)=n$.\ If $f$ and $%
g $ are non-zero polynomials, we have [42,43] 
\begin{equation*}
\deg (fg)=\deg (f)+\deg (g)
\end{equation*}
and 
\begin{equation*}
\deg (f+g)\leq \max (\deg (f),\deg (g)).
\end{equation*}
Thus, in view of the definitions given above, \emph{for the function fields
the valuations are} \textbf{always} \textbf{non-archimedean}.

We would like to take advantage now of the fact that valuations can be
introduced in many ways. In particular, the finite field \textbf{F}$_{p}$
with $p$ elements is obtained as a quotient \textbf{Z}/$p$\textbf{Z}\ \ with 
$p$ being a prime. If the coefficients $\alpha _{i}$ of a polynomial $\ f(T)$
of degree $n\ $\ belong to \textbf{F}$_{p},$\textbf{\ }then the quotient 
\textbf{F}$_{p}[T]/f(T)\equiv \mathbf{GF}(p^{n})\equiv \mathbf{F}_{q}$ forms
the Galois field of $q=$ $p^{n}$ elements (congruence classes), e.g. see
Ref.[49], page 422, and Appendix A of this work. The advantage of working
with finite fields lies in the fact that for each $n$ there is an
irreducible polynomial in \textbf{F}$_{p}[T],$ e.g. see Ref.[49]$,$ page 472$%
,$ and Appendix A of this work$.$ This means that if we use the irreducible
polynomials only, then the degree can be used instead of order. To
facilitate reader's understanding of these facts, a summary of relevant
number theoretic results is provided in the Appendix A. In view of these
results instead of Eq.(2.6) it is more advantageous to use $\left| f\right|
=q^{\deg (f)}$ as valuation and it can be checked [12,42,47] that such
defined valuation obeys axioms a), b) and Eq.(2.3) and, hence, is
non-archimedean.

\subsection{ Zeta functions}

\textbf{Definition 2.3}.The zeta function $\zeta _{\mathbf{F}_{q}}(s)$ of
the function field \textbf{F}$_{q}[T]$ is defined by 
\begin{equation}
\zeta _{\mathbf{F}_{q}}(s)=\sum\limits_{f\in F_{q}[T]}^{{}}\frac{1}{\left|
f\right| ^{s}}\text{ , where }f\text{ is the monic polynomial.}  \tag{2.13}
\end{equation}

Recall [46] that the monic polynomial $f(T)$ is such for which $\alpha
_{0}=1.$ Using Appendix A we find that there are exactly $q^{d}$ monic
polynomials of degree $d$ in \textbf{F}$_{q}$[T]. Hence, one obtains [43] 
\begin{equation}
\sum\limits_{\deg (f)<d}\frac{1}{\left| f\right| ^{s}}=1+\frac{q}{q^{s}}+%
\frac{q^{2}}{q^{2s}}+\cdot \cdot \cdot +\frac{q^{d}}{q^{ds}}  \tag{2.14}
\end{equation}
and, accordingly, 
\begin{equation}
\zeta _{\mathbf{F}_{q}}(s)=\frac{1}{1-q^{1-s}}.  \tag{2.15}
\end{equation}
In view of Eq.(1.23) we obtain as well $\zeta _{\mathbf{F}_{q}}(s)=\zeta
_{q}(s-1).$ Next, we define zeta function over the \textbf{rational}
function field $K$= \textbf{F}(T). For this, we need a notion of the
projecive line \textbf{P}$^{1}(F)$ for the function fields. Since the usual
projective line is just $\mathbf{R}$\textbf{\ }$\cup \{\infty \},$ evidently
for the field $\mathbf{F}_{q}$\textbf{\ }it is $K$= $\mathbf{F}_{q}\cup
\{\infty \}.$ In our case we have two types of polynomials, e.g. see
Eq.s(2.8) and (2.10). Clearly, $\zeta _{\mathbf{F}_{q}}(s)$ is applicable
for the first type. If we include the ''prime at infinity'', we obtain 
\begin{equation}
\zeta _{K}(s)=\frac{1}{1-q^{1-s}}\frac{1}{1-q^{-s}}.  \tag{2.16}
\end{equation}
Let now $h(f)$ be some function (to be discussed explicitly later) defined
on the set of monic polynomials. With help of this function \ we obtain the
following

\textbf{Definition 2.4.} The Dirichlet function $D_{h}(s)$ associated with $%
h(f)$ is given by 
\begin{equation}
D_{h}(s)=\sum\limits_{f\text{ }monic}\frac{h(f)}{\left| f\right| ^{s}}%
=\sum\limits_{n=0}^{\infty }H(n)u^{n}  \tag{2.17}
\end{equation}
where $u=q^{-s}$ and 
\begin{equation}
H(n)=\sum\limits_{\substack{ f\text{ }monic  \\ \deg (f)\leq n}}h(f)\text{ .}
\tag{2.18}
\end{equation}
The function $H(n)$ should possess the multiplicative property, that is $%
H(m)H(n)=H(mn)$. The Dirichlet function $D_{h}(s)$ sometimes is also called
as zeta function $Z_{K}(u)$ associated with the field $K$ [43]. One of the
most fundamental achievements of number theory is summarized in the
following theorem

\textbf{Theorem 2.5}. \textit{Let }$K$\textit{\ be a function field in one
variable }$T$\textit{\ with a finite constant field }\textbf{F}\textit{\
with }$q$\textit{\ elements. Suppose that the genus of }$K$\textit{\ is }$g$%
\textit{. Then there is a polynomial }$L_{K}(u)\in \mathbf{Z}[u]$\textit{\
of degree }$\mathit{2}g$\textit{\ such that} 
\begin{equation}
Z_{K}(u)=\frac{L_{K}(u)}{(1-qu)(1-u)}  \tag{2.19}
\end{equation}

$where$ $L_{K}(u)=$ $\prod\limits_{i=1}^{2g}(1-\alpha _{i}u)$ $with$ $\alpha
_{i}\alpha _{g+i}=q,1\leq i\leq g.$

The proof of this theorem, attributed to A.Weil, can be found in many
places, e.g. see Refs[17,43-45].

For the sake of space we refer our readers to the literature for details of
relation between the genus $g$ and the field $K.$ In short, we can embed $%
f(T)\in \mathbf{F}_{q}\mathbf{[}T\mathbf{]}$ into $n$-dimensional projective
space \textbf{P}$^{n}$(\textbf{F}$_{q})$ that is to convert it into the
homogenous polynomial and to equate such polynomial to zero. The genus is an
invariant of this homogenized polynomial with respect to some known
transformations. All this is beautifully explained by Raoul Bott in Ref.[50].

Combining Eq.s(2.17)-(2.19) produces 
\begin{equation}
Z_{K}(u)=\sum\limits_{n=0}^{\infty }H(n)u^{n}.  \tag{2.20}
\end{equation}
We shall need yet another interpretation of the same results. To facilitate
matters, in view of Eq.(2.15), following Rosen, Ref. [43], we write formally 
\begin{equation}
\frac{1}{1-qu}=\prod\limits_{l=1}^{\infty }(1-q^{-ls})^{-a_{l}},  \tag{2.21}
\end{equation}
where $a_{l}$ is number of monic irreducible polynomials of degree $l$ . To
obtain $a_{l}$ we need to take the logarithmic derivatives with respect to $%
u $ of both sides of Eq.(2.21). Then, multiplying the result by $u$ produces 
\begin{equation}
\frac{qu}{1-qu}=\sum\limits_{l=1}^{\infty }\frac{la_{l}u^{l}}{1-u^{l}}. 
\tag{2.22}
\end{equation}
Finally, upon expansion comparing the coefficients of $u^{l}$ produces the
following result 
\begin{equation}
\sum\limits_{l\mid n}la_{l}=q^{n}  \tag{2.23}
\end{equation}
which provides the missing link between $a_{l}$ and $q$.Looking at
Eqs(2.15),(2.17),(2.20) and (2.21) one can write as well for general case of
genus $g$ similar result: 
\begin{equation}
Z_{K}(u)=\prod\limits_{l=1}^{\infty }(1-u^{l})^{-a_{l}}.  \tag{2.24}
\end{equation}
This time, however, the connection between $a_{l}$ and $q$ is more
complicated. To obtain this connection we have to take the logarithm of both
sides of Eq.(2.24). By expanding this logarithm in a power series in $u$ we
obtain, 
\begin{equation}
\ln Z_{K}(u)=\sum\limits_{m=1}^{\infty }\frac{N_{m}}{m}u^{m},  \tag{2.25}
\end{equation}
Here, by definition, $N_{m}$ is the number of zeros of $\ f(T)$ in\textbf{\ P%
}$^{n}(\mathbf{F}_{q}).$ For the projective curve of genus $g$ associated
with such polynomial, it can be shown [17,43-45] that 
\begin{equation}
N_{m}=q^{m}+1-\sum\limits_{i=1}^{2g}\alpha _{i}^{n}.  \tag{2.26}
\end{equation}
For $g=0$ substitution of this result into Eq.(2.25) reproduces back
Eq.(2.16) as required. It should be noted, however, that the results just
presented are only valid for the so called non-singular curves, that is for
the Riemann surfaces without nodes and holes. For singular surfaces (curves)
the above formalism should be amended as discussed (albeit incompletely) in
Ref.[51].To bring all these results closer to those obtained in section 3 of
our earlier work, Ref.[4], it is helpful to provide the alternative (\textbf{%
dynamical}) interpretation of just obtained results.

To this purpose, we begin with simple observations. Consider a map $%
f:x\rightarrow y$ given by $y=f(x)$. The number of the fixed points of this
map can be easily obtained graphically by plotting together at the same plot
the straight line $y=x$ and $y=f(x).$Following Ref.[52] we define the 
\textit{global Lefshetz number} $\mathcal{L(}f\mathcal{)}$ as the \textbf{%
intersection} number $I$($\Delta ,graph(f))$ where $\Delta $ is diagonal $%
(x,x)$ in the direct product $(x,y)\in X\times Y.$ Next, we need a notion of
the \textit{local} Lefshetz number. It can be easily understood if we recall
how the linearization of dynamical systems is done. For instance, for two
dimensional case we have $f(x)=Ax$ + higher order terms where the dynamical
matrix $A$ can always be brought to the diagonal form and, hence, given by 
\begin{equation*}
A=\left( 
\begin{array}{cc}
\alpha _{1} & 0 \\ 
0 & \alpha _{2}
\end{array}
\right)
\end{equation*}
so that the local Lefshetz index $\mathcal{L}_{0}(f)$ is given by $sign\left[
(\alpha _{1}-1)(\alpha _{2}-1)\right] .$ For a source or a sink both
eigenvalues have the same sign while for a saddle they have the opposite
sign. Clearly, the local Lefshetz number is equal to the index of the vector
field, e.g. see our earlier work, Ref.[6], for illustrations and further
details. Hence, 
\begin{equation}
\mathcal{L(}f\mathcal{)=}\sum\limits_{f(x)=x}\mathcal{L}_{x}\mathcal{(}f%
\mathcal{)}  \tag{2.27}
\end{equation}
and this is yet another way of writing the Poincare-Hopf index theorem
extensively used in our work, Ref.[6]. In this reference the index $%
I(x)\equiv \mathcal{L}_{x}\mathcal{(}f\mathcal{)}$ of singularity \ at point 
$x$ was defined as 
\begin{equation}
I(x)=\frac{1}{2\pi }\func{Im}\oint\limits_{\mathcal{C}}\frac{df}{f} 
\tag{2.28}
\end{equation}
with $\mathcal{C}$ being a unit circle. This definition is valid for two
dimensional case only. It is easy to extend this result to higher
dimensions. Following Ref.[52] the local Lefshetz number can be identified
as well with the degree ($\deg $) of mapping $\varphi :$ 
\begin{equation}
z\rightarrow \frac{f(z)-z}{\left| f(z)-z\right| }.  \tag{2.29}
\end{equation}
More accurately, suppose $z$ is an isolated fixed point of \ $f(x)$ in 
\textbf{R}$^{k}$ (any manifold locally is \textbf{R}$^{k}).$ If $\mathcal{B}$
is a small closed ball centered at $z$ that contains no other fixed points,
then the map $\varphi $ is a smooth map defined by $\varphi :\partial 
\mathcal{B}\rightarrow S^{k-1}.$ For $k=2$ this is just mapping of one
circle onto another (or to itself).The number of times one circle winds
around another (the winding number at the point $z$) is $\deg _{z}\varphi .$
In view of these remarks, \ such definition of the degree clearly coincides
with Eq.(2.28) in two dimensions. Hence, $I(z)\equiv \mathcal{L}_{z}\mathcal{%
(}f\mathcal{)}=\deg _{z}\varphi .$ \ \ 

These results can be easily generalized now as follows. Following
Milnor[53], let us consider sequence of iterates $f^{2}(z)=f(f(z))$, $%
f^{3}(x)=f(f(f(z)))$, etc. and their fixed points. Clearly, we will obtain
the associated Lefshetz numbers $\mathcal{L(}f^{2}\mathcal{)},\mathcal{L(}%
f^{3}\mathcal{)},...,$ and, accordingly, we can construct the zeta function 
\begin{equation}
\ln Z_{f}(u)=\sum\limits_{m=1}^{\infty }\frac{\mathcal{L(}f^{m}\mathcal{)}}{m%
}u^{m}  \tag{2.30}
\end{equation}
whose meaning should be clear. Obtained results are to be used in the next
section.

\section{Variations on a theme by Fermat}

\subsection{Circle maps, function fields and the Veneziano amplitude}

We would like to extend connections between the number theory and the theory
of dynamical systems in order to reexamine results obtained earlier in
Ref.[4]. Let us begin with the observation that every polynomial is
completely defined by its coefficients $c_{i}$ (Appendix A).Suppose now that
these coefficients belong to the field \textbf{F}$_{p}$ and consider
multiplication of such polynomials. This multiplication can be performed as
follows. First, we remove the requirement that $p$ is the prime number.
Next, we introduce the space $\Omega _{N}$ (where $N\geq 2$ replaces $p$) 
\begin{equation}
\Omega _{N}\text{ :=\{}\omega =(c_{0},c_{1,}c_{2},...)\mid c_{i}\in \lbrack
0,1,...,N-1]\text{ for }i\in Z\}  \tag{3.1}
\end{equation}
of one sided sequences of $N$ symbols which is in one-to-one correspondence
with the set of polynomials. Multiplication of polynomials simply replaces $%
\omega $ by $\omega ^{\prime }$=$(c_{0}^{^{\prime }},c_{1,}^{^{\prime
}}c_{2}^{\prime },...)$ with coefficients $c_{i}^{\prime }\in \lbrack
0,1,...,N-1].$ That is polynomial multiplication corresponds to some
permutation in the sequence of $c_{i}^{^{\prime }}s$ i.e. to the
automorphism $:$ $\Omega _{N}\rightarrow \Omega _{N}.$ Such procedure can be
computerized and serves as coding/decoding system (cryptography) for
transmission of information [44,45]. A particular realization of such
automorphism known as $one$ $sided$ $N$-shift can be defined as follows 
\begin{equation}
\sigma _{N}^{R}\text{ : }\Omega _{N}^{R}\rightarrow \Omega _{N}^{R}\text{ \
, }\sigma _{N}^{R}\omega =\omega ^{\prime }=(c_{1}^{\prime },c_{2,}^{\prime
}c_{3}^{\prime },...).  \tag{3.2}
\end{equation}
The upper script $R$ stands for the ''right''. Surely, one can consider the
iterates of such defined shift, and, accordingly, one can look for the fixed
(periodic) points (orbits) for such iterates.

\textbf{Definition 3.1}. The periodic orbits for the one sided \ shift are
the periodic sequences $\left( \sigma _{N}^{R}\right) ^{m}\omega =\omega $
if and only if $c_{m+n}=c_{n}$.Every periodic sequence is uniquely
determined by its basis \ $c_{0},c_{1,}c_{2},...,c_{n-1\text{ }}.$

Using results of Appendix A we conclude that there are $N^{n\text{ }}$%
periodic sequences. In particular, if $N$ is the prime number we get $p^{n}$
periodic sequences. This is the number of irreducible monic polynomials as
discussed in Section 2. \emph{Hence, the dynamical zeta} \emph{function for
the one sided shifts coincides with that given by Eq.(2.15)}. This result is
in accord with that obtained in Ref.[54] ( e.g. see page 107, Eq.(3.1.5)) by
slightly different set of arguments. Being armed with such result we are
ready now to accomplish much more.

To this purpose we would like to remind our readers some results from the
Nielsen-Thurston theory of surface homeomorphisms. As we have discussed in
our previous works, Refs.[4-6], 2+1 gravity can be considered essentially as
physically reformulated Nielsen-Thurston theory. For convenience of our
readers the key points of this theory are summarized in the Appendix B. In
this subsection we will develop our formalism without interruption referring
to the Appendix B whenever it is necessary. In particular, using Proposition
B.2., it should be clear that any surface homeomorphism can be lifted to the
unit disk model of \textbf{H}$^{2}$. Then, the surface dynamics becomes
dynamics of maps of the circle. Using Appendix B and Ref.[55] it follows
that it is sufficient to restrict ourself by the periodic maps. If this is
the case, then the circle $S^{1}$can be identified with \textbf{F}$_{m}=%
\mathbf{Z}/m\mathbf{Z}$ and, if we are interested in the invertible
transformations, then the number $m$ should be some prime, i.e. $m=p$.
Hence, we arrive \ once again at the cyclotomic fixed point equation,
Eq.(A3), written now as 
\begin{equation}
f:x^{p}=x.  \tag{3.3}
\end{equation}
If we ignore the trivial solution $x=0,$ it can be rewritten as 
\begin{equation}
x^{p-1}=1  \tag{3.4}
\end{equation}
and will possess $p-1$ nontrivial fixed points. $n$ times iteration of
Eq.(3.3) produces 
\begin{equation}
f^{n}:x^{p^{n}}=x  \tag{3.5}
\end{equation}
and, again, it will have $p^{n}-1$ non trivial fixed points so that the
Lefshetz number associated with the degree of such mapping is $\mathcal{L}%
_{x}\mathcal{(}f^{n}\mathcal{)}=p^{n}-1$ in view of Eq.(3.4). This result
coincides with that obtained by different methods in Ref.[54], Proposition
8.2.4. Use of this number in Eq.(2.30) produces at once 
\begin{eqnarray}
Z_{f}(t) &=&\exp \left( \sum\limits_{n=1}^{\infty }\frac{p^{n}-1}{n}%
t^{n}\right)  \TCItag{3.6} \\
&=&\exp \{-\ln (1-pt)+\ln (1-t)\}  \notag \\
&=&\frac{1-t}{1-pt}.  \notag
\end{eqnarray}
Finally, by replacing $t$ with $p^{-s}$ \ in Eq.(3.6) we obtain the $p$-adic
version of the Veneziano amplitude, Eq.(1.16b), discussed in Section 1.

\subsection{From Brieskorn-Pham to Fermat and back}

Previous result, as good as it is, by itself \ provides no clues about its
extension to the multiparticle amplitudes. In this subsection we are going
to make the first step towards this goal. The rest of the paper depends
crucially on this first step.

To begin, consider a set of polynomials $P_{B-P}$ of the type 
\begin{equation}
P_{B-P}:f(\mathbf{z})=z_{0}^{a_{0}}+z_{1}^{a_{1}}+\cdot \cdot \cdot
+z_{n}^{a_{n}}  \tag{3.7}
\end{equation}
defined in $\mathbf{C}^{n+1}$ with $a_{0},...,a_{n}$ being an $n+1$ tuple of
positive integers greater than or equal to 2. The polynomial $f(z)$ will be
considered as the map $f$: $\mathbf{C}^{n+1}$ $\rightarrow \mathbf{C}.$\ The
Brieskorn-Pham (B-P) variety \ $V_{B}(f)$ is defined now as [25-27] 
\begin{equation}
V_{B-P}(f)=\{\mathbf{z}\in C^{n+1}\mid f(\mathbf{z})=0\}.  \tag{3.8}
\end{equation}
This variety has singularity only at one point: $\mathbf{z}=0$. The Fermat
variety is a special case of the B-P variety for which $a_{0}=a_{1}=\cdot
\cdot \cdot =a_{n}=a.$

\textbf{Remark 3.2. }The\textbf{\ }Fermat variety is the Calabi-Yau variety
when $a$=$n+1,$ Ref.[56].

\textbf{Corollary} \textbf{3.3.} Because of this observation it is clear
that if the Fermat variety has something to do with physics, then the
methods used in mirror symmetry calculations [57] in physics can be applied
directly to the Fermat varieties as well.

Fortunately this \textbf{is} the case as will be demonstrated in this
subsection. Our goals are broader than just to establish this correspondence
since we want to create an additional links with number and knot theories
which, to our knowledge, are absent in the existing treatments [57].

Consider an intersection $\mathbf{K}$=$V_{B-P}(f)\cap $ $S^{2n+1}$ (this
intersection is actually a knot or link as it will become apparent from the
discussed below) of the B-P variety with the boundary of $2n+2$ dimensional
ball $\mathcal{B}$ \ and define the mapping 
\begin{equation}
\phi (\mathbf{z})=\frac{f(\mathbf{z})}{\left| f(\mathbf{z})\right| }. 
\tag{3.9}
\end{equation}
It is clear that $\phi (\mathbf{z})$ is again the circle map of the type
discussed in the previous subsection. This time, it is more than just this
map since, actually, the mapping \ $\phi :$ $S^{2n+1}\backslash \mathbf{K}%
\rightarrow S^{1}$ defines a smooth fiber bundle Ref.[58], Lemma 19.10, with
base a circle $S^{1}$and a fiber the complement of a knot/link $\mathbf{K}$
in $S^{2n+1}$. Earlier, in Ref.[4], Section 3, we discussed in detail the
mapping torus construction of 3-manifolds fibering over the circle. We would
like to argue here that for the Brieskorn variety associated with the
polynomial $f(z)=z_{0}^{a_{0}}+z_{1}^{a_{1}}$ we can recover back results
obtained earlier.

To this purpose let us recall the mapping torus construction first. Consider
an orientation preserving surface homeomorphism $h:S\rightarrow S$. In the
case of punctured (holed) torus this homeomorphism should respect the
presence of a hole. The circumference of the hole is our base space $S^{1}($%
e.g. read Ref.[59]) which is our knot $\mathbf{K,}$ since the knot is just a
circle embedded into $S^{3}.$The Seifert surface of the knot is a fiber and
the 3-manifold is just a fiber bundle constructed in a following way. Begin
with the product $S\times 0$ (the initial state) and S$_{h}\times 2\pi $
(the final state) so that for each point $x\in S$ we have $(x,0)$ and $%
(h(x),2\pi )$ respectively. The interval $I=(0,2\pi )$ can be closed (to
form a circle $S^{1})$ by identifying $0$ and $2\pi $ causing the
identification: 
\begin{equation}
i\text{ : }(x,0)=(h(x),2\pi ).  \tag{3.10}
\end{equation}
The mapping torus $T_{h}$ fiber bundle is just the quotient 
\begin{equation}
T_{h}=\frac{S\times I}{i}.  \tag{3.11}
\end{equation}
It is 3-manifold which fibers over the circle and is complementary to the
fibered knot (lying at the boundary of such (cusped) 3-manifold as explained
in Section 5 of Ref.[4]).

In the present case let us consider the monodromy map $h(z)$ given by 
\begin{equation}
h_{t}(z_{0},...,z_{n})=(\exp \{it/a_{0}\}z_{0},...,\exp \{it/a_{n}\}z_{n}). 
\tag{3.12}
\end{equation}
If the point $z_{0},...,z_{n}$ belongs to the variety $V_{B-P}(f)$ then, the
identification analogous to that given by Eq.(3.10) takes place by
construction. And, hence, the rest of the fiber bundle related arguments
also goes through. Therefore, indeed, the inverse map $\phi ^{-1}(\mathbf{z}%
) $ produces the fiber bundle known as Milnor fibration. It is $n+2$
dimensional manifold embedded into $S^{2n+1}.$ Let $n=1$, then we are
dealing with the usual knots embedded in $S^{3}$ and their complementary $3-$
manifolds$.$ It is of interest for us to reobtain the Alexander polynomial $%
\Delta _{\mathbf{K}}$ for such knots. This task can be accomplished if the
monodromy matrix $\mathbf{M}$ is known as it is explained in detail in our
earlier work, Ref.[4]. Then, the polynomial $\Delta _{\mathbf{K}}$ is given
by 
\begin{equation}
\Delta _{\mathbf{K}}(t)=\det (t\mathbf{E}-\mathbf{M})  \tag{3.13}
\end{equation}
with $\mathbf{E}$ being the unit matrix. Although we had discussed in
Ref.[4] how the matrix $\mathbf{M}$ can be obtained, here we need yet
another way of obtaining this matrix. To this purpose following Pham [25]
and further explained by Milnor [26], let us look at Eq.(3.9) and select
such fiber for which 
\begin{equation}
\phi (\mathbf{z})=\frac{f(\mathbf{z})}{\left| f(\mathbf{z})\right| }=1. 
\tag{3.14}
\end{equation}
This can be achieved, for example, by requiring $\ f(\mathbf{z})=1$ or 
\begin{equation}
z_{0}^{a_{0}}+z_{1}^{a_{1}}+\cdot \cdot \cdot +z_{n}^{a_{n}}=1.  \tag{3.15}
\end{equation}

\textbf{Remark 3.4.} For $n=1$ and $a_{0}=a_{1}=m$ this is just the standard
Fermat equation written in the affine form.

Following Pham again, let us replace $z_{i}^{a_{i}}$ by $t_{i}$ so that we
get $\sum\nolimits_{i=0}^{n}t_{i}=1.$ If we impose an extra requirement : $%
t_{i}\geq 0,$ then such coordinates become the barycentric coordinates for
simplicial complexes used in algebraic topology for homological calculations
[60]. It is clear, that the only choice for $z_{i}^{a_{i}}$ to become
complex while still to obey Eq.(3.15) is to replace $t_{i}$ by $z_{i}=$ $%
t_{i}^{\tfrac{1}{a_{i}}}\exp (\pm \frac{2\pi i}{a_{i}})$ .

Next, looking at Eq.s(3.12) and (3.15) let us consider the homeomorphism $%
h_{2\pi }:$ $\phi ^{-1}(1)\rightarrow \phi ^{-1}(1)$ , i.e. 
\begin{equation}
h_{2\pi }(\mathbf{z})=(\exp \{i2\pi /a_{0}\}z_{0},...,\exp \{i2\pi
/a_{n}\}z_{n}).  \tag{3.16}
\end{equation}
This homeomorphism can be represented as a composition of homeomorphisms
acting on the homology basis: one generator for one basis element. The basis
elements are just circles, more exactly, the quotients $\omega \in $ \{%
\textbf{Z}/$a_{0}\mathbf{Z,...,}$\textbf{Z}/$a_{n}\mathbf{Z\}.}$ The
generators $r_{a}$ are associated with rotations, i.e. 
\begin{equation}
r_{a_{i}}(\omega )=\exp (\pm \frac{2\pi i\nu }{a_{i}})\omega \equiv \zeta
^{\nu }\omega  \tag{3.17a}
\end{equation}
with $\nu $ being in the range $1\leq \nu \leq $ $a_{i}-1$.\ Hence, $\zeta
^{\nu }$\ \ is just an eigenvalue of the rotation (homology) operator. The
reader should consult Pham [25] and Milnor [26] for more details. For the
case of torus we have just two elements in the homology basis so that we
have 
\begin{equation}
\lbrack r_{a_{1}}\otimes r_{a_{2}}](\omega _{1},\omega _{2})=\xi ^{\nu }\eta
^{\mu }(\omega _{1},\omega _{2}).  \tag{3.17b}
\end{equation}
With help of above information the Alexander polynomial can be written now
according to Milnor [26] (and Brieskorn [27] ) as 
\begin{equation}
\Delta _{\mathbf{K}}(t)=\prod\limits_{\substack{ \omega _{1}^{p}=1,\omega
_{2}^{q}=1  \\ \ \omega _{1},\omega _{2}\neq 1}}(t-\omega _{1}\omega _{2}), 
\tag{3.18}
\end{equation}
where all Greek letters denote the corresponding roots of unity.

To get a feeling of this result, let us consider the simplest example of the
trefoil knot $\mathcal{T}$. It was discussed in our earlier work, Ref.[4],
and it is a typical example of torus knot [59,61]. More specifically, it is
2,3-type of torus knot. Therefore, for such knot we have two ''homological''
cyclotomic equations: $\omega ^{2}=1$ and $\omega ^{3}=1.$The first produces
just one relevant root $\omega _{1}=-1$, while the second has two: $\omega
_{2}=\rho =\frac{1}{2}(-1\pm \sqrt{-3}),$ and the complex conjugate $\bar{%
\rho}.$The Alexander polynomial, Eq.(3.18), can be calculated now
momentarily, 
\begin{equation}
\Delta _{T}=(t+\rho )(t+\bar{\rho})=t^{2}-t+1.  \tag{3.19}
\end{equation}
The result coincides with earlier obtained, Eq.(3.9) of Ref.[4], as
required. Earlier, in Ref.[4], we argued, in accord with Ref.[59], that
there are only two fibered knots for Seifert surfaces of genus one. The
figure 8 knot is the only knot other than trefoil which has punctured torus
as its Seifert surface. The Alexander polynomial for the figure 8 knot 
\textbf{cannot} be obtained from the polynomial given by Eq.(3.18) since the
eigenvalues of the monodromy matrix always have modulus equal to one by
construction [62]. We had explained in Ref.[4] that at least one of the
eigenvalues of the monodromy matrix should be strictly larger than one in
order to reproduce the figure 8 Alexander polynomial. The question arises:
what happens if we use numbers other than 2 and 3 describing the trefoil ?
This issue was also addressed by Milnor [26] and Brieskorn [27]. The
resolution of the apparent paradox lies in considering the Alexander
polynomial for \textbf{links} instead of single knots. But this should be
done with some care since links may require multicomponent Alexander
polynomial for their description. This topic is discussed further in section
5. In principle, the beauty of this approach to knots and links lies in the
fact that it is not restricted to knots and links embedded in $S^{3}.$
Multidimensional analogs of knots and links can be readily considered, e.g.
with help of the generalized Alexander polynomial 
\begin{equation}
\Delta (t)=\prod\limits_{\omega _{k}^{i_{k}}=1;\text{ }\omega _{k}\neq
1.}(t-\omega _{0}\omega _{1}\cdot \cdot \cdot \omega _{n}),  \tag{3.20}
\end{equation}
where $k=0-n$ and each $i_{k}$ runs between $0<i_{k}$ \TEXTsymbol{<} $a_{k}$%
.This more general case is relevant for calculations associated with
multiparticle Veneziano amplitudes. It is discussed in Section 5.

Being armed with these results we would like to reobtain them now in another
way. \emph{This} \emph{will bring us new physical interpretation of
mathematically known results}. Let us begin with the Fermat equation written
in its standard form 
\begin{equation}
x^{N}+y^{N}=z^{N}.  \tag{3.21}
\end{equation}
For $N=2$ this is just the Pythagorean equation. The problem of finding all
integer solutions for this equation is known as problem about finding the
Pythagorean triples. We had discussed this problem in Section 2 of Ref.[4].
This problem is completely solvable. For $N>2$ the problem of proving that
the above equation has only trivial solutions (the Fermat's last theorem)
was solved by Andrew Wiles only in 1995 [63].Some physically relevant
results dependent on his proof are discussed in Section 4.3. At the same
time, for finite fields the above equation has many solutions. To realize
that such Fermat equation has actually many solutions it is sufficient to
look at solutions of cyclotomic equations (using results on Appendix A) 
\begin{equation}
x^{N}=x,\text{ }y^{N}=y,\text{ }z^{N}=z.  \tag{3.22a}
\end{equation}
With their help Eq.(3.21) can be replaced by the linear equation in the
projective space $\mathbf{CP}^{2}$%
\begin{equation}
x+y=z.  \tag{3.22b}
\end{equation}
Eq.(3.22b) represents a hyperplane in such space. This hyperplane can be
embedded into the Grassmanian associated with such projective space. This is
done with help of Pl\"{u}cker embedding \ so that each hyperplane (that is
the particular solution of the Fermat's equation) produces a point in the
Grassmanian. The situation in the present case becomes analogous to that we
had discussed earlier in connection with the Witten-Kontsevich model,
Ref.[64]. With each point in the Grassmanian associated particular Veneziano
amplitude as we shall demonstrate below and in Section 5. Mathematically,
such an amplitude is interpreted as \textbf{one of the} \textbf{periods} of
the Fermat's curve. Section 5 provides generalization of this result to
Fermat hypersurfaces.

In his fundamental work, Ref.[65], Andre Weil had made estimates of number
of solutions \ of Eq.(3.21) in cyclotomic fields and came up with his famous
zeta function, Eq.(2.19). Below, in the next subsection, we shall
demonstrate additional \ physical uses of such zeta function. For the time
being, we would like to focus attention of our readers on issues of
immediate relevance.

In particular, Eq.(3.21) is written in the so called \emph{projective} form.
In such form it represents the Fermat curve in the complex projective space $%
\mathbf{PC}^{2}$. It may be also convenient to write it in the $\emph{affine}
$ form, i.e. 
\begin{equation}
x^{N}+y^{N}=1,  \tag{3.23}
\end{equation}
by dividing both sides of Eq.(3.21) by $z^{N}.$ For $N>2$, following Gross
[34], we denote such affine Fermat curve as $F(N)$. Clearly, $F(N)$ is just
a special case of Eq.(3.15). This time, however, our treatment of Eq.(3.23)
is going to be different. From the algebraic geometry it is known [66] that
genus $g$ of $F(N)$ is equal to $\frac{1}{2}(N-1)(N-2).$ That is the Fermat
curve is the Riemann surface of genus $g$ . We associate with such surface
its Jacobian \ $J(N)$ and subsequently we compute, following Gross and
Rohrlich [34], the periods of differential 1-forms (to be constructed
momentarily) on $J(N)$. Let us call a double of integers $(r,s)$ \textit{%
admissible} if $1\leq r,s\leq N-1$ (e.g. compare with Eq.(3.17a))$.$ To any
admissible double we associate the differential 
\begin{equation}
\eta _{r,s}=x^{r-1}y^{s-1}\frac{dx}{y^{N-1}}  \tag{3.24}
\end{equation}
of the \textbf{second} kind ''living'' on $F(N).$ If $\zeta $ is a primitive 
$N^{th}$ root of unity, let $A$ and $B$ be the automorphisms of the curve
given by the formulas 
\begin{equation}
A(x,y)=(\zeta x,y)\text{, }B(x,y)\text{=}(x,\zeta y).  \tag{3.25}
\end{equation}
The group generated by $A$ and $B$ \ acts naturally on $\eta _{r,s}.$ The
forms $\eta _{r,s}$ are the \textbf{eigenforms} for this action, with
eigenvalues $\zeta ^{rj+sk}$\ . That is\footnote{%
See Section 5.3.2. for additional details} 
\begin{equation}
\left( A^{j}B^{k}\right) \eta _{r,s}=\zeta ^{rj+sk}\eta _{r,s}.  \tag{3.26}
\end{equation}
This result should be compared with Eq.(3.17b). Based on the result by
Rohrlich (Appendix of Ref.[34]) it is possible to show that this similarity
is not just coincidental. For the sake of space we only provide a sketch of
an argument leaving details aside. These details are given in Refs[35, 67].

The affine version of the Fermat curve, Eq.(3.23), can be parametrized as
follows 
\begin{equation}
x=t,\text{ }y=(1-t^{N})^{\tfrac{1}{N}}.  \tag{3.27}
\end{equation}
Such parametrization naturally limits the range of $t$ to the segment $[0,1]$%
. Hence, we have effectively the mapping $\gamma _{0}:[0,1]\rightarrow F(N)$
of the segment $[0,1]$ to the Fermat curve. By combining this result with
Eq.(3.24) we obtain, 
\begin{equation}
I_{r,s}=\int\limits_{\gamma _{0}}x^{r-1}y^{s-1}\frac{dx}{y^{N-1}}%
=\int\limits_{0}^{1}t^{r-1}(1-t^{N})^{\tfrac{s-1}{N}}\frac{dt}{(1-t^{N})^{%
\tfrac{N-1}{N}}}=\frac{1}{N}B(\frac{r}{N},\frac{s}{N}).  \tag{3.28}
\end{equation}
Let $\frac{r}{N}=a,\frac{s}{N}=b,$ then we obtain the nonsymmetrized
Veneziano amplitude, Eq.(1.2), 
\begin{equation}
NI_{r,s}=\frac{\Gamma (a)\Gamma (b)}{\Gamma (a+b)}.  \tag{3.29}
\end{equation}
Clearly, the symmetrized Veneziano amplitude can be obtained now without any
problems.

\textbf{Remark 3.5}.The numbers $a,b$ and $c=1-a-b$, can be made negative in
accord with Veneziano formula, Eq.(1.6).The details are explained in Section
5.

To obtain the symmetrized amplitude and also to make connection with knots
and links ( in view of Eq.s(3.17b), (3.26)), we need to make few additional
steps. Following Lang [35], let $\zeta =\exp (\frac{2\pi i}{N}).$ Using the
affine version of the Fermat curve, Eq.(3.23), it can be shown that the
chain 
\begin{equation}
\kappa =\gamma _{0}-(1,\zeta )\gamma _{0}+(\zeta ,\zeta )\gamma _{0}-(\zeta
,1)\gamma _{0},  \tag{3.30}
\end{equation}
where $(1,\zeta ),etc.$ have the same meaning as in Eq.(3.25) and $\kappa $
is a closed path, i.e. a cycle on the Fermat curve. When lifted to the
Jacobian $J(N)$ such a curve becomes one of the toral periods. Hence, using
Eq.(3.30) and making trivial changes of variables one finds 
\begin{equation}
\int\limits_{\kappa }\eta _{r,s}=(1-\zeta ^{r})(1-\zeta ^{s})\frac{1}{N}B(%
\frac{r}{N},\frac{s}{N}).  \tag{3.31}
\end{equation}
Since 
\begin{equation*}
B(\frac{r}{N},\frac{s}{N})=\frac{\Gamma (\frac{r}{N})\Gamma (\frac{s}{N})}{%
\Gamma (\frac{r}{N}+\frac{s}{N})}=\frac{1}{\pi }\Gamma (\frac{r}{N})\Gamma (%
\frac{s}{N})\Gamma (\frac{t}{N})\sin (\pi \frac{r+s}{N}),
\end{equation*}
where $t=N-r-s$, we can rewrite Eq.(3.31) (up to a constant) in a manifestly
symmetric form: 
\begin{equation}
\int\limits_{\kappa }\eta _{r,s,t}=-\zeta ^{\frac{r}{2}}\zeta ^{\frac{s}{2}%
}\Gamma (a)\Gamma (b)\Gamma (c)[\sin \pi a\sin \pi b\sin \pi c].  \tag{3.32}
\end{equation}
Finally, by multiplying both sides by $\zeta ^{\frac{t}{2}}$\ \ we obtain
(up to constant again) 
\begin{eqnarray}
\zeta ^{\frac{t}{2}}\int\limits_{\kappa }\eta _{r,s,t} &=&\Gamma (a)\Gamma
(b)\Gamma (c)[\sin \pi a\sin \pi b\sin \pi c]  \TCItag{3.33} \\
&\doteq &\Gamma (a)\Gamma (b)\Gamma (c)[\sin 2a+\sin 2b+\sin 2c].  \notag
\end{eqnarray}
The r.h.s. of Eq.(3.33) looks almost the same as the Veneziano amplitude
Eq.(1.12). It will have exactly the same particle spectrum. Surely, it can
be made to look exactly the same if we use the symmetrized form of
Eq.(3.28). Accordingly, we obtain the following

\textbf{Corollary 3.6.}With accuracy up to a root of unity\textbf{, }the%
\textbf{\ }Veneziano amplitude is just a period of the Jacobian variety for
the Fermat curve.

The meaning of the phase factors $\zeta ^{r}$ , $\zeta ^{s}$ will be
explained from another point of view in Section 5.

Next, it can be shown [34,35] that 
\begin{equation}
\int\limits_{A^{j}B^{k}\kappa }\eta _{r,s}=\int\limits_{\kappa }\left(
A^{j}B^{k}\right) \eta _{r,s}=\zeta ^{rj+sk}\int\limits_{\kappa }\eta _{r,s}
\tag{3.34}
\end{equation}
and, therefore, the periods of the Fermat curve are eigenfunctions of the
tensor product of the rotational (monodromy) operators, e.g. see Eq.s
(3.17a),(3.17b), acting on the homology basis generated by closed $\kappa $
contours. From here we obtain yet another

\textbf{Corollary 3.7}. Because the parameters $r$ and $s$ in Eq.(3.31) \
have the same meaning as $p$ and $q$ in Eq.(3.18) for the Alexander
polynomial, the systematics of particle mass spectrum is in one-to one
correspondence with the systematics of torus knot/links described by the
Alexander polynomial, Eq.(3.18). More general polynomial, Eq.(3.20),
describing multidimensional knots/links is discussed further in Section 5%
\footnote{%
In Section 5 we argue that, actually, it is more advantageous to relate the
particle mass spectrum to the Hodge spectrum (defined in Section 5 and
Appendix D)}.

Relevance and connections of these results with knots/links will be
discussed further in Section 5 devoted to calculation of the multiparticle
Veneziano amplitudes. In the meantime, we would like to discuss some
additional links between the number theory and physics by reinterpreting
mathematical work of Koblitz and Gross, Ref. [68], in physical terms.

\subsection{Physical applications of Gross-Koblitz results}

In Section 2 we introduced Weil's zeta function, Eq.(2.19). From the
associated discussion it should be clear that this function is effective
measure of complexity of Riemann surfaces of genus $g\geq 1$ as compared to
the ''round sphere''. This complexity is reflected in the polynomial $%
L_{K}(u)=$ $\prod\limits_{i=1}^{2g}(1-\alpha _{i}u)$ whose inverse roots $%
\alpha _{i}$ carry all information about this complexity. The task of
finding these roots was left unaccomplished in Section 2. Although the task
of explicit finding of $\alpha _{i}^{\prime }s$ can be rather complicated
for arbitrary algebraic curve, for the Fermat curves this task can be
brought to completion as was demonstrated by Gross and Koblitz, Ref.[68],
see also [69]. In order to connect their results with the results discussed
earlier in this paper let us consider the simplest nontrivial case of the
elliptic curve, i.e. the case when $g=1$. Then Eq.(2.19) is reduced to 
\begin{equation}
Z_{K}(u)=\frac{(1-\alpha u)(1-qu/\alpha )}{(1-qu)(1-u)}.  \tag{3.35}
\end{equation}
It should be clear that $\alpha $ and $q/\alpha $ are associated with
homology basis for torus. They might me thought as two periods of the
elliptic curve, more exactly two q-adic periods (as explained below in
Section 4.3). To bring all this ''down to earth'' consider, following Landau
and Lifshitz, Ref.[70], a simple problem about dynamics of heavy pendulum.
The period $T$ of such pendulum is determined essentially by the complete
elliptic integral of the first kind : $T=4K(k),$ where $k$ is known
parameter and 
\begin{equation}
K(k)=\int\limits_{0}^{\frac{\pi }{2}}\frac{dx}{\sqrt{1-k^{2}\sin ^{2}x}}. 
\tag{3.36}
\end{equation}
Let, for example, $k=1/\sqrt{2}.$ Then, it can be shown [72] that 
\begin{equation}
K(1/\sqrt{2})=\frac{\sqrt{2}}{4}\frac{\Gamma (1/4)\Gamma (1/2)}{\Gamma (3/4)}%
=\frac{\sqrt{2}}{4}B(1/4,1/2)  \tag{3.37}
\end{equation}
to be compared with Eq.(3.28).

Two questions immediately arise: a) is this a pure coincidence that
mathematical expression for the period of the pendulum coincides with that
obtained earlier for the Fermat curve? b) can heavy pendulum possess the
second period and if it can what it means physically? To answer the first
question we invoke the important result of Gerhard Frey [71] who discovered
that the arithmetical properties of the elliptic curve $E_{A,B,C}$ defined
by the Weierstrass equation 
\begin{equation}
y^{2}=x(x-A)(x+B)  \tag{3.38}
\end{equation}
are related to the arithmetic (diophantine) properties of the Fermat
equation 
\begin{equation}
a^{p}+b^{p}+c^{p}=0  \tag{3.39}
\end{equation}
which is reduced to equation 
\begin{equation}
A+B+C=0  \tag{3.40}
\end{equation}
if $A=a^{p},B=b^{p}$and $C=c^{p}.$ This observation paved the way for
proving the Fermat last theorem by Wiles [63].To answer the second
question(s) we follow Mark Kac (as described in Ref.[72], page 77). \ The
argument goes as follows. Suppose we change time in Newton's equation for
pendulum : $t\rightarrow \sqrt{-1}t.$ Such substitution is equivalent to
reversing of direction of the gravitational force and leads to a
complementary periodic motion of the pendulum. That is the solution of the
pendulum problem has not only its obvious \textbf{real} temporary period but
also \textbf{purely imaginary} period as well. It is given explicitly by $%
\sqrt{-1}K$($k^{\prime })$ where $k^{\prime }=\sqrt{1-k^{2}}.$ For $k=1/%
\sqrt{2}$ we get $k=k^{\prime }$ so that the ratio of two periods is just $%
\sqrt{-1}.$ But $\sqrt{-1}$ is the Gaussian prime (e.g.read our earlier
work, Ref.[4]) that is it plays the same role as $q$ in Eq.(3.35) and,
therefore, indeed, our conjecture about the meaning of $\alpha ^{\prime }s$
in Eq.(3.35) is apparently correct. Gross and Koblitz went much further in
their analysis. We believe, that it is worthwhile to discuss their results
in some detail since there is some physics associated with them.

To accomplish this task we need to introduce some definitions. For more
details reader is urged to read pedagogically written book by Winnie Li
[17]. Let \textbf{F}$_{q}$ be a finite field. For $a\in $\textbf{F}$_{q}$
define the \textbf{additive} character $\psi ^{a}$of the field \textbf{F}$%
_{q}$ as follows: $\psi ^{a}(x)=\psi (ax)$ $\forall x\in $\textbf{F}$_{q}.$%
The \textbf{multiplicative} character $\chi (x)$ is determined by the
cyclotomic equation, Eq.(A.5). It is just one of the q-th \ nontrivial roots
of unity and there are $q-1$ of them. With help of these characters one can
construct the Gaussian and the Jacobi sums. In particular, the Gauss sum is
defined as 
\begin{equation}
g(\chi ,\psi )=-\sum\limits_{x\in \mathbf{F}_{q}^{\ast }}\chi (x)\psi (x) 
\tag{3.41}
\end{equation}
where \textbf{F}$_{q}^{\ast }$\ =\textbf{F}$_{q}\setminus 0$,\ \ while the
Jacobi sum is defined as 
\begin{equation}
J(\chi _{1},\chi _{2})=-\sum\limits_{\substack{ x\in \mathbf{F}_{q}^{{}}  \\ %
x\neq 0,1}}\chi _{1}(x)\chi _{2}(1-x).  \tag{3.42}
\end{equation}
The Gauss sum is analog in \textbf{F}$_{q}$ of the gamma function 
\begin{equation}
\Gamma (s)=\int\limits_{0}^{\infty }x^{s}e^{-x}\frac{dx}{x}  \tag{3.43}
\end{equation}
since this is ''sum'' (i.e. integral with respect to its Haar measure $\frac{%
dx}{x})$ over the multiplicative character x$^{s}$ and the additive
character $e^{-x}.$ The Jacobi sum is analog of Euler's beta function since 
\begin{equation}
J(\chi _{1},\chi _{2})=\frac{g(\chi _{1},\psi )g(\chi _{2},\psi )}{g(\chi
_{1}\chi _{2},\psi )}.  \tag{3.44}
\end{equation}
The p-adic gamma function (not to be confused with Eq.(1.21)) $\Gamma
_{p}(n) $ for an integer $n\geq 2$\ is defined by 
\begin{equation}
\Gamma _{p}(n)=(-1)^{n}\prod\limits_{\substack{ j=1  \\ (p,j)=1}}^{n-1}j. 
\tag{3.45}
\end{equation}
It possesses almost standard properties, e.g.$\Gamma _{p}(x+1)=\left\{ 
\begin{array}{c}
-x\Gamma _{p}(x)\text{ if }x\in \mathbf{Z}_{p}^{\ast } \\ 
-\Gamma _{p}(x)\text{ if }x\in p\mathbf{Z}_{p}
\end{array}
\right. .$ Koblitz had demonstrated in Ref.[69] that for the Fermat curve,
Eq.(3.23), the zeta function of Weil is given by [see also[111], ch-r 6], 
\begin{equation}
Z(F(n);u\mid \mathbf{F}_{q})=\frac{1}{(1-qu)(1-u)}\prod\limits_{\substack{ %
1\leq r,s<N-1  \\ r+s\neq N}}^{N-1}(1-J(\chi _{{}}^{r},\chi _{{}}^{s})u) 
\tag{3.46}
\end{equation}
In addition, he shows that 
\begin{equation}
J(\chi _{{}}^{r},\chi _{{}}^{s})=\frac{\Gamma _{p}(\frac{r}{N})\Gamma _{p}(%
\frac{s}{N})}{\Gamma _{p}(\frac{r+s}{N})}  \tag{3.47}
\end{equation}
to be compared with Eq.(3.28). From here it follows that Eq.(3.47) \textit{%
is the p-adic} \textit{analogue of the 4 particle Veneziano amplitude}. In
proving the Theorem 4.13 of Ref.[68] Gross and Koblitz show that knowledge
of the p-adic beta function allows \textbf{to restore} the full beta
function, i.e. the full Veneziano amplitude and vice versa. This observation
provides the strongest support to the conjecture of Raoul Bott who said,
Ref.[50], page 159, that ''I would not be too surprised if discrete mod $p$
mathematics and the $p$-adic numbers would eventually be of use in the
building of models for very small phenomena''.

\textbf{Remark 3.8.}It is interesting to notice at this point that if
instead of the complete elliptic integral, Eq.(3.36), we would consider an
indefinite elliptic integral,e.g. 
\begin{equation*}
z=\int \frac{d\mathcal{P}}{4\mathcal{P}^{3}-g_{2}\mathcal{P}-g_{3}}
\end{equation*}
then, its inverse is the stationary solution of the KdV equation, i.e. 
\begin{equation*}
\left( \mathcal{P}^{^{\prime }}\right) ^{2}=4\mathcal{P}^{3}-g_{2}\mathcal{P}%
-g_{3}.
\end{equation*}
Mulase [72] had shown how one can easily obtain the time-dependent solution
from the time-independent one. Hence, the dual of KdV leading to the
Virasoro algebra (as we had discussed in our earlier work, Ref.[73]) can be
obtained as well.

Plausible as they are, these results can be considerably extended
strengthened and generalized beyond the existing string theory formalism.
Steps in these directions are presented in the following sections.

\section{The Chowla-Selberg formula, the Fermat's last theorem, the
conformal field theory and the Veneziano amplitude}

\subsection{\protect\bigskip Statement of the problem}

Originally, the Chowla-Selberg formula was announced in 1949, Ref.[74], but
was left practically unnoticed. This caused the authors to provide more
details. The revised and extended paper received \ the widespread attention
among mathematicians \ but was published only in 1967 [31]. Ironically, the
Veneziano's paper had appeared in 1968 [1] while the Annals of Mathematics
paper by Ramachandra [75] published in 1964 containing detailed derivation
of the Chowla-Selberg main results (to appear only in 1967 !) \ also was
left largely unnoticed.

In this work we would like to make connections between these historical
papers. We begin with the description of the problem Chowla and Selberg
wanted to solve. Looking at Eq.s(3.36) and (3.37) they wanted to know if the
result, Eq.(3.37), is mere coincidence or if it is general rule. To make
things more interesting, let us recall that the classical theory of elliptic
functions provides us with the following results. Periods $K$ and $K^{\prime
}$ are given through the elliptic theta functions [76] 
\begin{equation}
K=\frac{\pi }{2}\theta _{3}^{2}(\tau ),  \tag{4.1}
\end{equation}
\begin{equation}
K^{\prime }=\frac{\pi }{2}\theta _{3}^{2}(-\frac{1}{\tau }),  \tag{4.2}
\end{equation}
while $\tau $ is determined implicitly through 
\begin{equation}
k^{2}=\frac{\theta _{2}^{4}(\tau )}{\theta _{3}^{4}(\tau )}  \tag{4.3}
\end{equation}
with 
\begin{equation}
\theta _{2}=2q^{\frac{1}{4}}H_{0}H_{1}^{2}\text{ \ and \ }\theta
_{3}=H_{0}H_{2}^{2}  \tag{4.4}
\end{equation}
provided that $q=\exp (\pi i\tau )$ and functions $H_{0}$, $H_{1}$ and $%
H_{2} $ are given by 
\begin{equation}
H_{0}=\prod\limits_{k=1}^{\infty }(1-q^{2k})\text{ , }H_{1}=\prod%
\limits_{k=1}^{\infty }(1+q^{2k})\text{ , }H_{2}=\prod\limits_{k=1}^{\infty
}(1+q^{2k-1}).  \tag{4.5}
\end{equation}
Based on these results, it seems like a hopeless task to calculate either of
periods exactly. Needless to say the connection between Eqs.(3.36) and
(3.37) looks totally mysterious. Nevertheless, Chowla and Selberg had
demonstrated that this connection is \textbf{not} an accident. Under certain
conditions to be discussed below, the representation of periods as
products/ratios of gamma functions is the only possibility. In view of the
Veneziano formula, Eq.(1.2) (or (1.6)), the conditions under which such
representation is possible certainly is of physical interest.

Traditionally, calculation of the periods is done with help of the
hypergeometric function as we had discussed earlier in Ref.[77]. It is
appropriate to remind some aspects of this connection now. We have $K$=$%
\frac{\pi }{2}F(\frac{1}{2},\frac{1}{2},1;k^{2}),$ where the hypergeometric
function is obtained as solution of the hypergeometric equation 
\begin{equation}
z((1-z)F^{^{\prime \prime }}+[c-(a+b+1)z]F^{\prime }-abF=0.  \tag{4.6}
\end{equation}
Since both $K$ and $K^{\prime }$ can be obtained as solutions of Eq.(4.6) we
can form their ratio 
\begin{equation}
\frac{y_{1}}{y_{2}}=w(\tau )=\frac{K^{\prime }(\tau )}{K(\tau )},  \tag{4.7}
\end{equation}
where $y_{1}$and $y_{2}$ are solutions of the Fuchsian -type equation 
\begin{equation}
y^{\prime \prime }+\frac{1}{2}\{w,\tau \}y=0  \tag{4.8}
\end{equation}
with the \textit{Schwarzian} derivative $\{w$,$\tau \}$ known to be as 
\begin{equation}
\{w,\tau \}=\left( \frac{w^{\prime \prime }}{w^{\prime }}\right) ^{\prime }-%
\frac{1}{2}\left( \frac{w^{\prime \prime }}{w^{\prime }}\right) ^{2} 
\tag{4.9}
\end{equation}
with $w=w(\tau ),w^{\prime }=\frac{dw}{d\tau }$, etc. Clearly, everybody who
is familiar with string and conformal field theories will recognize at this
moment connections between mathematics and physics. We are \textbf{not}
going to develop such connections in this paper nevertheless. The only
reason we have brought the hypergeometric equation to the attention of our
readers is to emphasize that \textbf{this is equation for the periods of the
elliptic} \textbf{curve}. As it is well known, the Knizhnik-Zamolodchikov
(K-Z) equations are basic for all conformal statistical mechanical models
[78-80] and are effectively reducible to the equations of hypergeometric
type. \emph{Hence, all results of conformal field theories will remain} 
\emph{unchanged in what follows\footnote{%
Strong additional support of this claim is presented in Section 5.3.2 below.}%
}. In particular, the expression for the nearest neighbor spin-spin
correlator for the Ising model involves $F$($\frac{1}{2},\frac{1}{2}%
,1;k^{2}) $ [81], page 69 . But, much more can be actually accomplished
thanks to results obtained by Chowla and Selberg.

To begin, we would like to remind our readers about the partition function
for the free bosons on the torus described, for example, in Ref.[82], pages
340-344. The case of free fermions (Ising model) technically is almost the
same. Indeed, in the first place the whole computation depends upon
calculation of the sum 
\begin{equation}
G(s)=\sum\limits_{\substack{ m,n  \\ m\neq n}}^{{}}\frac{1}{\left| m+n\tau
\right| ^{2s}},  \tag{4.10}
\end{equation}
while in the second, of the sum 
\begin{equation}
G_{\mu ,\nu }(s)=\sum\limits_{m,n}^{{}}\frac{1}{\left| m+n\tau +\left( \mu
+\nu \tau \right) \right| ^{2s}}.  \tag{4.11}
\end{equation}
The purpose of the entire calculation in both cases lies in explicitly
obtaining $G^{\prime }(0)$ and $G_{\mu ,\nu }^{\prime }(0)$ where the prime
means differentiation with respect to $s$ variable. It is clear that in both
cases obtained expressions are $\tau -$ dependent. They also should be
invariant with respect to modular transformations: $\tau \rightarrow \tau +1$
and $\tau \rightarrow \dfrac{-1}{\tau }$ . It is well known, e.g. see our
earlier work, Ref.[4], that all transformations of the type 
\begin{equation}
\tau ^{\prime }=\frac{a\tau +b}{c\tau +d}  \tag{4.12}
\end{equation}
with $a,b,c,d$ being some integers subject to restriction $ad-bc=1$ can be
obtained by successive applications of the above two. In the same work we
also explained what these transformations mean topologically. In view of
generalizations which follow we need to present these results from the
different perspective now.

\subsection{\protect\bigskip Complex multiplication}

To this purpose it is helpful to recall that, according to the theorem by
Jacobi [76], the modular lattice $L,$ 
\begin{equation}
L=\mathbf{Z}\omega _{1}+\mathbf{Z}\omega _{2}\text{ or, symbolically, }L%
\text{=[}\omega _{1},\omega _{2}],  \tag{4.13}
\end{equation}
has periods such that $\tau =\func{Im}\dfrac{\omega _{2}}{\omega _{1}}>0,$%
that is one of the periods should be complex. Eq.(4.13) looks the same as
Eq.(A.2) and, actually, has the same meaning. Moreover, from the algebraic
point of view, $L$ is just an ideal. In the Appendix B of our work, Ref.[4],
we explained the notion of an ideal. For the sake of uninterrupted reading
we would like to recall few things from that reference.

Suppose we have some set $I$ and another set $L,$then $\forall \alpha ,\beta
\in I$ and $\forall \xi \in L$, we have 
\begin{equation}
\alpha +\beta \in I\text{ (module property), }\alpha \xi \in L\text{ (ideal
property).}  \tag{4.14}
\end{equation}
In the present case, we have $\alpha $ and $\beta $ $\in \mathbf{Z}$ and $L$%
=[$\omega _{1},\omega _{2}].$ Clearly, if $\omega _{1}$ and $\omega _{2}$
constitute the basic parallelogram of such lattice then, any point $\mathbf{x%
}$=$\alpha \omega _{1}+\beta \omega _{2}$ also belongs to the lattice.
Surely, for a given lattice we can make many sublattices using equivalence
relation $x\equiv y$ $(\func{mod}m)$, that is $x-y=mk$ for some integer $k$%
.Specifically, if we initially had lattice $L$=[$\omega _{1},\omega _{2}]$
we can consider lattice $L^{\prime }$=[$\omega _{1}^{\prime },\omega
_{2}^{\prime }]$ where, in view of the results of the Appendix A, we have to
demand 
\begin{eqnarray}
\omega _{2}^{\prime } &=&a\omega _{2}+b\omega _{1}  \TCItag{4.15} \\
\omega _{1}^{\prime } &=&c\omega _{2}+d\omega _{1}.  \notag
\end{eqnarray}
Eq.(4.12) is obtainable from these two equations by forming their ratio. We
refrain from doing this for the moment since we want to consider the
possibility $\omega _{1}^{\prime }=\alpha \omega _{1}^{{}},\omega
_{2}^{\prime }=\alpha \omega _{2}^{{}}.$ That is we want to study the
eigenvalue problem for the matrix $\mathbf{A}$ given by 
\begin{equation}
\mathbf{A}=\left( 
\begin{array}{cc}
a & b \\ 
c & d
\end{array}
\right) .  \tag{4.16}
\end{equation}
According to Eq.(B.3) of Appendix B the problem of finding the eigenvalues
of matrix $\mathbf{A}$ is reduced to the problem of finding eigenvalues of
the quadratic equation 
\begin{equation}
\alpha ^{2}-tr\mathbf{A}\text{ }\alpha +\det \mathbf{A}=0.  \tag{4.17}
\end{equation}
This equation should be supplemented with some physical conditions in order
to make sense of its solutions. In particular, let $a(L)$ denote the area of
the period parallelogram associated with the lattice $L$. Evidently, [83], 
\begin{equation}
a(L)=\frac{1}{2}\left| \omega _{1}\bar{\omega}_{2}-\omega _{2}\bar{\omega}%
_{1}\right| ,  \tag{4.18}
\end{equation}
where the overbar means the complex conjugation. In particular, for the
lattice $L_{\tau }=[1,\tau ]$ Eq.(4.18) produces $a(L)=\left| \func{Im}(\tau
)\right| .$ Now, if we rescale $\omega ^{\prime }s$, we obtain $a(\alpha
L)=\left| \alpha \right| ^{2}a(L).$ If we require that, upon such rescaling,
the area $a(L)$ remains unchanged this leaves us with the only one option: $%
\left| \alpha \right| ^{2}=1.$ This result should be dealt with in
connection with Eq.(4.17). This leaves us with not too many choices. One
option is to have $tr\mathbf{A}=0$ thus producing rather trivial result: $%
\alpha =\pm 1.$ The rest of options comes from the standard number-theoretic
results about units in the quadratic fields as discussed for example by
Hardy and Wright [84] (see also Appendix B of our earlier work, Ref.[4]).
Hence, in addition to $\pm 1,$ we obtain also $\alpha =\pm i$ and $\alpha
=\pm \rho ,\alpha =\pm \rho ^{2}$ with $\rho =\frac{1}{2}(-1+\sqrt{-3}).$
The last result we had encountered earlier in connection with the Alexander
polynomial for trefoil knot, Ref.[4], Section 3.These are the only options
available for the quadratic fields. In the light of these results the
following definition is useful:

\textbf{Definition 4.1}. Two complex tori $T=\mathbf{C}/L$ and $T^{\prime }=%
\mathbf{C}/L^{\prime }$ are \textit{isomorphic} (respectively \textit{%
isogenous} ) provided that there is an isomorphism (respectively isogeny) $%
\alpha :T\rightarrow T^{\prime }$ induced by multiplication $\alpha :\mathbf{%
C}\rightarrow \mathbf{C}$ for $\alpha \in \mathbf{C}$ with $\alpha
L=L^{\prime }$ (respectively $\alpha L\subseteq L^{\prime }).$

From this definition it follows that isogeny is an equivalence relation
because $L^{\prime }$is homothetic to $L$. By establishing such relation we
are effectively embedding our tori into the projective space. Homothetic
lattices are associated with isomorphic elliptic curves [85]. This feature,
known in literature as \emph{complex multiplication (CM)} [85], admits
generalization to Riemann surfaces of higher genus[18].

It is important to realize that in general an arbitrary torus $T$ may 
\textbf{not} possess \emph{CM. }In the $\emph{traditional}$ (and, hence,
more general) case there is \emph{no need} to require $\omega _{1}^{\prime
}=\alpha \omega _{1}^{{}},\omega _{2}^{\prime }=\alpha \omega _{2}^{{}}.$ In
this case Eq.s(4.15) are reduced to Eq.(4.12), provided that $ad-bc=1$.
Eq.(4.12) might be interpreted as equation representing motion in the
Teichm\"{u}ller space of the punctured torus: Different initial points in
the moduli space for $T$ will produce different orbits in the
Teichm\"{u}ller space of the torus. All this is explained in our earlier
work, Ref.[4].

Let us now go back to Eq.s (4.15). The second of these equations can be
rewritten as $\alpha =c\tau +d$ and, since $\tau $ is a complex number, $%
\alpha $ should be a complex number as well.

\textbf{Remark 4.2.} At this point we would like to notice that if $\tau $
would be real, the multiplier $\alpha $ becomes real as well. Then, instead
of CM theory one would have $\emph{real}$ \emph{multiplication} (RM) theory.
Such theory would make no sense from the point of view of traditional
algebraic geometry but would make sense from the point of view of non
commutative geometry of Connes according to the recent papers by Manin
[86,87]. In our work such possibility is \textbf{not} considered since CM is
sufficient for description of physically relevant phenomena. Surely, we do
not exclude the possibility that RM might have some physical significance
and leave this option open for further study.

\bigskip

Going back to the subject of CM, we notice, that the first of Eq.s(4.15) can
be written as ($c\tau +d)\tau =a\tau +b.$ This result can also be written in
the form of Eq.(4.12), i.e. 
\begin{equation}
\tau =\frac{a\tau +b}{c\tau +d}.  \tag{4.19}
\end{equation}
This is equation for the geodesic in \textbf{H }as we had discussed in
Section 2 of Ref\textbf{.}[4]. Explicitly written, it looks like 
\begin{equation}
c\tau ^{2}+(d-a)\tau -b=0.  \tag{4.20}
\end{equation}
It can be equivalently rewritten as 
\begin{equation}
\left( c\tau \right) \tau =b+(d-a)\tau .  \tag{4.21}
\end{equation}
If we denote $c\tau $ as $\beta $ then, the above equation can be presented
as $\beta \omega _{2}=b\omega _{1}+(d-a)\omega _{2}$ \ and, in addition, we
have $\beta \omega _{1}=c\omega _{2}.$ These equations are of the same kind
as Eq.s(4.15) as required. Moreover, using the same equations and
eliminating $\tau $ we obtain the following equation for $\alpha :$%
\begin{equation}
\alpha ^{2}-(a+d)\alpha +ad-bc=0  \tag{4.22}
\end{equation}
in agreement with Eq.(4.17) or, taking into account that $ad-bc=1,$ we
obtain 
\begin{equation}
\alpha ^{2}-(a+d)\alpha +1=0.  \tag{4.23}
\end{equation}
This is equation for an integer in the quadratic field (Appendix B,
Ref.[4]). But earlier we have obtained $\alpha =c\tau +d$ and, hence, such
integer must belong to the \emph{imaginary} quadratic field. Therefore, $%
\tau $ also belongs to the imaginary quadratic field but is not an integer
in general since $\tau =(\alpha -d)/c.$ As we had discussed in Ref.[4], the
imaginary quadratic fields are being characterized by their class numbers $h(%
\sqrt{-d}).$These numbers directly associated with the unique factorization
property for the ideals of these fields. Every number $\varkappa $ in
imaginary quadratic field can be presented as $\varkappa =a+b\sqrt{-d}$,
where $a$ and $b\in \mathbf{Q.}$ Unique factorization producing class number 
$h(\sqrt{-d})=1$ is possible if $-d=1,2,3,7,11,19,43,67$ and $163.$%
Sometimes, instead of 1 and 2 numbers 4 and 8 being used but, surely, these
produce the same results. We mention this fact for the sake of comparison
with the existing literature. If we would like to restrict ourself by these
numbers, the parameter $\tau $ in Eq.s(4.10) and (4.11) \textbf{cannot be
arbitrary anymore}. \textbf{It should be} \textbf{discrete}. Nevertheless,
because we can vary integers $d$ and $c$ in the equation $\tau =(\alpha
-d)/c,$this discreteness does not lead to just one value for $\tau $.

Now we would like to illustrate all these statements by simple but important
examples. These examples serve to underscore the differences between what is
known so far in the conformal field and string theories and new elements
which, in our opinion, should be included. Inclusion of these new elements
will enable us to use powerful number-theoretic methods in the conformal and
string theories\footnote{%
Usefulness of these ''new elements'' to conformal and ''string'' theories
can be seen from both: the references provided at the end of Section 1.3 and
from actual developments presented in Sections 5.2. and 5.3.}.

Going back to Eq.(4.23) we obtain the following expressions for $\alpha
^{\prime }s:$%
\begin{equation}
\alpha _{1,2}=\frac{a+d}{2}\pm \frac{1}{2}\sqrt{\left( a+d\right) ^{2}-4}. 
\tag{4.24}
\end{equation}
Since $\alpha $ should be an integer in the imaginary quadratic field we are
left with the following set of options:

a) $a=d=0$, thus producing $\alpha _{1,2}=\pm i;$

b) $a=\pm 1,b=0$ (or $a=0,b=\pm 1$), thus producing $\alpha _{1,2}=\frac{1}{2%
}(\pm 1\pm \sqrt{-3})$

c) $a=d=\pm 1,$thus producing $\alpha _{1,2}=\pm 1.$

We have obtained all these results earlier and, surely, we would not
reproduce them once again should these be the only options. Fortunately,
there are other less trivial options leading to far reaching consequences.

Consider $a(L^{\prime })=\frac{1}{2}\left| \omega _{1}^{^{\prime }}\bar{%
\omega}_{2}^{\prime }-\omega _{2}^{\prime }\bar{\omega}_{1}^{\prime }\right| 
$ where $\omega _{1}^{\prime }$ and $\omega _{2}^{\prime }$ are given by the
l.h.s. of Eq.(4.15). This expression can be rearranged with the help of
Eq.(4.15) as follows: 
\begin{eqnarray}
a(L^{\prime }) &=&\frac{1}{2}\left| \omega _{1}^{^{\prime }}\bar{\omega}%
_{2}^{\prime }-\omega _{2}^{\prime }\bar{\omega}_{1}^{\prime }\right| 
\TCItag{4.25} \\
&=&\frac{1}{2}\left| (ad-bc)(\omega _{1}\bar{\omega}_{2}-\omega _{2}\bar{%
\omega}_{1})\right|  \notag \\
&=&\frac{n}{2}\left| \omega _{1}\bar{\omega}_{2}-\omega _{2}\bar{\omega}%
_{1}\right| =na(L)  \notag
\end{eqnarray}
provided that $ad-bc=n.$ We would like to argue now that such relation is
legitimate. Indeed, the relation $ad-bc=1$ comes as result of the
requirement for the \textbf{inverse} of the matrix \textbf{A}, Eq.(4.16),
should have only integer coefficients [88]. This is a legitimate requirement
as long as we are dealing with \textbf{the same} lattice and just trying to
change the basis. But if we are dealing with sublattices, relation $ad-bc=1$
can be relaxed to $ad-bc=n.$ This can be rigorously proven, e.g. see Ref.
[89], ch-r1, and leads to a very deep results summarized in the Appendix C.
The scaling result $a(\alpha L)=\left| \alpha \right| ^{2}a(L)$ discussed
earlier can now be rewritten as $a(L^{\prime })=a(\alpha L)=na(L).$ This
suggests that if $L$ is sublattice of \ $L^{\prime },$ then these
sublattices are homothetic and therefore are isogenic and, hence, are
equivalent. Accordingly, Eq.(4.22) should be amended leading to 
\begin{equation}
\alpha _{1,2}=\frac{a+d}{2}\pm \frac{1}{2}\sqrt{(a+d)^{2}-4n}.  \tag{4.26}
\end{equation}
Thus, indeed, the complex multiplication is associated with integers in the
imaginary quadratic field. From the above discussion it follows that there
are only $9$ imaginary quadratic fields whose ring of integers has class
number $h(\sqrt{-d})=1.$ Actually, more can be said. Following Silverman,
Ref.[85], we notice that the $j(L)$ invariant of the elliptic curve (modular
form of weight zero, e.g. see Eq.(5.8) below) is just some rational number $%
\in \mathbf{Q}$ provided that $j(L)$ characterizes the equivalence classes
of elliptic curves with CM whose class number is 1. Moreover $j(L)\in 
\mathbf{Q}$ only in 3 other cases so that there are exactly 13 classes of
elliptic curves with complex multiplication. Such curves are known as \emph{%
Weil curves}. They possess remarkable properties to be briefly discussed in
the next subsection. It should be noted, however, that we are concerned
about $h=1$ case only because it is technically much simpler as compared to
other quadratic fields for which $h>1$. \ Altogether, quadratic imaginary
fields are much simpler to handle than the real quadratic fields [90].

\subsection{Fermat's last theorem, periods of elliptic curves and Veneziano
amplitudes}

Zeta function of the elliptic curve $Z_{K}(u)$ is given by Eq.(3.35). Now we
would like to connect this function with modular forms of Appendix C.
According to famous \emph{Taniyama-Shimura (T-S) conjecture} (now proven,
thanks to works of Wiles [63] and Taylor-Wiles [91]) \textbf{every elliptic
curve over} $\mathbf{Q}$\textbf{\ is modular.} Naturally, we would like to
explain what this notion means and how it is connected with $Z_{K}(u).$
Since the results to be discussed depend crucially on (now proven) $T-S$
conjecture and since the proof\ of this conjecture effectively implies,
according to works by Frey [71] and Ribet [92], the Fermat's last theorem,
e.g. see Eq.s(3.38)-(3.40), it happens, that \textbf{the mathematics related
to Veneziano amplitudes is also related to the} \textbf{Fermat's \ last
theorem}. Accordingly, the multiparticle generalization of these amplitudes,
to be discussed in Section 5, is dependent to some extent also on the
validity of the Fermat's last theorem.

Let $\mathcal{E}$ be an elliptic curve over $\mathbf{Q}.$ Analytically, it
can be written in the so called (affine) Weierstrass form as 
\begin{equation}
y^{2}+a_{1}xy+a_{3}y=x^{3}+a_{2}x^{2}+a_{4}x+a_{6}  \tag{4.27}
\end{equation}
with coefficients in $\mathbf{Q}$. Let $\Delta (a_{1},...,a_{6})$ be the
discriminant of the elliptic curve $\mathcal{E}$ [83,85] . It carries
information about the smoothness of $\mathcal{E}$. If $\Delta =0$ the curve
is singular, that is it may contain a node or a cusp. Suppose that our curve
is nonsingular in the field $\mathbf{Q}$\textbf{.}Suppose as well that we
would like to know if such curve will still remain nonsingular if we \emph{%
reduce} it, that is, if we would like to consider instead of Eq.(4.27) its 
\textbf{F}$_{p}$ analog (Appendix A) 
\begin{equation}
y^{2}+a_{1}xy+a_{3}y=x^{3}+a_{2}x^{2}+a_{4}x+a_{6}\text{ (}\func{mod}p\text{)%
}  \tag{4.28}
\end{equation}
\textbf{Remark 4.3.}Such reduction allows us to use without change all
results of Section 2.2.

Surely, to do so we need to get rid of denominators in Eq.(4.27) so that all
coefficients belong to $\mathbf{Z}$. Upon reduction the curve $\mathcal{E}%
_{p}$ may become singular (even though it was not singular initially!). This
happens when $\Delta (a_{1},...,a_{6})\equiv 0(\func{mod}p)$ $\ $and
represents the case of $\emph{bad}$ reduction. Otherwise the reduction is
called \emph{good}. Clearly, $\mathcal{E}$ has bad reduction only for a 
\emph{finite} number of primes. The bad reduction can be measured by means
of the $\emph{c}$\emph{onductor}. The conductor is a number $N$ such that 
\begin{equation}
Cond(\mathcal{E})\equiv N=\prod\limits_{p}p^{f_{p}}.  \tag{4.29}
\end{equation}
Here $f_{p}=0$ if $p\nmid \Delta $ and $f_{p}\geq 1$ othervice. Let $N_{1}$
be the number of solutions of Eq.(4.28) in the projective space (e.g.see
Eq.(2.26)) then, the Hasse-Weil $L$ function of $\mathcal{E}$ can be defined
as 
\begin{equation}
L(\mathcal{E},s)=\prod\limits_{p}L_{p}(p^{-s}),  \tag{4.30}
\end{equation}
where, in the case of good reduction, $\left[ L_{p}(p^{-s})\right]
^{-1}=(1-pu)(1-u)Z_{p}(u)\mid _{u=p^{-s}}$with Z$_{p}(u)$ defined by
Eq.(3.35). In general case we have as well 
\begin{equation}
L_{p}(p^{-s})=1-a_{p}p^{-s}+\psi (p)p^{1-2s}  \tag{4.31}
\end{equation}
with $a_{p}=1+p-N_{1}(p)$ and $\psi (p)=0$ or $1$ depending upon $p$
dividing or not $Cond(\mathcal{E}).$ Hence, in general we can write 
\begin{equation}
L(E,s)=\prod\limits_{p\mid N}\frac{1}{1-a_{p}p^{-s}}\prod\limits_{p\nmid N}%
\frac{1}{1-a_{p}p^{-s}+p^{1-2s}}.  \tag{4.32}
\end{equation}
Now it is time to connect this result with the modular forms discussed in
the Appendix C. More exactly, we need to make connections with the cusp
forms of weight 2. The cusp form $\ f(\tau )$ \ is defined by its $q$
expansion: $f(\tau )=\sum\nolimits_{n=1}^{\infty }c(n)q^{n}$, $q=\exp (2\pi
i\tau ).$ Using the properties of Hecke operators (Appendix C) and, in
particular, Eq.s(C.9) and (C.16) (for $k$=2), we obtain (upon acting on $%
f(\tau ))$ the following set of recursions for $c(n)^{\prime }s:$%
\begin{equation}
c(p^{e})c(p)=c(p^{e+1})+pc(p^{e-1})\text{ for }p\text{ prime, }p\nmid N, 
\tag{4.33a}
\end{equation}
\begin{equation}
c(p^{e})=\left[ c(p)\right] ^{e},\text{for }p\text{ prime and }p\mid N,\text{%
and}  \tag{4.33b}
\end{equation}
\begin{equation}
c(m)c(n)=c(mn).  \tag{4.33c}
\end{equation}
The Mellin transform of the cusp form can be defined now as 
\begin{eqnarray}
F(s) &=&\int\limits_{0}^{\infty }\frac{d\xi }{\xi }f(i\xi )\xi ^{s}=\Lambda
(f,s):=\left( 2\pi \right) ^{-s}\Gamma (s)\sum\limits_{n=1}^{\infty }\frac{%
c(n)}{n^{s}}  \TCItag{4.34} \\
&\equiv &\left( 2\pi \right) ^{-s}\Gamma (s)L(f,s).  \notag
\end{eqnarray}
It can be easily demonstrated, e.g. see Ref. [93], that $L(f,s)=L(\mathcal{E}%
,s)$ provided that the cusp form (of weight $k=2$) is also an eigenform for
the involution operator $W_{N}$ defined by 
\begin{equation}
W_{N}f(\tau )=\pm N\tau ^{-2}f(\frac{-1}{N\tau }).  \tag{4.35}
\end{equation}
More exactly, \ it is expected that the cusp form of weight 2 is an
eigenform of two operators $T_{2}(n)$ and $W_{N}:$%
\begin{equation}
T_{2}(n)f(\tau )=c(n)f(\tau ),  \tag{4.36a}
\end{equation}
\begin{equation}
W_{N}f(\tau )=\eta f(\tau )  \tag{4.36b}
\end{equation}
with $\eta =\pm 1$(fermionic property). If $f(\tau )$ lies in one of the two
eigenspaces for $W_{N}$ and satisfies Eq.(4.36a) then, $L(f,s)=L(\mathcal{E}%
,s).$

\textbf{Definition 4.4.} An elliptic curve over $\mathbf{Q}$ is said to be 
\emph{modular} if there exist a cusp form of weight 2 on $\mathcal{S}_{N}$
(Appendix C) for some $N$ such that $L(f,s)=L(\mathcal{E},s).$

\textbf{Remark 4.5.} This identification means that $a_{p}^{\prime }s$ in
Eq.(4.32) should be replaced by $c(p)^{\prime }s$

Shimura and Taniyama conjectured that every elliptic curve over $\mathbf{Q}$
is modular and Wiles [63] proved correctness this conjecture. Now it remains
to explain what all these results have to do with the Veneziano amplitudes.
We shall provide the answer based on results of the Eichler-Shimura theory
(surely only those aspects that are relevant to our immediate needs) and
keeping in mind further uses of the Chowla-Selberg formula (to be discussed
later in Subsection 4.5.). In addition, the entire Section 5 is devoted to
elaborations of these results from yet another perspective.

We begin with observation that the combination $f$($\tau )d\tau $ is
invariant with respect to action of modular transformations which belong to $%
\mathcal{S}_{N}.$ Define now a ''path integral'' 
\begin{equation}
\Phi _{f}(\gamma )=\int\limits_{\tau _{0}}^{\gamma \left( \tau _{0}\right)
}f(\xi )d\xi .  \tag{4.37}
\end{equation}
This integral is actually independent of $\tau _{0}$ for $\gamma \in 
\mathcal{S}_{N}.$ If $\gamma \in \mathcal{S}_{N}$ is elliptic ($\left|
tr\gamma \right| <2)$ or parabolic ($\left| tr\gamma \right| =2)$ it can be
shown [93] that $\Phi _{f}(\gamma )=0.$ This is consistent with Eq.(4.20)
which is expected to have two solutions, that is to produce the geodesic
generated by the hyperbolic transformations. Hence, only hyperbolic
transformations should be taken into account when calculating $\Phi .$ Since 
$\Phi _{f}(\gamma )$ is independent of $\tau _{0}$ this is possible only if $%
\Phi _{f}(\gamma )$ describes \textbf{closed} geodesic on the corresponding
Riemann surface $R=$ $\mathbf{H/\mathcal{S}_{N}.}$ But closed geodesic on
the Riemann surface $R$ corresponds to the period (e.g. see Eq.(3.31)) when
it is lifted to the Jacobian $J(R)$ of $R$. Surely, if $R$ has more than one
cusp form, there will be more than one period. Since these cusp forms are
independent of each other, the corresponding periods are also independent.

\textbf{Remark 4.6}.It is remarkable that such periods are defined on
surfaces with cusps. In the case of dynamics of 2+1 gravity (equivalent to
the dynamics of train tracks) the simplest non trivial Riemann surface on
which the train track dynamics can be studied is represented by once
punctured torus [4]. The simplest non trivial correlator for the
Kontsevich-Witten model of 2 dimensional topological gravity also involves
integration over the moduli space of once punctured torus [4].

The intersection pairing \ form $<C$, $\omega >$ can be defined now as 
\begin{equation}
\int\limits_{\tau _{0}}^{\gamma \left( \tau _{0}\right) }f(\xi )d\xi
=\int\limits_{C}\omega =<C,\omega >\equiv \Omega .  \tag{4.38}
\end{equation}
This form can be easily defined for the Riemann surfaces of higher genus.
Many interesting results related to Selberg trace formula, dynamics of
hyperbolic flows, etc. can be found in Refs. [94,95] and references therein.
In this work we are concerned with different issues however. In particular,
we would like to explain why $\alpha $ and $q/\alpha $ defined in Eq.(3.35)
can be called ''periods'', more exactly, $p$-adic periods. We also would
like to connect these $p$-adic periods with $\Omega .$

Ignoring for the moment bad reduction terms in Eq.(4.32) it is useful to
rewrite Eq.(3.35) in the following equivalent form 
\begin{equation}
Z(p^{-s})=\frac{1+(N_{p}-p-1)p^{-s}+p^{1-2s}}{(1-p^{-s})(1-p^{1-2s})}\equiv 
\frac{\zeta _{p}(s)\zeta _{p}(s-1)}{L_{p}(s)}  \tag{4.39}
\end{equation}
where, according to Eq.(4.31), $N_{p}-p-1=-c(p)$ (see Remark 4.5). The
global zeta function $\hat{\zeta}(s)$ is \ obtained now as 
\begin{equation*}
\hat{\zeta}(s)=\prod\limits_{p}\frac{\zeta _{p}(s)\zeta _{p}(s-1)}{L_{p}(s)}%
\equiv \frac{\zeta (s)\zeta (s-1)}{L(\mathcal{E},s)}
\end{equation*}
with 
\begin{equation}
L(\mathcal{E},s)=\prod\limits_{p}\left[ 1+(N_{p}-p-1)p^{-s}+p^{1-2s}\right]
^{-1}.  \tag{4.40}
\end{equation}
Let $s=1$ in Eq.(4.40). Then, we obtain a very remarkable result, Ref.[96], 
\begin{equation}
L(\mathcal{E},1)=\prod\limits_{p}\left( \frac{p}{N_{p}}\right) .  \tag{4.41}
\end{equation}
Moreover, using Eq.(4.34), we obtain as well 
\begin{equation}
L(\mathcal{E},1)=-2\pi i\int\limits_{0}^{i\infty }f(x)dx.  \tag{4.42}
\end{equation}
Next, following Manin, Ref.[97], we would like to combine Eq.s(4.35),
(4.36b) in order to get 
\begin{equation}
L(\mathcal{E},1)=(1\pm 1)\sum\limits_{n=1}^{\infty }\frac{c(n)}{n}\exp
(-2\pi n/N)  \tag{4.42}
\end{equation}
so that if $c(n)$ are known, $L(\mathcal{E},1)$ can be calculated
numerically very efficiently. Next, the function $L(f,s)$, Eq.(4.34),
actually coincides with the Hecke zeta function $L(s,\chi )$ with
Gr\"{o}ssencharakter $\chi .$ It is known, that such function has the
following presentation and the associated with it Euler product 
\begin{equation}
L(s,\chi )=\sum\limits_{\QTR{sl}{a}}\frac{\chi (\QTR{sl}{a})}{N(\QTR{sl}{a}%
)^{s}}=\prod\limits_{\QTR{sl}{p}}(1-\chi (\QTR{sl}{p})N(\QTR{sl}{p}%
)^{-s})^{-1},  \tag{4.43}
\end{equation}
where the\ summation is taking place over all nonzero integral ideals with $%
N(\QTR{sl}{a})$ being the norm of the ideal (e.g. see Ref.[98] for a quick
introduction to these concepts) while the product is taken over all prime
ideals. In Refs.[99], ch-r.18, and [100], ch-r.8, it is shown (using some
examples) that, actually, $L(s,\chi )=L(f,s).$ Next, given this fact one
needs to invoke\ the famous theorem by Weil [101] which identifies Hecke's
Gr\"{o}ssencharakters with the Jacobi sums, e.g. see Eq.(3.42). Once this is
done, Eqs.(3.35),(3.38), (3.46),(3.47) and (4.31) should be used along with
Definition 4.4. and Remark 4.5. in order to claim that, 
\begin{equation}
a(p)=\alpha (r)+\bar{\alpha}(s)\simeq J(\chi _{{}}^{r},\chi _{{}}^{s})=\frac{%
\Gamma _{p}(\frac{r}{N})\Gamma _{p}(\frac{s}{N})}{\Gamma _{p}(\frac{r+s}{N})}%
\equiv \omega _{p}\text{ .}  \tag{4.44}
\end{equation}
Since according to Eq.(2.19) $\alpha \cdot \bar{\alpha}=p,$we obtain as
well, $\bar{\alpha}=p/\alpha $. \ Because the r.h.s. of Eq.(4.44) is a $p$%
-adic period, the l.h.s., that is $\alpha (r)$ (or $a(p)$ ) is also a $p$%
-adic period. This proves the claim we made after Eq.(3.35) about the
physical nature of $\alpha .$ Moreover, this proves the claim made earlier
that to reach the conclusion about $\alpha $ the last Fermat's theorem
should be used.

\textbf{Remark 4.7.}The\textbf{\ }connection between $\alpha ^{\prime }s$
and $\ $the $p$-adic periods was first made by Dwork. Good \ and readable
summary of this and related papers by Dwork is given in Ref.[102].Relevant
results are also summarized briefly in the book by Koblitz , Ref.[69]. The
arguments presented above differ however from those of Dwork. They rely on
the fundamental result of Andre Weil [101].

We would like now to connect these facts with Eq.(4.41) and also with
Eq.s(1.23), (1.24). To do this let us consider, following Birch and
Swinnerton-Dyer, Ref.[96], (and also Cassels, Ref.[103]) the product 
\begin{equation}
\prod\limits_{p}\omega _{p}\simeq \tau  \tag{4.45}
\end{equation}
where $\tau $ denotes the Tamagawa number. Calculation of Tamagawa numbers
is described for example in the book by Weil [104 ]\footnote{%
More recent \ useful reference is the monograph by Maclachlan and Reid,
Ref.[105], where the Tamagawa numbers are discussed along with many other
relevant facts.}. Nevertheless, the problem of finding $\tau $ in Eq.(4.45)
is not \ completely solved to our knowledge. By analogy with Eq.(1.20), the\
above result remains unchanged if we replace $\omega _{p}$ by $\left|
a\right| _{p}\omega _{p}$ since $\prod\limits_{p}\left| a\right| _{p}=1.$
That is the product, Eq.(4.45), is invariant with respect to some scale
changes. This is important in view of Eq.(4.44). The product in Eq.(4.45) is
taken over all places, including $\infty ,$ therefore, 
\begin{equation}
\tau \simeq \Omega \prod\limits_{finite\text{ }p}\omega _{p}  \tag{4.46}
\end{equation}
Surely, in view of Eq.(4.44) we expect $\Omega =$ $B(r,s)$ with $B(r,s)$
being the usual beta function that is the Veneziano amplitude. Eq(4.46) is%
\textbf{\ the exact analog} of Eq.(1.25) of the p-adic string theory [11].
To make all this connected with the number theory we have to look at
Eq.(4.38) and to assume that 
\begin{equation}
\int\limits_{C}\omega \rightleftarrows \int\limits_{C_{p}}\hat{\omega}_{p}. 
\tag{4.47}
\end{equation}
More accurately, this suggests that $\int\limits_{C_{p}}\hat{\omega}%
_{p}\simeq c(p)$ and, since $N_{p}-p-1=-c(p)$ (according to
Eq.s(4.31),(4.39)), one can divide both sides by $p$ in order to obtain 
\begin{equation*}
\frac{N_{p}}{p}=\frac{p+1-c(p)}{p}.
\end{equation*}
In view of Eq.s(4.41),(4.44), we finally obtain 
\begin{equation}
L(\mathcal{E},1)=\prod\limits_{p}\left[ \frac{p+1-\omega _{p}}{p}\right]
^{-1}  \tag{4.48}
\end{equation}
with the product being taken again over all places, including $\infty .$ The
ramifications of this identity lead to the so called Birch-Swinnerton-Dyer
conjecture (still unproven). It can be stated formally as 
\begin{equation}
\lim_{s\rightarrow 1}(s-1)^{-r}L(\mathcal{E},s)=C\cdot \Omega
\prod\limits_{finite\text{ }p}\omega _{p}  \tag{4.49}
\end{equation}
with $C$ and $r$ being some constants (details can be found in Ref.[85] ).
Evaluation of the constant $C$ is the major stumbling block in proving the
conjecture. The constant $r$ is assumed to be known. Since this conjecture
involves the Veneziano amplitude $\Omega ,$ Eq.(4.49) provides yet another
link between the ''string'' and number theories. We would like to discuss
now links between the number theory and the conformal field theories.

\subsection{Number-theoretic formulation of statistical mechanical models of
conformal field theory}

The idea about deep links between the number theory and statistical
mechanics is not new. It was discussed, for example, in the review paper by
Tracy [106]. More recent works relating number theory and statistical
mechanics can be found in Refs.[107-109]. Unlike these authors, we would
like to address here different type of issues. In particular, we would like
to understand better what makes two dimensional models so special from the
point of view of number theory. Once the answer to this question is found,
there is an opportunity to think about higher dimensional exactly solvable
models using number theory.

In Section 4.1.two functions $G(s)$ and $G_{\mu ,\nu }(s)$ were introduced
in Eq.s (4.10) and (4.11). These are the basic functions to which one can
relate many two dimensional exactly solvable statistical models at
criticality. We would like to rewrite these functions in the
number-theoretic form using simple example presented in detail below.

Consider a ring of complex numbers $A:=\{a+bi\mid a,b,\in \mathbf{Z}\}.$This
is the ring of Gaussian integers. Elements of $A$ correspond to points on
the lattice in the complex plane. For the complex number $x=a+bi$ the norm $%
\left\| x\right\| $is defined as usual by 
\begin{equation}
\left\| x\right\| =a^{2}+b^{2}=x\bar{x}.  \tag{4.50}
\end{equation}
We define now the $\zeta -$function for $A$: 
\begin{equation}
\zeta _{A}(s):=\sum\limits_{\substack{ a\in A  \\ a\neq 0}}\frac{1}{\left\|
a\right\| ^{s}}.  \tag{4.51}
\end{equation}
Since the ring of Gaussian integers has 4 units (that is there are 4 numbers
for which $\left\| a\right\| =1)$ this should be taken into account when $%
\zeta _{A}(s)$ is presented as the Euler product 
\begin{equation}
\zeta _{A}(s)=4\prod\limits_{\QTR{sl}{q}}(1-\left\| \QTR{sl}{q}\right\|
^{-s})^{-1}  \tag{4.52}
\end{equation}
to be compared with the Hecke $L$ function, Eq.(4.43). The product in the
r.h.s. of Eq.(4.52) is over prime ideals. To find these ideals, consider the
following chain of arguments. Suppose $a+bi$ belong to the prime ideal, then 
$a-bi$ should also belong to the prime ideal because if it were possible to
decompose $a-bi$ into two factors, the same would be possible for $a+bi.$
Hence, $a^{2}+b^{2}$ has the following decomposition into prime factors 
\begin{equation}
a^{2}+b^{2}=\left( a+bi\right) \left( a-bi\right) .  \tag{4.53}
\end{equation}
This suggests that either $a^{2}+b^{2}=p$ or $a^{2}+b^{2}=p^{2}$ where $p$
is the prime number. If $p\equiv 3\func{mod}4,$then $a^{2}+b^{2}=p$ cannot
be solved, that is $p$ is prime element of $A$.If, however, $p\equiv 1\func{%
mod}4,$then $a^{2}+b^{2}=p$ is solvable so that $p=q_{1}q_{2}$ with $%
q_{1}=a+bi$ and $q_{2}=a-bi.$These statements can be trivially checked for $%
p=2.$This discussion provides a nice illustration to general results stated
in Appendix B of our earlier work, Ref.[4].Using these results we obtain the
following result for $\zeta _{A}(s),$ Ref.[99]$:$%
\begin{equation}
\zeta _{A}(s)=4\left( \frac{1}{1-2^{-s}}\right) \prod\limits_{p\equiv 1\func{%
mod}4}\left( \frac{1}{1-p^{-s}}\right) ^{2}\prod\limits_{p=3\func{mod}%
4}\left( \frac{1}{1-p^{-2}}\right) .  \tag{4.54}
\end{equation}
Here the last term on the right comes from prime elements of $A$ with norm $%
p^{2},$the middle comes from two factors \emph{both} with norm $p$. Finally,
the factor containing 2 comes from the element $1+i$ whose norm is $2$.
Eq.(4.54) can be rewritten now as 
\begin{equation}
\zeta _{A}(s)=4\zeta (s)L_{\chi }(s)  \tag{4.55}
\end{equation}
with $\zeta (s)$ being the usual Riemann zeta function and $L_{\chi }(s)$
being the Dirichlet $L$-function 
\begin{equation}
L_{\chi }(s)=\sum\limits_{n=1}^{\infty }\frac{\chi (n)}{n^{s}},  \tag{4.56}
\end{equation}
where 
\begin{equation*}
\chi (n)=\left\{ 
\begin{array}{c}
0,\text{ if }n\text{ is even} \\ 
1,\text{ if }n\equiv 1\func{mod}4 \\ 
-1,\text{ if }n\equiv \text{-}1\func{mod}4
\end{array}
\right. .
\end{equation*}
There is an alternative way of presenting the same thing, e.g. see
Eq.(2.17). Namely, we can write as well 
\begin{equation}
\zeta _{A}(s)=\sum\limits_{n=1}^{\infty }\frac{h(n)}{n^{s}}  \tag{4.57}
\end{equation}
where the factor $h(n)$ denotes the number of ways of representing $n$ as
the sum of two squares $x^{2}+y^{2},x,y\in \mathbf{Z}.$ Elementary exercise
in number theory[99] produces the following result for $h(n)\footnote{%
Here for $m=1\func{mod}4,$ $\chi (m)=1,$ for $m=3\func{mod}4,$ $\chi (m)=-1,$
etc., just like in Eq.(4.56).}:$%
\begin{equation}
h(n)=4\sum\limits_{m\mid n}\chi (n).  \tag{4.58}
\end{equation}
Clearly, if instead of the ring of Gaussian integers we would have a ring of
integers of the imaginary quadratic field, that is $a+\sqrt{-d}b$, we could
repeat almost word for word all the preceding arguments. This allows us to
rewrite $G(s)$ of Eq.(4.10) as 
\begin{equation}
G(s)=\sum\limits_{n=1}^{\infty }\frac{h(n)}{n^{s}}  \tag{4.59}
\end{equation}
where the function $h(n)$ counts the number of integer solutions to the
equation $m^{2}+\tau ^{2}l^{2}=n.$ Hence, it can be presented as well in the
form of zeta function analogous to $\zeta _{A}(s).$ As for the function $%
G_{\mu .\nu }(s),$this is nothing but the special case of the function $%
S_{a}(x,y,s)$ (for $y=a=0)$ introduced by Kronecker, e.g. read
Chapter7(section 4) of the famous book by Weil, Ref.[110]. Already Kronecker
knew how to connect this function with the theta functions and, since these
functions can be made to satisfy the KdV equation [72] whose dual is related
to the Virasoro algebra (e.g see Remark 3.8) it is clear that all results of
the conformal field theory can be obtained now. One can do much better,
however. Looking at Eq.s(4.50) and (4.51) one can easily realize that
Eq.(4.51) is not restricted to numbers of quadratic imaginary field provided
that one defines the norm for the number field. In the case of quadratic
field the norm $\left\| x\right\| =x\bar{x}$. To generalize this result to
more general number field it is sufficient to define the norm $\left\|
x\right\| $as the product $\left\| x\right\| =x_{1}\cdot x_{2}\cdot \cdot
\cdot x_{n}$ where $x_{2}$,$...,x_{n}$ are ''conjugates'' of $x_{1}$.To
understand what all this means it is sufficient to realize that the
algebraic equation 
\begin{equation}
c_{0}+c_{1}x+\cdot \cdot \cdot +c_{n}x^{n}=0  \tag{4.60}
\end{equation}
normally will have $n$ roots. These roots are conjugates of each other. Let $%
c_{i}$ belong to the field $K$ while the roots to the \emph{field extension}
of $K$, say $E$. Let $\sigma $:$E\rightarrow E$ be an automorphism of $E$
which leaves $K$ $($that is $c_{i}$ ) fixed. If $x_{1}$is the root of
Eq.(4.60), then $\sigma (x_{1})$ is also the root of the same equation. The
group $\sigma $of automorphisms of $E$ is known as \textbf{Galois group}, $%
Gal(E/K).$The Kronecker-Weber theorem states that every finite abelian
extension of $\mathbf{Q}$ (that is finite extension of $E/\mathbf{Q}$ with
Gal($E/\mathbf{Q}$) abelian) can be embedded into a \textbf{cyclotomic}
extension \textbf{Q}($\omega )$ with $\omega $ being some root of unity.
This result underscores the importance of the cyclotomic fields discussed in
the Appendix A.

Going back to Eq.(4.51) and replacing $A$ with $E$ we obtain the
multidimensional generalization of $G(s)$. The actual computations will
depend crucially on the solvability of the Diophantine equation of the type 
\begin{equation}
\left\| x\right\| _{E/K}=n.  \tag{4.61}
\end{equation}
Hence, the multidimensional extension of the results of conformal field
theories (CFT) depends on our ability to solve the Diophantine equations of
the type shown above. The theory of Diophantine equations is sufficiently
developed, e.g. see Ref.[99] and references therein. This fact provides hope
that multidimensional extension of CFT can be developed gradually.

Before leaving this subsection, several comments are still in place. First,
we would like to rewrite the Dirichlet $L$-function, Eq.(4.56), in a
different form. To this purpose, we need to know that the Dirichlet
character $\chi $ is actually some root of unity. If we identify Gal($%
\mathbf{Q}$($\zeta _{n})/\mathbf{Q})$ with $\left( \mathbf{Z}/n\mathbf{Z}%
\right) ^{\ast }$ (where $\zeta _{n}$ is root of unity and * denotes the
multiplicative group of residue classes (such group does not contain zero)),
then the Dirichlet character $\func{mod}n$ is the Galois character. Since
all characters are of the form $\exp (2\pi iut/n)=\chi _{u}(t)$ it could
happen that $\chi _{u}(t_{1})=\chi _{u}(t_{2})$ if $(t_{1},n)=(t_{2},n)=1$
and $t_{1}=t_{2}\func{mod}f$ for some $f$ $\emph{smaller}$ than $n$.Such $f$
is called $\emph{conductor}$ of $\chi \footnote{%
Not to be confused with that defined earlier in Eq.(4.29) !}.$ It is
important to realize that $f$ divides $n$ so that $\chi $ is periodic with
period $f$. With this information in our hands, the Dirichlet $L$-function,
Eq.(4.56), can be rewritten as [111] 
\begin{equation}
L_{\chi }(s)=\frac{1}{f^{s}}\sum\limits_{a=1}^{f}\chi (a)H(\frac{a}{f},s) 
\tag{4.62}
\end{equation}
with the Hurwitz zeta function being defined as 
\begin{equation}
H(b,s)=\sum\limits_{n=0}^{n}\frac{1}{(b+n)^{s}},\text{ \ }0<b\leq 1. 
\tag{4.63}
\end{equation}
Using this result, Eq.(4.11) can be rewritten with help of Eq.(22) (Chapter
7, section 9 of Ref.[110]) in a similar form (which needs some minor
modifications for higher dimensional lattices) 
\begin{equation}
G_{\mu ,\nu }(s)=H(x,2s)+H(1-x,2s)  \tag{4.64}
\end{equation}
where $x=\mu +\nu \tau .$Using results of Weil's book, Ref. [110], it is not
difficult to figure out the higher dimensional analogue of Eq.(4.64). In
fact, the reader may want to consult Refs.[112,113] \ where this topic is
discussed. Their work is to be considered also in the next subsection.

Actually, physically interesting results are only those obtainable as limits 
$G^{\prime }(0)$ and $G_{\mu ,\nu }^{\prime }(0)$ with prime being
differentiation with respect to $s$ [82]. Calculation of these limits leads
us directly to the Chowla-Selberg formula for periods to be considered in
the next subsection.

\subsection{The Chowla-Selberg formula and the Veneziano amplitude}

The Chowla-Selberg formula, Ref.[31], is essentially the same thing as the
1st Kronecker's limit formula as discussed, for example, in the books by
Weil, Ref.[110], and Lang, Ref.[114]. It deals with the unusual
reinterpretation of this formula due to results of Lerch (Eq.(23) of Chapter
8, section 9 of Ref.[110]). Such reinterpretation does not come across
easily upon reading Ref.[31]. More directly it was obtained in the paper by
Ramachandra [75] and was subsequently reproduced in Weil's book, Ref.[110].

Let $K$= \textbf{Q}($\sqrt{-d}$) be the quadratic imaginary field, i.e.
imaginary quadratic extension of $\mathbf{Q}$ of discriminant -$d$. Let $%
r_{K}$ be its ring of integers and suppose that $r_{K}=\mathbf{Z}+\mathbf{Z}%
z $ with $z\in K.$ For such field the Dirichlet function can be written as 
\begin{equation}
L_{\chi }(s)=\frac{1}{d^{s}}\sum\limits_{n=1}^{d}\chi (n)H(\frac{n}{d},s) 
\tag{4.65}
\end{equation}
to be compared with Eq.(4.62). By analogy with Eq.(4.55) we consider the
Dedekind zeta function for the field $K$. Differentiating \ this function
with respect to $s$ we obtain: 
\begin{equation}
R(s)\equiv \frac{\partial }{\partial s}\left[ \zeta (s)L_{\chi }(s)\right]
=\zeta ^{\prime }(s)L_{\chi }(s)+\zeta (s)L_{\chi }^{\prime }(s).  \tag{4.66}
\end{equation}
Now we have to take the limit $s\rightarrow 0$ in this equation. Calculation
of $\zeta ^{\prime }(0)$ is easy and produces ln2$\pi .$ Calculation of $%
L_{\chi }^{\prime }(0)$ is somewhat more involved. Indeed, using Eq.(4.65)
we get for $L_{\chi }^{\prime }$ the following result: 
\begin{equation}
L_{\chi }^{\prime }(s)=\left[ -\ln d\right] L_{\chi }(s)+\frac{1}{d^{s}}%
\sum\limits_{n=1}^{d}\chi (n)H^{\prime }(\frac{n}{d},s).  \tag{4.67}
\end{equation}
The Lerch formula obtained by Lerch in 1894 ( Eq.(23) on page 60 of Weil's
book, Ref.[110]) comes to the rescue thus producing 
\begin{equation}
H^{\prime }(x,s=0)=\ln \frac{\Gamma (x)}{\sqrt{2\pi }}.  \tag{4.68}
\end{equation}
The value of $L_{\chi }(0)$ was known already to Dirichlet: 
\begin{equation}
L_{\chi }(0)=\frac{2h(\sqrt{-d})}{w_{0}},  \tag{4.69}
\end{equation}
where the class number $h(\sqrt{-d})$ was introduced in Subsection 4.2 and $%
w_{0}$ is the number of roots of unity in $K.$ Finally, the value of $\zeta
(0)$=$-1/2$ while that for $\zeta ^{\prime }(0)=-2\ln 2\pi $. Collecting all
these results in Eq.(4.66) we obtain: 
\begin{equation}
R(s=0)=-\frac{4h}{w_{0}}\ln 2\pi +\left[ \frac{1}{2}\ln d\right] -\frac{1}{2}%
\sum\limits_{n=1}^{d}\chi (n)\ln \frac{\Gamma (n/d)}{\sqrt{2\pi }}. 
\tag{4.70}
\end{equation}
Specializing to those imaginary quadratic fields for which $h=1$ the 1st
limit formula of Kronecker (Eq.(17) on page 75 of Weil's book) is given by 
\begin{equation}
\zeta _{K}(s\rightarrow 0)=1-\frac{s}{12w_{0}}\ln (\mathcal{F}(L)) 
\tag{4.71}
\end{equation}
with $\mathcal{F}(L)$ denoting Siegel-Ramachandra invariant: $\mathcal{F}%
(L)=\left( 2\pi \right) ^{-24}\left[ N(L)\right] ^{12}\left| \Delta (\tau
)\right| ^{2},$ Ref.$[114],$ch-rs$19,21.$ Here $\Delta (\tau )=\left( 2\pi
\right) ^{12}q\prod\limits_{n=1}^{\infty }(1-q^{2n})^{24}$ $\equiv $\ $%
\left( 2\pi \right) ^{12}\eta (\tau )^{24}$\ \ \ with $\eta $($\tau )$ being
\ the usual Dedekind eta-function while $N(L)$ being the norm of the
principal ideal. This norm can be easily calculated for the quadratic fields
[115].

Indeed, if the ideal $L$=[$\omega _{1},\omega _{2}],$then the $\emph{%
different}$ $\mathcal{D}$ can be calculated as 
\begin{equation*}
\mathcal{D=}\left| 
\begin{array}{cc}
\omega _{1} & \omega _{2} \\ 
\bar{\omega}_{1} & \bar{\omega}_{2}
\end{array}
\right| =\omega _{1}\bar{\omega}_{2}-\omega _{2}\bar{\omega}_{1}\equiv
\omega _{1}\bar{\omega}_{1}(\bar{\tau}-\tau )
\end{equation*}
while the same different for the basic lattice $\hat{L}=[1,(\bar{\tau}-\tau
)]$ is given by $\mathcal{\hat{D}=}(\bar{\tau}-\tau )$. The norm $N(L)$ is
determined by the rule : $N(L)=\mathcal{D}/\mathcal{\hat{D}}$ so that $%
N(L)=\omega _{1}^{2}.$ Keeping in mind that the Dedekind eta function is the
modular form of weight 12 and using Eq.(C.4)one obtains easily: $\mathcal{F}%
(L)=\left( \func{Im}\tau \right) ^{12}\left| \Delta (\tau )\right| ^{2}.$

Differentiating Eq.(4.71) with respect to $s$ and equating this result with
Eq.(4.70) (for $h=1$), we obtain (omiting unimportant constants): 
\begin{equation}
\left( \func{Im}\tau \right) ^{12}\left| \Delta \right| ^{2}\simeq
\prod\limits_{n=1}^{d}\Gamma (\frac{n}{d})^{6w_{0}\chi (n)}.  \tag{4.72}
\end{equation}
We would like to compare this result with that known from the CFT. In
particular, the free boson field theory on the torus is discussed in
Ref.[82] , ch-r 10, section 10.2. If we let $w_{0}$ in Eq.(4.71) to be one
then, using this formula we obtain at once 
\begin{equation}
\frac{\partial }{\partial s}\zeta _{K}(s)\mid _{s=0}=-2\ln \left[ \func{Im}%
\tau \left( \eta (\tau )\right) ^{2}\right] =G^{\prime }(0)  \tag{4.73}
\end{equation}
with $G^{\prime }(0)$ being defined earlier by Eq.(4.10). This result
coincides exactly with Eq.(10.29) obtained with help of the the path
integral for the boson field theory, Ref.[82]. The path integral for such
theory is effectively the Euclidean version of the bosonic string in the
conformal gauge [116]. More details on this topic are given in the next
subsection.

In the meantime, use of Eq.(4.72) enables us to provide different
interpretation of the results known in physics. First, using Eq.s(4.1)-(4.5)
and following Chowla and Selberg, Ref.[31], we write $\Delta (\tau )=\varpi
K^{12}$ with $K$ being one of periods of the elliptic curve and $\varpi $
being some (known in principle) algebraic number. Taking into account that $%
\func{Im}\tau $ is also an algebraic number (e.g. read Section 4.2) we
notice that the combination $\left( \func{Im}\tau \right) ^{12}\left| \Delta
\right| ^{2}\simeq \left( K\right) ^{24}.$ Therefore, \ using Eq.(4.72) we
obtain 
\begin{equation}
K=C\sqrt{\pi }\prod\limits_{n=1}^{d}\Gamma (\frac{n}{d})^{\frac{w_{0}}{4}%
\chi (n)}  \tag{4.74}
\end{equation}
where $C$ is some constant (algebraic number). This is \textbf{exactly} the
Chowla-Selberg formula (for $h=1$), Eq.(4), page 110 of Ref.[31]. The factor
of $\sqrt{\pi }$ comes in view of Eq.s(4.1) and (4.70).

\textbf{Remark.4.8. }At this point we would like to remind to our readers
that \textbf{exactly the} \textbf{same} Dedekind zeta function $\zeta
_{K}(s) $ used in obtaining the Chowla-Selberg formula was obtained earlier
in our work, Ref.[4], and was interpreted as dynamical partition function of
2+1 gravity.

We would like to demonstrate now how the Veneziano-like amplitudes can be
obtained with help of such partition function.

To do so, several things should be kept in mind. First, the number $w_{0}$ \
is equal to $4$ for $d=1$ (or d=4), it is equal $6$ for $d=3$ and it is
equal to $2$ in the rest of situations as it was explained earlier, e.g. see
discussions after Eq.s(4.18) and (4.24). Next, the character $\chi (n)$ in
Eq.(4.74) is equal to $\pm 1$ since it is the Kronecker symbol that is $\chi
(n)=\left( \dfrac{d}{n}\right) $\ according to p.109 of Ref.[31]. With this
information, let us first look at the simplest case $d=4$ having in mind to
reproduce Eq.(3.37). In this case we need to evaluate the following
Kronecker's symbols: ($\frac{4}{1}),(\frac{4}{2})$ and ($\frac{4}{3})$
(since ($\frac{4}{4})=0$ by definition). Using Ref.[115], pages 141-143, we
find: ($\frac{4}{1})=1,(\frac{4}{2})=0,$($\frac{4}{3})=-1.$ In arriving at
these results we took into account that the discriminant of the field $d$ is
not a square-free integer while $d/4$ is. Taking into account that $\sqrt{%
\pi }=\Gamma (\frac{1}{2})$ we obtain back the result given in Eq.(3.37).
Following Chowla and Selberg, consider now another example. This time, let $%
d=7$ so that $w_{0}=2.$ Then, similar calculations produce 
\begin{equation}
K=C\sqrt{\pi }\left[ \frac{\Gamma (\frac{1}{7})\Gamma (\frac{2}{7})\Gamma (%
\frac{4}{7})}{\Gamma (\frac{3}{7})\Gamma (\frac{5}{7})\Gamma (\frac{6}{7})}%
\right] ^{\frac{1}{2}}.  \tag{4.75}
\end{equation}
Using Eq.(1.9) it is possible to transform Eq.(4.75) into more
comprehensible form: 
\begin{equation}
K=\left( \hat{C}/\pi \right) \Gamma (\frac{1}{7})\Gamma (\frac{2}{7})\Gamma (%
\frac{4}{7}).  \tag{4.76}
\end{equation}
Looking at the Veneziano amplitude, Eq.(1.12), it is evident that the
bracket containing sinus factors will not contribute to physically
interesting poles. Hence, we may rewrite this amplitude as $A(a,b,c)\simeq
\Gamma (a)\Gamma (b)\Gamma (c)$ keeping in mind that $a+b+c=1$ $(or-1)$. In
Ref.[117] Gross had shown that it is always possible to reexpress $K$ as the
product of gamma functions for other $d^{\prime }s$ with $h=1$. \emph{In
this sense, the Chowla-Selberg formula, Eq.(4.74), could be considered as
the Veneziano amplitude.} In view of the Remark 4.8., it follows that the
partition function of 2+1 gravity is capable of producing the Veneziano
amplitudes and, since this function was not obtained in 26 dimensions,
accordingly, our Veneziano-like amplitude(s) ''live'' in normal space-time
dimensions. This conclusion is in accord with earlier obtained based on
Eq.(3.6).

Obtained result, Eq.(4.74), is inconvenient, however, for use in particle
physics since it is restricted to a very specific values of parameters,
strongly depends upon the toroidal geometry and cannot be easily generalized
to the Riemann surfaces of higher genus in spite of several attempts to do
so [112,113].We provided here all these derivations nevertheless because we
believe that the number-theoretic methods discussed in this work could be
potentially useful for development of CFT models either in dimensions higher
than two or for models defined on Riemann surfaces of higher genus. This is
especially important in view of the fact that such an extension for higher
genus surfaces in mathematics literature is only possible for those Riemann
surfaces ( and, hence, the Abelian varieties, Jacobians, etc.) which admit
CM. The theory of CM was developed by Taniyama and Shimura [18] and can be
found also in the monograph by Lang, Ref.[118]. To our knowledge, the work
of Shimura and Taniyama on CM had not been used in physics literature so far.

Before closing this section, we would like to discuss yet another reason for
discussing the Chowla-Selberg results.

\subsection{Heights of elliptic curves, the Chowla-Selberg formula and
string theory}

In this subsection we would like to discuss still another connection between
the Veneziano amplitudes and the number theory. It is based on realization
that the l.h.s. of Chowla-Selberg formula, Eq.(4.72), can be written in a
different way. This is possible because the function $\Delta (\tau )$ is the
modular form and, hence, following ideas of Eichler-Shimura theory discussed
in Section 4.3., we can construct the invariant differential form $\alpha
=\left( \Delta (\tau )\right) ^{\frac{1}{12}}dz$ and, with help of such
form, the intersection pairing [119]: 
\begin{eqnarray}
\frac{i}{2}\int\limits_{\mathbf{C}/L}\alpha \wedge \bar{\alpha} &=&\frac{i}{2%
}\int\limits_{\mathbf{C}/L}\left| \Delta (\tau )\right| ^{\frac{1}{6}%
}dz\wedge d\bar{z}  \TCItag{4.77} \\
&=&\left| \Delta (\tau )\right| ^{\frac{1}{6}}\int\limits_{A}dx\wedge dy 
\notag \\
&=&\left| \Delta (\tau )\right| ^{\frac{1}{6}}\func{Im}\tau .  \notag
\end{eqnarray}
Following Silverman [89,119], denote the height $h(\mathcal{E}/\mathbf{Q})$
of the elliptic curve $E$ as 
\begin{equation}
h(\mathcal{E}/\mathbf{Q})=-\frac{1}{2}\ln \frac{i}{2}\int\limits_{\mathbf{C}%
/L}\alpha \wedge \bar{\alpha}.  \tag{4.78}
\end{equation}
The above definition is not the most complete. Pedagogically clear
explanation of the concept of height and its usefulness in arithmetic
algebraic geometry is given in the review paper by Mazur [120]. Rigorous
connection between the height and the Chowla-Selberg formula is presented in
relatively recent Annals of Mathematics paper by Colmez [121]. We
deliberately avoid all intricacies of this concept\footnote{%
These can be found easily in the literature just cited, e.g. see Ref.[122]
in addition.} since our goal is more modest: we want to connect Eq.(4.78)
with string theory.

Following Ref.[8], the Weil-Petersson fundamental (1,1) K\"{a}hler form $%
\omega _{W-P}$ on the Teichm\"{u}ller space of the torus (and/or the
punctured torus[9]) is given by 
\begin{equation}
\omega _{W-P}=\frac{i}{4\left( \func{Im}\tau \right) ^{2}}d\tau \wedge d\bar{%
\tau}.  \tag{4.79}
\end{equation}
At the same time, the genus one partition function for the bosonic string in
26 space-time dimensions is given by [123,124]: 
\begin{eqnarray}
Z_{1} &=&const\int\limits_{\mathcal{M}_{1,1}}\frac{\omega _{W-P}}{\left( 
\func{Im}\tau \right) ^{12}\left| \Delta \right| ^{2}}  \TCItag{4.80} \\
&=&const\int\limits_{\mathcal{M}_{1,1}}\omega _{W-P}\exp (24h(\mathcal{E}/%
\mathbf{Q}))  \notag
\end{eqnarray}
with integration domain taken over the moduli space $\mathcal{M}_{1,1}$of
once punctured torus. In our previous work, Ref.[64], we had discussed
extensively the above integral for the case when the height function is
equal to zero. This leads immediately to the Kontsevich-Witten matrix
models, etc. Since the height function is closely connected with the
Arakelov theory [119,121], naturally, extension of Eq.(4.80) to the Riemann
surfaces of higher genus involves elements of this theory. This indeed was
accomplished by Manin [125] and Bost and Jolicoeur [126] ( see also paper by
Smit [127] ). The above formula and its higher genus generalizations
contains several important drawbacks. First, the modular form $\Delta (\tau
) $ is a cusp form. That is, it can exist only if the torus is cusped, i.e.
it has at least one puncture.(The dimension (and, hence, the number) of cusp
forms surely depends upon the actual number of cusps). This topology is the
simplest possible for the case of 2+1 gravity as it was extensively
discussed in our earlier work, Ref.[4]. But, in string theory the puncture
is normally associated with some particle via vertex operator insertion. 
\textbf{By definition}, the string partition function \textbf{without} 
\textbf{external} \textbf{sources (or sinks)} should \textbf{not} contain
punctures. Second, the expression for the height makes sense in the
Diophantine geometry dealing with finite number of points on the algebraic
curves. The Chowla-Selberg formula, Eq.(4.72), is in perfect agreement with
such statement since it involves \textbf{discrete} $\tau ^{\prime }s$ coming
from the imaginary quadratic fields. Use of Eq.(4.80) destroys the complex
multiplication option and, hence, disconnects the l.h.s. of Eq.(4.72) from
the r.h.s.\footnote{%
This can be repaired if the traditional Teichmuller/moduli space is replaced
by its $p$-adic analogue [128]. Even then, as the recent developments
indicate [129,130], one still will end up with the p-adic version of the
results presented in Section 5 below.} If, however, the complex
multiplication is destroyed (not used, ignored) then, according to Weil,
page 40, paragraph 7, Ref.[110 ], the very basic transformation law,
Eq.(C.7), of Appendix C for the modular functions is violated. Third, the
genus zero partition function seemingly \textbf{does not} require
integration over the moduli space [7]. This, however, is suspicious. Indeed,
the open string world sheet for genus zero case is represented by the upper
Poincare halfplane \textbf{H} which is a model of hyperbolic (that is non
Euclidean) space. The minimal number of open external states is four and
they appear as insertions at the boundary of \textbf{H} (e.g. see
Fig.1.15(c) of Ref.[7]). Technically speaking, these are just the cusp
points. The four times punctured sphere is a \textbf{hyperbolic} surface
whose cover is \textbf{H} with four cusps. The \ locations of cusps is not
fixed. The motion of the cusp points is described by the motion in the
parameter, i.e. in the Teichm\"{u}ller, space. Nag had demonstrated [9] that
the Teichm\"{u}ller space of the four times punctured sphere is \textbf{the
same} as of once punctured torus. Notice, that in view of these arguments
there is no need to discuss separately open and closed strings. This
explains why the Veneziano amplitude makes more sense than the
Shapiro-Virasoro. If one accepts the above arguments, then, 2+1 gravity and
string theory are in agreement with each other so that formally Eq.(4.80) is
a partition function for ''time-reduced'' 2+1 gravity. This was noticed
already in our earlier work, Ref.[64] and further developed by Krasnov [131]%
\footnote{%
For more recent references, please, see the footnote \# 3.}. Fortunately,
there is much more efficient alternative route of obtaining \ the
multiparticle Veneziano amplitudes. It is presented in the next section.

\section{Fermat's hypersurfaces and Veneziano amplitudes}

\subsection{The Picard-Fuchs equations}

In Section 3 we had shown that the standard Veneziano four particle
amplitude is just one of the periods of the Jacobian variety for the Fermat
curve. For physical applications it is of interest to obtain the
multiparticle amplitudes. This can be done in two ways: either by explicit
calculation based on generalization of the results presented in Section 3 or
by considering some sort of equations whose solutions produce the desired
periods. The situation in the present case is very similar to that
encountered in mirror symmetry calculations [57]. Because of this similarity
we will be brief in discussing the second option. But, we feel, that this
option should be mentioned explicitly in the text because it provides some
additional insight into physical and mathematical meaning of periods and
their calculation.

We begin with the simplest example borrowed from the pedagogically written
paper by Griffiths [132]. He considers calculation of the period of the
following integral along the closed contour $\Gamma $ in the complex
z-plane: 
\begin{equation}
\pi (\lambda )=\oint\limits_{\Gamma }\frac{dz}{z(z-\lambda )}.  \tag{5.1}
\end{equation}
Since this integral depends upon parameter $\lambda $ the period $\pi
(\lambda )$ is some function of $\lambda $ which can be determined as
follows. Differentiate $\pi (\lambda )$ in Eq.(5.1) with respect to $\lambda
.$This produces: $\pi ^{\prime }(\lambda )=\oint\limits_{\Gamma }\frac{dz}{%
z(z-\lambda )^{2}}$ . The combination 
\begin{equation}
\lambda \pi ^{\prime }(\lambda )+\pi (\lambda )=0  \tag{5.2}
\end{equation}
produces the desired differential equation which enables us to calculate $%
\pi (\lambda ).$This simple result can be vastly generalized to cover the
case of periods of integrals of the type 
\begin{equation}
\Omega (\lambda )=\oint\limits_{\Gamma }\frac{P(z_{1,}...,z_{n})}{%
Q(z_{1,}...,z_{n})}dz_{1}\wedge dz_{2}...\wedge dz_{n}  \tag{5.3}
\end{equation}
The equation $Q(z_{1,}...,z_{n})=0$ determines the variety. It may contain
parameter (or parameters) $\lambda $ so that the polar locus of values of $%
z^{\prime }s$ satisfying equation $Q=0$ depends upon this parameter. By
analogy with Eq.(5.2) one obtains 
\begin{equation}
\sum\limits_{n=0}^{k}P_{n}(\lambda )\frac{d^{n}}{d\lambda ^{n}}\Omega
(\lambda )=0.  \tag{5.4}
\end{equation}
This is the Picard-Fuchs equation for periods. To solve this equation one
needs to know the explicit form of polynomials $P_{n}(\lambda )$ . In
general this is not an easy task as it was demonstrated by Manin [133] many
years ago. And, because this is not an easy task, we believe, that
calculation of periods using generalization of section 3 is more efficient.
Nevertheless, method of differential equations teaches us things which are
not immediately apparent when direct calculations of periods is made.

In Section 4.1. we noticed that for the elliptic curves the periods can be
calculated as solutions of the hypergeometric equation, Eq.(4.6). Surely,
this equation is of Picard-Fuchs type. However, in Eq.(4.6) there is no
explicit differentiation with respect to parameter $\lambda $ and, hence,
now we would like to correct this deficiency. Fortunately, this task was
accomplished by Manin [133]. We follow Ref.[134], however, which is more
elementary. The Legendre form, Ref.[83], page 179, of the elliptic curve $%
\mathcal{E}$ is given by $y^{2}=x(x-1)(x-\lambda ).$ The invariant
differential form $\omega $ on $\mathcal{E}$ is given by 
\begin{equation}
\omega (\lambda )=\frac{dx}{2y}=\frac{dx}{2\sqrt{x(x-1)(x-\lambda )}}. 
\tag{5.5}
\end{equation}
Differentiation with respect to $\lambda $ produces 
\begin{equation}
\lambda (\lambda -1)\frac{\partial ^{2}}{\partial \lambda ^{2}}\omega
(\lambda )+(2\lambda -1)\frac{\partial }{\partial \lambda }\omega (\lambda )+%
\frac{1}{4}\omega (\lambda )=d(\frac{-\sqrt{x(x-1)(x-\lambda )}}{2(x-\lambda
)^{2}}).  \tag{5.6}
\end{equation}
Integrating this relation along the contours 2$\int\limits_{\lambda
}^{1}=\oint\nolimits_{\alpha }$ and/or 2$\int\limits_{0}^{1}=\oint%
\nolimits_{\beta }$ we obtain the equation for periods 
\begin{equation}
\lambda (\lambda -1)\frac{\partial ^{2}}{\partial \lambda ^{2}}\Omega
(\lambda )+(2\lambda -1)\frac{\partial }{\partial \lambda }\Omega (\lambda )+%
\frac{1}{4}\Omega (\lambda )=0  \tag{5.7}
\end{equation}
in accord with Manin, Ref.[133]. Let us recall [83] that for the elliptic
curve in the Legendre form the $j-$invariant is known to be 
\begin{equation}
j(\lambda )=2^{8}\frac{(\lambda ^{2}-\lambda +1)^{3}}{\lambda ^{2}(\lambda
-1)^{2}}.  \tag{5.8}
\end{equation}
Use of this invariant allows us to compare elliptic curves between each
other. In particular, if two curves are isogenous they have the same $j-$%
invariant. This means that $j-$invariant (or parameter $\lambda )$ describes
the family of all elliptic curves and, hence, $\lambda $ can be identified
with the point in the moduli space of elliptic curves. This means that the
differential operator in Eq.(5.7) is defined on the moduli space of elliptic
curves and \ it effectively describes the deformation of complex structure
for such curves. Since the moduli space is non Euclidean, the differential
operator in Eq.(5.7) is actually defined in some ''curved'' space
characterized by some connection (Gauss-Manin connection). The commutator of
covariant derivatives on such space produces zero curvature (so that
connection is flat but not trivial nevertheless). Surely all this machinery
can be extended to more sophisticated algebraic varieties. A good summary of
results applicable to general case can be found in Refs.[36,134,135]. In
this paper we are not going to use these results however and, hence, the
above discussion is only meant to indicate the alternative route to reach
the same destination.\emph{The necessity to mention about such a route comes
from the fact }$\emph{that}$\textbf{\ }$\mathbf{it}$ $\mathbf{actually}$ $%
\mathbf{provides}$ $\mathbf{the}$\textbf{\ }$\mathbf{unified}$ $\mathbf{%
description}$ $\mathbf{of}$\textbf{\ }$\mathbf{both}$ $\mathbf{conformal}$ $%
\mathbf{and}$ ''$\mathbf{string"}$\textbf{\ }$\mathbf{theories}$ as
mentioned briefly already in Section 4.1.Although the Picard-Fuchs equation
for the B-P type singularities can be easily derived (as discussed in the
Appendix D), this route for obtaining the multiparticle Veneziano amplitudes
is not unique. In the next subsection we discuss another route closer in
spirit to that envisioned by Landau [24] and further developed by Pham [25]
and others.

\textbf{Conjecture}: Based on the results of Pham [25], Lefshetz [136] and
Griffiths [132], and those presented in this work, it is natural to expect
that \textbf{any} scattering amplitude (correlation function) in particle
physics (in conformal field theories) should come out as solution of some
kind of the Picard-Fuchs equation and, hence, should be interpreted as one
of the periods of differential forms ''living'' on the moduli space
appropriate for the particular scattering problem. In the case of conformal
field theories this conjecture could be considered as actually proven in
view of the work of Schehtman and Varchenko [137] (to be discussed briefly
below in connection with hyprplane arrangements) while in the case of
particle physics this work along with those on mirror symmetry [57] (and
references therein) could \ be considered as first steps towards its proof.

\subsection{ Veneziano amplitudes from Fermat hypersurfaces}

To obtain the multiparticle Veneziano amplitude we have to generalize
results of Section 3.2. It is appropriate to mention at this point that
development of the multiparticle Veneziano amplitudes leading to string
theories (as discussed in detail in Ref.[19]) differs markedly from
presentation which follows. Our presentation is based on works of Deligne
[37] and Milnor [26]. Additional mathematical details and references can be
found in Ref.[38].

In section 3.2 the main object of interest was the differential 1-form,
Eq.(3.24), leading to the Veneziano amplitude, Eq.(3.33). Now we need to
find the analogous form for the Fermat hypersurface, Eq.(3.15), with $%
a_{0}=\cdot \cdot \cdot =a_{n}=N.$ To do so, we would like to remind our
readers about some facts from the theory of Riemann surfaces to be
generalized to more complicated case of Fermat hypersurfaces. In the case of
Riemann surface $R$ there are three types of differential forms $\omega $.
For the first and second type the period integrals $\int\limits_{\gamma
}\omega $ are non-zero only for \textbf{non-null} homologous cycles $\gamma $
on $R$ whereas for the third type one obtains the so called \emph{residual
periods} associated with poles of differential forms $\omega .$They can
occur for null homologous cycles as well and, hence, are of no interest to
us [138]. Suppose now that we have a $p$-form $\omega $ on a manifold 
\textbf{X}. If we integrate such a form over $p$-cycle, then one obtains a
period (usually a complex number) depending only on a homology class of the
cycle provided that $\omega $ is closed (that is $d\omega =0).$The homology
classes form $p^{th}$ homology group of \textbf{X} with complex
coefficients, H$_{p}$($p$,\textbf{C}).The dual to it is $p^{th}$ cohomology
group H$^{p}$($p$,\textbf{C})$\simeq $H$_{p}$($p$,\textbf{C})$^{\ast }.$ A
closed $p$-form $\omega $ yields a cohomology class $[\omega ]\in $H$^{p}$(X,%
\textbf{C}). The Stoke's theorem shows that this class remains unchanged if
we add to it an exact form, that is differential $d\psi $ of the $p-1$ form $%
\psi .$This circumstance allows us to introduce the equivalence relation
and, hence, to study the quotients, e.g. the $\emph{de}$ $\emph{Rahm}$ $%
p^{th}$ \emph{cohomology group} H$_{DR}^{p}(\mathbf{X})=C^{P}(\mathbf{X}%
)/E^{p}(\mathbf{X}),$ with $C^{p}$ and $E^{p}$ being respectively the vector
spaces of closed and exact $p$-forms with complex coefficients. According to
de Rham, Ref.[138,139], there is a canonical homomorphism: H$_{DR}^{p}(%
\mathbf{X})\rightarrow $H$^{p}$(\textbf{X},\textbf{C}).Consider first how
all this applies to the Riemann surface $R$ which is complex one-dimensional
manifold \textbf{X}. For such manifold H$_{DR}^{1}(\mathbf{X})$ is
isomorphic to the vector space of holomorphic 1-forms. Each holomorphic
1-form $\omega $ is closed. This is so because $d\omega $ is a holomorphic
2-form which is equal to zero automatically since \textbf{X} is one
dimensional. From topology it is known that for the Riemann surface of genus 
$g$ the cohomology group H$^{1}$(\textbf{X},\textbf{C}) has dimension 2g .
Meanwhile the dimension of holomorphic 1-forms is g so that the holomorphic
1-forms yield only 1/2 of cohomology. To repair this situation one needs to
consider not only the cohomology classes coming from the holomorphic forms
but also coming from the \emph{anti holomorphic} forms. For a general
manifold \textbf{X}, following Hodge, one introduces the differential $(p,q)$%
- form $\omega $ (in local complex coordinates $z_{1},...,z_{n})$ such that 
\begin{equation}
\omega =\sum\limits_{{}}a_{i_{1,}\cdot \cdot \cdot ,i_{p}},_{j_{1,}\cdot
\cdot \cdot ,j_{q}}dz_{1}\wedge \cdot \cdot \cdot \wedge dz_{i_{p}}\wedge d%
\bar{z}_{j_{1}}\wedge \cdot \cdot \cdot \wedge d\bar{z}_{j_{q}}  \tag{5.9}
\end{equation}
Each $m$-form on \textbf{X} can be expressed uniquely as a sum of $(p,q)$%
-forms with $p+q=m.$

\bigskip The following theorem plays a very important role.

\bigskip

\textbf{Theorem 5.1}. \emph{If \textbf{X} is compact complex manifold, then
it is} \emph{Hodge }$\emph{manifold}$ $\emph{(or}$ \emph{Hodge-type manifold}%
) \emph{if and only if it is a projective algebraic manifold at the same
time.}

\bigskip

This statement is known as the Kodaira's embedding theorem, Ref.[139], ch-r
6. Evidently, the projective algebraic manifold must be complex.

\bigskip

\textbf{Definition 5.2}. A complex projective manifold \textbf{X} is a
nonsingular subvariety in \textbf{CP}$^{n}$ consisting of the common zeroes
of a system of homogenous polynomial equations in coordinates $%
(z_{0},...,z_{n})$ of \textbf{C}$^{n+1}.$

\bigskip

On \textbf{C}$^{n+1}$one can introduce the Hermitean scalar product (which
is just the finite dimensional analog of what is being used for the Hilbert
space of quantum mechanics): 
\begin{equation}
<z,z>=\sum\limits_{i=0}^{n}z_{i}\bar{z}_{i}.  \tag{5.10}
\end{equation}
With help of such defined scalar product it is possible to recover the
metric and the curvature for such spaces. Moreover, with little efforts this
can be extended to \textbf{CP}$^{n}[139].$ If the Hermitian metric $h$ of
the projective manifold \textbf{X} embedded in \textbf{CP}$^{n}$ is given by 
\begin{equation}
h=\frac{i}{2}\sum h_{\mu \nu }(z)dz_{\mu }\otimes d\bar{z}_{\mu }  \tag{5.11}
\end{equation}
then, the fundamental (1,1) form \ $\Omega $ is given by 
\begin{equation}
\Omega =\frac{i}{2}\sum h_{\mu \nu }(z)dz_{\mu }\wedge d\bar{z}_{\mu }. 
\tag{5.12}
\end{equation}

\textbf{Definition 5.3}. The Hermitian metric $h$ on \textbf{X} is of
K\"{a}hler type if $\Omega $ is closed, i.e. $d\Omega =0.$

Not all complex manifolds admit the K\"{a}hler metric\footnote{%
Comprehensive and readable account of K\"{a}hler manifolds is given in Andre
Weil's book, Ref.[140 ]. The book by Wells, Ref.[139], is also an excellent
source.}. Fortunately, those which admit have direct physical relevance. In
particular, all known physically meaningful symplectic manifolds of exactly
integrable finite dimensional classical dynamical system admit K\"{a}hler
type metric and $\Omega $ form.[40,141,142]. Using Kodaira's embedding
theorem we conclude that \textbf{all physically interesting projective
manifolds of K\"{a}hler type are} \textbf{also manifolds of Hodge type%
\footnote{%
For additional details, please read Section 6.}}.

In particular, let us consider the case of Riemann surface of genus g. It
can be mapped into g dimensional complex torus (the Jacobian) under
appropriate conditions (e.g. it must be a polarized Abelian variety)
[118,134,139,140]. And such torus surely admits K\"{a}hler metric and, hence
it is a Hodge-type manifold with Hodge decomposition: 
\begin{equation}
\text{H}^{1}\text{(\textbf{X},\textbf{C})}=\text{H}^{1,0}\text{(\textbf{X})}%
\oplus \text{ H}^{0,1}\text{(\textbf{X}).}  \tag{5.13}
\end{equation}

\bigskip

\textbf{Remark 5.4.} The condition that an abelian variety is of \emph{%
complex multiplication type} can be interpreted \ as a choice of basis for
the above decomposition, Eq.(5.13),[143].

\textbf{Remark 5.5}.The situation becomes more complicated if complex
hypersurface possess some kind of singularities, e.g B--P type discussed in
Section 3.2. In this case the \emph{mixed} Hodge structures should be
considered instead of \emph{pure} Hodge structures as shown by Deligne [37].
\ Systematic up to date exposition of this topic can be found in Ref.[135].
\ Mathematically rigorous and physically accessible treatment of these
structures can be found in excellent review paper by Varchenko [144]
(Appendix D).

\bigskip

Complex multiplication (discussed in Section 4.2) is closely associated with
manifolds of Hodge type. Fortunately, this is indeed the case for both the
Fermat curves considered in Section 3 [34] and for the Fermat hypersurfaces
[37] to be considered below in this section. Construction of Hodge type
manifolds admitting complex multiplication is concisely described in the
paper by Weil [33] and also in his book, Ref.[140]. More recent and much
more advanced presentation of this topic involving theories of motives,
Tannakian categories, crystalline cohomology, etc. can be found in Refs.
[38,145].

\bigskip

Consider now the projective form of Eq.(3.15) adopted to the Fermat case,
i.e. 
\begin{equation}
z_{0}^{N}+\cdot \cdot \cdot +z_{n}^{N}-z_{n+1}^{N}=0.  \tag{5.14}
\end{equation}
Let us consider it as a variety over the cyclotomic number field $K$=\textbf{%
Q}(exp\{2$\pi i/N\}).$ Then, using results of Appendix A we can rewrite it
as a hyperplane in \textbf{PC}$^{n},i.e.$%
\begin{equation}
z_{0}+\cdot \cdot \cdot +z_{n}-z_{n+1}=0.  \tag{5.15}
\end{equation}
Such a hyperplane can be presented as a point in the Grassmanian as we had
discussed in our earlier work, Ref.[64]. Each hyperplane, Eq.(5.15), is just
a result of a particular solution of the associated system of cyclotomic
equations (e.g. see Eq.(A.3)). Naturally, there are many other solutions and
each of them will be represented by its own point in the Grassmanian. One
can think about the transformations which connect different hyperplanes.
Such transformations are presented by the matrices $G$ belonging to the
rotation group. Among these transformations might be those, call them $N$,
which have the fixed point. If this is the case, then, one should consider a
qotient $G/N$. This quotient is essentially what is known in the literature
as the \emph{Griffiths domain} [139,146,147].The \textbf{Griffiths domain is
the moduli space for periods}. We shall demonstrate momentarily that each
hyperplane is associated with its period (that is with the\ multiparticle
Veneziano amplitude) so that the totality of hyperplanes are in one to one
correspondence with the totality of periods. This fact is discussed further
in Section 5.3.2. and Appendix D. In principle, if one can construct the
matrix of periods then,one can determine the Griffiths domain as well. To
keep focus of attention of our readers on physics, mathematically relevant
issues of physical importance are discussed in the Appendix D. This allows
us to come to the main purpose of this work- calculation of multiparticle
Veneziano amlitudes.

To this purpose, we need to use the Corollary 2.11. from Griffiths work,
Ref.[132]. It is related to the integral, Eq.(5.3).To explain things better,
we adopt the explanation of this Corollary using discussion from the
Brieskorn's book, Ref.[138], page 647.

Let ($x_{0},...,x_{n})$ be homogenous coordinates of the point in the \emph{%
projective} space and ( $z_{1},...,z_{n})$ be the associated with them
coordinates in the \emph{affine} space with $z_{i}=x_{i}/x_{0}$ , then the
rational $n$-form is given by 
\begin{equation}
\frac{p(z_{1,}...,z_{n})}{q(z_{1,}...,z_{n})}dz_{1}\wedge dz_{2}...\wedge
dz_{n}  \tag{5.16}
\end{equation}
with rational function $p(z)/q(z)$ being a quotient of two homogenous
polynomials of \textbf{the same} degree. Under substitution $%
z_{i}=x_{i}/x_{0}$, the form $dz_{1}\wedge dz_{2}...\wedge dz_{n}$ changes
to 
\begin{equation}
dz_{1}\wedge dz_{2}...\wedge
dz_{n}=x_{0}^{-(n+1)}\sum\limits_{i=0}^{n}(-1)^{i}x_{i}dx_{0}\wedge
...\wedge d\hat{x}_{i}\wedge ...dx_{n}  \tag{5.17}
\end{equation}
where the hat on top of $x_{i}$ means that it is excluded from the product.
Define now the form 
\begin{equation}
\omega _{0}:=\sum\limits_{i=0}^{n}(-1)^{i}x_{i}dx_{0}\wedge ...\wedge d\hat{x%
}_{i}\wedge ...dx_{n}.  \tag{5.18}
\end{equation}
In order to rewrite the form, Eq.(5.16), in terms of $x$ coordinates it is
sufficient to multiply the denominator of $p(z)/q(z)$ by $x_{0}^{n+1}$ and
to replace $z$ by $x$ in the quotient. Thus one obtains a homogenous
expression 
\begin{equation}
\omega =\frac{P(x)}{Q(x)}\omega _{0}  \tag{5.19}
\end{equation}
in which the (degree of $Q$)=(degree of $P$) +($n$+1). This is Griffiths
result, Corollary 2.11.of Ref.[132]. Conversely, each homogenous form,
Eq.(5.19), yields an affine rational form, Eq.(5.16), by substitution: $%
x_{0}=1,x_{i}=z_{i}.$

Next, consider the set $X(S^{1})$ of numbers which we call the \textbf{%
spectrum} of the critical point (Appendix D) of the B-P singularity, Eq.
(3.7). It is given by 
\begin{eqnarray}
X(S^{1}) &=&\{\text{\b{a}}\in (\mathbf{Z}/N\mathbf{Z})^{n+2}\equiv (\mathbf{Z%
}/NZ)\times \cdot \cdot \cdot \times (\mathbf{Z}/N\mathbf{Z})\mid \text{\b{a}%
=(}a_{0},...,a_{n+1}),\tsum\nolimits_{i}a_{i}=0\func{mod}N\}  \notag \\
&&  \TCItag{5.20a}
\end{eqnarray}
Let $<a_{i}>$ denote \ the representative of $a_{i}$ in \textbf{Z} \ such
that $1\leq <a_{i}>\leq N-1$(e.g. see Eq.(3.17a))$.$ Also, define the
average 
\begin{equation}
<\text{\b{a}}>=\frac{1}{N}\tsum\nolimits_{i}<a_{i}>  \tag{5.20b}
\end{equation}
then, for the Fermat hypersurface the differential form $\omega $,
Eq.(5.19), can be written as 
\begin{equation}
\omega =\frac{x_{0}^{<a_{0}>-1}...x_{n+1}^{<a_{n+1}>-1}}{\left(
x_{0}^{N}+\cdot \cdot \cdot -x_{n+1}^{N}\right) ^{<\text{\b{a}}>}}\omega _{0}
\tag{5.20c}
\end{equation}
with $\omega _{0}$ defined by Eq.(5.18) (with $n$ replaced by $n+1$). With
such form at our disposal it will be advantageous for us to use the affine
representation of this form using results just described. To this purpose,
by analogy with developments in Section 3.2, e.g. see the discussion after
Eq.(3.15), we make change of variables : $z_{i}$ $=t_{i}^{\frac{1}{N}}\exp (%
\frac{2\pi i}{N})\equiv t_{i}^{\frac{1}{N}}\zeta $ , provided that $%
\sum\nolimits_{i}t_{i}=1,t_{i}\geq 0,$ thus forming the $n+1$ simplex $%
\Delta .$ Such change of variables allows us to consider a map $f(\Delta )$: 
$\Delta \rightarrow V(F)$ from the simplex $\Delta $ to the affine variety $%
V(F)$%
\begin{equation}
Y_{0}^{N}+\cdot \cdot \cdot +Y_{n}^{N}=1,\text{ }Y_{i}=x_{i}/x_{n+1}\equiv
z_{i}.  \tag{5.21}
\end{equation}
This map can be made in many ways: for example we can suppress all $\zeta
^{\prime }s$ , then, suppress all, except one, then, all, except two, etc.
and still will get the desired mapping. To avoid guessings, we need to
reobtain back the result, Eq.(3.31), for the simplest four particle
Veneziano amplitude. Clearly, such amplitude should be obtainable from
general result, Eq.(5.20), i.e. it should come up as period of the integral $%
\int\nolimits_{\gamma }\omega .$ Such an integral contains a pole. There
should be a procedure generalizing that known in the standard complex
analysis of one variable of calculating residues of multidimensional complex
integrals. Fortunately such procedure exist and was developed by Leray, Pham
and others [136,148] in connection with Landau's work on analytical
properties of scattering amplitudes of Feynman's diagrams [24].With such
information at out disposal, we would like to reobtain the simplest
Veneziano amplitude now. Although results of Deligne [37] are of great help
in accomplishing this task, unfortunately, we cannot borrow them entirely
for (physical) reasons which will become clear upon comparing of our
derivations (below) with those by Deligne.

We begin with Eq.(3.31) written (up to constant factor of $N$) as follows: 
\begin{equation}
I=\frac{\left( -1\right) }{2\pi i}\zeta ^{\tfrac{-t}{2}}(1-\zeta
^{t})(1-\zeta ^{s})(1-\zeta ^{r})\Gamma (a)\Gamma (b)\Gamma (1-a-b) 
\tag{5.22}
\end{equation}
This form clearly calls for elimination of the factor $\zeta ^{\tfrac{-t}{2}%
} $for reasons of symmetry. This was actually done in Eq.(3.33). This time,
we would like to do such elimination more systematically. Hence, for the
time being we shall keep this factor dropping it at the end. Eq.(5.22) can
be presented equivalently in the following form: 
\begin{eqnarray}
\Gamma (a+b)I &=&(1-\zeta ^{s})(1-\zeta ^{r})\int\limits_{0}^{\infty
}\int\limits_{0}^{\infty }dx_{1}dx_{2}x_{1}^{a-1}x_{2}^{b-1}\exp
(-x_{1}-x_{2})  \TCItag{5.23} \\
&=&(1-\zeta ^{s})(1-\zeta ^{r})\int\limits_{0}^{\infty }dtt^{a+b-1}\exp
(-t)\int\limits_{0}^{1}d\hat{x}_{1}\hat{x}_{1}^{a-1}(1-\hat{x}_{1})^{b-1} 
\notag
\end{eqnarray}
In going from the first line to the second we have introduced new variables
: $x_{1}=\hat{x}_{1}t$ , $x_{2}=\hat{x}_{2}t$ , $x_{1}+x_{2}=t$ implying
that $\hat{x}_{1}+\hat{x}_{2}=1$ and the Jacobian of transformation equal to
one. The above can be rewritten therefore as 
\begin{equation}
I=(1-\zeta ^{s})(1-\zeta ^{r})\int\limits_{0}^{1}d\hat{x}_{1}\hat{x}%
_{1}^{a-1}(1-\hat{x}_{1})^{b-1}  \tag{5.24}
\end{equation}
Looking at Eq.s (5.17) and (5.20) and comparing with Eq.(5.24) we can
formally write: $I=(1-\zeta ^{s})(1-\zeta ^{r})\int\nolimits_{\gamma }\omega 
$ . And, indeed, this expression coincides with Eq.(3.31) in view of
Eq.(3.28).These results can be vastly generalized now.\ First, we would like
to notice that if we would make a change: $r\rightarrow -r$ , $s\rightarrow
-s,$ the arguments leading to Eq.(5.24) will remain unchanged. Therefore,
this means that this result can be used for calculation of \textbf{physical}
Veneziano amplitude, e.g. see Eq.s.(1.6) and (3.33). Second, based on this
observation, we can also change $\zeta ^{s}$ into $\zeta ^{-s}$ and $\zeta
^{r}$ into $\zeta ^{-r}$ without changing $a$ and $b$ in the integral of
Eq.(5.24). Vice versa : we can change $a$ and $b$ without change of phase
factors. To illustrate why this is possible consider as before change of
variables: $z_{i}$ $=t_{i}^{\frac{1}{N}}\exp (\frac{2\pi i}{N}).$ In the
integral, Eq.(5.24), we choose $z_{1}$ =$\hat{x}_{1}=t_{1}^{\frac{1}{N}}\dot{%
\zeta}$(where $\dot{\zeta}$ means that the $\zeta $ factor may or may not be
present. It may be present if we bring the external phase factors inside the
integral, Eq.(5.24)) and, accordingly, $z_{2}=$ $\hat{x}_{2}=t_{2}^{\frac{1}{%
N}}\dot{\zeta}$ so that for the Fermat curve in the affine form: $%
z_{1}^{N}+z_{2}^{N}=1,$ we obtain back equation for the simplex: $%
t_{1}+t_{2}=1.$Hence, we may consider the most general case of the integral
of the Veneziano-type: 
\begin{equation}
I_{V(F)}=\int\limits_{\Delta }Y_{1}^{a_{1}}\cdot \cdot \cdot
Y_{n+1}^{a_{n+1}}\frac{dY_{1}}{Y_{1}}\wedge \cdot \cdot \cdot \wedge \frac{%
dY_{n}}{Y_{n}}.  \tag{5.25}
\end{equation}
Naturally, it is determined up to a constant. The constant is easily
determined according to the following rules:

a) if the number of phase factors is even the overall sign of the integral
is ''+'', otherwise, it is ''-'';

b) the total phase factor is just $\zeta ^{\dot{a}_{1}+\dot{a}_{2}+\cdot
\cdot \cdot +\dot{a}_{n+1}},$ where $\dot{a}_{i}$ means that this factor may
be either zero or some rational fraction;

c) the sum $\sum\nolimits_{i=0}^{n+1}a_{i}=0\func{mod}N$ ;

d) upon change of variables:$Y_{i}=\dot{\zeta}t^{\frac{1}{N}}$ , the overall 
$c$-factor $c=\left( \frac{1}{N}\right) ^{n}$

\ \ emerges \ (and can be dropped);

f) the resulting integral 
\begin{equation*}
I_{V(F)}\dot{=}\int\limits_{\Delta }t_{1}^{b_{1}}\cdot \cdot \cdot
t_{n+1}^{b_{n+1}}\frac{dt_{1}}{t_{1}}\wedge \cdot \cdot \cdot \wedge \frac{%
dt_{n}}{t_{n}}
\end{equation*}
with $b_{i}=a_{i}/N$ can be calculated in the spirit of ingenious trick by
Deligne [37] which, of course, can be as well inferred directly from
Eq.(5.23).To this purpose, we multiply both sides of the last equation by $%
\Gamma (\sum\nolimits_{i=1}^{n+1}b_{i})I_{V(F)}=\Gamma
(\sum\nolimits_{i=1}^{n+1}b_{i})\int\limits_{\Delta }t_{1}^{b_{1}}\cdot
\cdot \cdot t_{n+1}^{b_{n+1}}\frac{dt_{1}}{t_{1}}\wedge \cdot \cdot \cdot
\wedge \frac{dt_{n}}{t_{n}}$ .\ For the r.h.s. we get 
\begin{eqnarray*}
&&\Gamma (\sum\nolimits_{i=1}^{n+1}b_{i})\int\limits_{\Delta
}t_{1}^{b_{1}}\cdot \cdot \cdot t_{n+1}^{b_{n+1}}\frac{dt_{1}}{t_{1}}\wedge
\cdot \cdot \cdot \wedge \frac{dt_{n}}{t_{n}} \\
&=&\int\limits_{0}^{\infty }\int\limits_{\Delta }\frac{dt}{t}%
t^{\sum\nolimits_{i=1}^{n+1}b_{i}}\exp (-t)t_{1}^{b_{1}}\cdot \cdot \cdot
t_{n+1}^{b_{n+1}}\frac{dt_{1}}{t_{1}}\wedge \cdot \cdot \cdot \wedge \frac{%
dt_{n}}{t_{n}}
\end{eqnarray*}
Let now $s_{i}=tt_{i}$ so that, as before, $t=\sum%
\nolimits_{i=1}^{n+1}s_{i},s_{i}\geq 0.$This then produces the following
integral 
\begin{eqnarray*}
&&\int\limits_{0}^{\infty }\cdot \cdot \cdot \int\limits_{0}^{\infty }\exp
(-\sum\nolimits_{i=1}^{n+1}s_{i})\text{ }s_{1}^{b_{1}}\cdot \cdot \cdot
s_{n+1}^{b_{n+1}}\frac{ds_{1}}{s_{1}}\wedge \cdot \cdot \cdot \wedge \frac{%
ds_{n}}{s_{n}}\wedge \frac{ds_{n+1}}{s_{n+1}} \\
&=&\Gamma (b_{1})\cdot \cdot \cdot \Gamma (b_{n+1})
\end{eqnarray*}
so that finally we obtain the contribution to the Veneziano amplitude 
\begin{equation}
I_{V(F)}\dot{=}\frac{\prod\limits_{i=1}^{n+1}\Gamma (b_{i})}{\Gamma
(\sum\nolimits_{i=1}^{n+1}b_{i})}  \tag{5.26}
\end{equation}
The total amplitude is the sum of the above with the appropriate phase
factors.

It should be clear by now why it is permissible to multiply both sides of
Eq.(3.33) by factor of $\zeta ^{\frac{t}{2}}($and, of course, to do the same
in the most general case).

\bigskip

\textbf{Remark 5.6}. Eq.(5.27) can be found in Gross paper, Ref.[34], where
it is given without derivation. The role of complex multiplication in
obtaining this result is strongly emphasized in his paper. The same equation
was formally obtained by Edwards [29] already in 1922.

\textbf{Remark 5.7} In accord with Section 3.2, different combinations of $%
a_{i}^{\prime }s$ correspond to different periods of Fermat hypersurface.
Each such period corresponds to a point in the Grassmanian so that different
periods correspond to different points. This makes calculation of the
Veneziano amplitudes \ similar to our earlier calculations of the
Witten-Kontsevich averages, Ref.[64].

\textbf{Remark 5.8}. There is yet another method of obtaining the results of
this subsection. It is based on the asymptotic analysis of complex
oscillatory integrals as discussed in the book by Arnol'd, Gusein-Zade and
Varchenko, Ref. [36]. Fortunately, the results of this reference not only
useful for reobtaining the multiparticle Veneziano amplitudes (as discussed
in the Appendix D) but also they provide a very economical way of getting
all the results associated with the mixed Hodge structures. These results
will be utilized in Section 5.3.2. in connection with the Hodge spectrum
which is in one-to-one correspondence with the particle mass spectrum.

\subsection{Dynamics of Seifert fibered phase and Veneziano amplitudes}

\subsubsection{Connections between zeta function and Alexander polynomial}

In our previous work, Ref.[4], we had considered dynamics of the Seifert
fibered phase of 2+1 gravity (e.g. see Appendix 3 for definitions)
associated with solution of the Witten-Kontsevich model. In this work we
have reobtained the partition function (the p-adic version of this function)
of 2+1 gravity in Section 3.1., Eq.(3.6), working with dynamics of gravity
in this phase. Earlier, we argued that such partition function is associated
also with the p-adic analogue of the four particle Veneziano amplitude,
Eq.s(1.15),(1.16b),(1.17). In Section 3.2. we introduced the generalized
Alexander polynomial, Eq.(3.20). According to Milnor [26] and Brieskorn
[27], such polynomial describes the multidimensional fibered knots and
links. Since according to the Corollary 3.7.the spectrum of particle masses
\ emerging as poles in four particle scattering amplitude is in one- to- one
correspondence with the appropriate Alexander polynomial describing fibered
knots and/or links in 3 dimensions, it is natural to expect that the same
will be true for the multiparticle scattering\footnote{%
As the development presented below indicates, this option is less convenient
than that emerging from the correspondence of mass spectrum with the Hodge
spectrum.}.

\textbf{Remark 5.9}. It should be noted at this point that the Alexander
polynomial may describe more than one knot/link. This happens in the case of
cable knots/links, Ref.[149], pages 121-122. In such cases one might think
of one-to-one correspondence between the mass spectrum and knot/link cable
families. This option is not going to be explored in this paper in view of
the remarks made in the footnote below.

To proceed, we need to explain why two apparently different equations,
Eq.(3.13) and (3.18), describe the same Alexander polynomial for the
Seifert-fibered knots/links. The connection between the above equations
could be thought as a corollary to a very deep result of Grothendieck [62]
which can be formulated in the form of the following

\bigskip

\textbf{Theorem 5.10}. \emph{The Alexander polynomial for \textbf{algebraic}
knots/links embedded into }$S^{2n+1}is$\emph{\ equal to the product of
cyclotimic polynomials.}

\bigskip

\textbf{Remark 5.11.}For the purposes of this work it is sufficient to keep
in mind that the link \textbf{K}=$V_{B-P}(f)\cap S^{2n+1}$ introduced in
Section 3.2. is algebraic [149].

\textbf{Remark 5.12}. The Alexander polynomial $\Delta _{\mathbf{K}}(t)$,
Eq.(3.18), can be obtained from the ratio of cyclotomic polynomials given
below 
\begin{equation}
\Delta _{\mathbf{K}}(t)=\frac{(t^{pq}-1)(t-1)}{(t^{p}-1)(t^{q}-1)}. 
\tag{5.27}
\end{equation}
Indeed, using equations (A.6)-(A.8) \ of Appendix A we obtain for instance, 
\begin{equation*}
t^{q}-1=(t-1)(t^{q-1}+\cdot \cdot \cdot )=\prod\limits_{i}^{q}(t-\xi _{i}),
\end{equation*}
with $\xi _{i}$ being one of the roots of unity (including one). Use of such
type of products in Eq.(5.27) produces, indeed, the desired result,
Eq.(3.18). For 3-dimensional Seifert-fibered knots /links the result,
Eq.(5.27), had been obtained by Le Trang [150] and, independently, by
Sumners and Woods [151]. The result, Eq.(5.27), leads to the following
additional observation originally made by Milnor [26]. It appears, that with
help of Eq.s (2.30),(3.16), (3.17) one can connect the Alexander polynomial
and the Weil-type zeta function. Earlier, in Eq.(3.6) we had obtained such
zeta function for the p-adic version of Veneziano amplitude. Moreover, for
the torus the Weil-type zeta function, Eq.(3.35), was obtained earlier by
Dwork [102], e.g. see Remark.4.7. Looking at Eq.(3.35) it is educational to
rewrite it in the following form: 
\begin{equation}
\frac{1}{Z_{K}(u)}=\frac{(u-1)(qu-1)}{(\alpha u-1)(qu/\alpha -1)}. 
\tag{5.28}
\end{equation}
Assume now that $q$=$t^{mn},\alpha =t^{m}$ so that $q/\alpha =t^{n}.$ In
addition, assume that $u=u^{n}($ or $u=u^{m},etc.).$ Then, using
cyclotomic-type expansion ( dispayed above) and letting $u=1$ at the end of
calculations produces back the Alexander polynomial, Eq.(3.18), for the
torus knots. This then implies that, at least for torus knots, $%
Z_{K}(t)\Delta _{\mathbf{K}}(t)=1.$ This result is not totally surprising if
one recalls [152] that for a well behaving matrix \textbf{A }one has the
following identity 
\begin{equation}
\left[ \det (\mathbf{I}-t\mathbf{A})\right] ^{-1}=\exp
\{\sum\limits_{n=1}^{\infty }\frac{t^{n}}{n}\left( tr\mathbf{A}^{n}\right) \}
\tag{5.29}
\end{equation}
Looking at Eq.(3.13) and recalling that det \textbf{M}=$\pm 1$ (e.g. read
Section 3 of Ref.[4]) and comparing the r.h.s. of the above identity with
Eq.(2.25) one obtains the desired relationship between the zeta function and
the Alexander polynomial. In Section 3 of Ref.[4] we have demonstrated that
the monodromy matrix \textbf{M} is indeed involved in surface dynamics.

\textbf{Remark 5.13.} The cyclotomicity (Theorem 5.9) implies that such
dynamics is\emph{\ }characteristic \emph{only} for the Seifert fibered phase
(regime). This is in accord with results of Milnor's book [26], e.g read
chapter 10.

The question remains: will such relationship hold for surfaces more
complicated than torus? Although the answer to this question can be found in
Milnor's book [26], we prefer to approach it differently. To this purpose
let us notice that the arguments used in obtaining of Eq.(5.28) can be also
applied to zeta function, Eq.(3.6), the p-adic Veneziano amplitude. In this
case we obtain 
\begin{equation}
\frac{1}{Z_{f}(t)}=\frac{1-pt}{1-t}=\prod\limits_{\xi ^{n}=1,\xi \neq
1}(t-\xi )=\Delta _{\mathbf{K}_{0}}(t),  \tag{5.30}
\end{equation}
where $\Delta _{\mathbf{K}_{0}}(t)$ is the Alexander polynomial for the
''empty'' knot [58].This result follows from Milnor's general theory
discussed in Section 3.2. Indeed, let in Eq.(3.7) there is only one term ,
i.e.$f(z)=z^{n}.$We can still construct a circle mapping, Eq.(3.9), and
after applying the same type of arguments leading to the Alexander
polynomial, Eq.(3.20), we will get $\Delta _{\mathbf{K}_{0}}(t).$ Just made
observation can be vastly generalized with help of the fundamental theorem
of Thom and Sebastiani [153]. The arguments by presented in Ref.[154] are
based on this theorem and summarized in the following \ additional theorem
(Oka, Ref.[154]):

\bigskip

\textbf{Theorem 5.14.} \emph{Let f be a polynomial in }$\mathbf{C}^{n}\times 
\mathbf{C}^{m}$\emph{\ such that f(z,w)=g(z)+h(w) for each} \emph{(z,w)}$\in 
\mathbf{C}^{n}\times \mathbf{C}^{m}$\emph{\ where g(z) and h(z) are weighted
homogenous polynomials in }\textbf{C}$^{n}$\emph{\ and }\textbf{C}$^{m}$%
\emph{\ respectively. Let }$X=f^{-1}(1)\subset \mathbf{C}^{n}\times \mathbf{C%
}^{m},y=g^{-1}(1)\subset \mathbf{C}^{n}$\emph{\ and }$\emph{Z=h}^{-1}\emph{%
(1)}\subset \mathbf{C}^{m}.$ \emph{Then there is a natural homotopy
equivalence }$\alpha $\emph{\ between X and Y}$\star Z$ \emph{where Y}$\star
Z$\emph{\ is the join of Y and z with strong topology}.

\bigskip

As a corollary, the same author obtains the following result:

\bigskip

\textbf{Theorem 5.15}.\emph{\ Let g, h and f be the same as in Theorem 5.14.
and assume that g and h have isolated critical points at the origin. Then
the characteristic\ (Alexander) polynomials of the associated fiberings }$%
\Delta _{g}(t)$\emph{, }$\Delta _{h}(t)$\emph{\ and }$\Delta _{f}(t)$\emph{\
satisfy the equation} 
\begin{equation}
\Delta _{f}(t)=\Delta _{g}(t)\star \Delta _{h}(t).  \tag{5.31}
\end{equation}
\emph{That is letting} $\Delta _{g}(t)=\prod\nolimits_{i}(t-\lambda _{i})$
and $\Delta _{h}(t)=\prod\nolimits_{i}(t-\mu _{i})$ \emph{we should obtain} $%
\Delta _{f}(t)=\prod\nolimits_{i,j}(t-\lambda _{i}\mu _{j}).$

\textbf{Remark} \textbf{5.16}. The concept of a join could be found in any
book on topology, e.g. see Ref.[60]. We actually have effectively used it
already when we had considered the map $\Delta \longrightarrow V(F)$ as
discussed before Eq.(5.21). Rougly speaking, the concept of a join could be
understood based on the following example taken from Ref.[155] (see also
Pham[25] and Milnor[26]). Let \textbf{Z}/$p$ and \textbf{Z}/$q$ be finite
cyclic groups consisting of all p-th, respectively q-th roots of unity. The
join \textbf{J}=\textbf{Z}/$p\star $\textbf{Z}/$q$ could be thought as
totality of all vectors $(s\xi ,t\eta )\in \mathbf{C}^{2}$ with $s,t$ $\geq
0,s+t=1,$and $\xi \in $\textbf{Z}/p, $\eta \in $\textbf{Z}/$q$ . For \ the
circle map of the type 
\begin{equation*}
\psi (z,w)=\frac{z^{p}+w^{q}}{\left| z^{p}+w^{q}\right| }
\end{equation*}
it could be shown [26], that $\mathbf{J}\ $is the deformation retract of the
fiber $\psi ^{-1}(1).$

\textbf{Remark 5.17.}From Theorems 5.14. and 5.15 it follows that the
Alexander polynomial of the multidimensional \ fibered knot/ link can be
constructed with help of the Alexander polynomials for empty knots, e.g. $%
\Delta _{g}(t)=\prod\nolimits_{i}(t-\lambda _{i}),$ Eq.(5.30), associated
with zeta functions of circle maps related to periodic surface automorphisms
characteristic for the Sefert-fibered phase (Appendix B)\footnote{%
Essentially the same conclusions can be reached using the laguage of
singularity theory (Appendix D), e.g. see the Fubini's theorem, Eq.(D.7),
and the discussion which follows.}.

With all useful information just discussed we still have not provided an
answer to the question we had posed earlier about the relationship between
the Alexander polynomial and zeta function for multicomponent links.
Fortunately, such problem was discussed by A'Campo [156] so that it remains
only to illustrate his findings. In particular, for the B-P type
singularity, e.g. see Eqs.(3.7),(3.8), the Alexander polynomial $\Delta (t)$
should be 
\begin{equation}
\Delta (t)=[(t-1)^{-1}(t^{N}-1)^{\chi (S_{N})}]^{(-1)^{n}}  \tag{5.32}
\end{equation}
with $\chi (S_{N})$ being the Euler characteristic of the complement of the
Fermat hypersurface in \textbf{CP}$^{n}.$ A'Campo provides the following
result for the characteristic : $\chi (S_{N})=\frac{1-(1-N)^{n+1}}{N}.$
Next, if the Milnor number (that is the multiplicity of the critical point)
is given by 
\begin{equation}
\mu =(-1)^{n}[-1+N\chi (S_{N})]=(N-1)^{n+1},  \tag{5.33}
\end{equation}
then the zeta function $Z(t)$ and the Alexander polynomial $\Delta (t)$ are
connected through 
\begin{equation}
\Delta (t)=t^{\mu }[\frac{t-1}{t}Z(\frac{1}{t})]^{(-1)^{n+1}}.  \tag{5.34}
\end{equation}
Consider now a special case : $N=3$, $n=2$, treated by A'Campo. It is
relevant for 4-particle Veneziano amplitude discussed in Section 3.2. The
treatment of general case will become obvious upon completion of discussion
for this special case. The Alexander polynomial for this \ special case can
be written at once as follows: 
\begin{eqnarray}
\Delta (t) &=&(1-t)^{-1}(t^{3}-1)^{3}  \notag \\
&=&(1-t)^{2}\left[ (t-\rho )(t-\bar{\rho})\right] ^{3}  \notag \\
&=&(1-t)^{2}(t^{2}+t+1)^{3}  \TCItag{5.35}
\end{eqnarray}
here $\rho $ is the same as in Eq.(3.19). The reader familiar with knot
theory might recognize immediately in the above expression the Hosokawa-type
Alexander polynomial for 2 component link. To prove that the above
expression is, indeed, valid for a link, it is sufficient to invoke the
theorem by Murasugi, Ref.[157], page 117:

\bigskip

\textbf{Theorem 5.18}. \emph{Let }$\mathcal{L}$\emph{\ be a link with number
of components }$\mu \geq 2,$\emph{\ then the Alexander polynomial }$\Delta _{%
\mathcal{L}}(1)=0.$

\bigskip

To prove that the above polynomial is of Hosokawa type (denote it by $\hat{%
\Delta}_{\mathcal{L}}(t))$ we use the following facts from knot theory
[158]: if \ the Alexander polynomial is of Hosokawa type then, a) the
multiplicity of the factor $(t-1)$ is even, b) $\hat{\Delta}_{\mathcal{L}%
}(t) $ is a Laurent polynomial of even degree (up to multiplication by
factors $t^{m}$ with $m=0,\pm 1,\pm 2,...)$, c) the number of components\ $%
\mu $\ of the link should be greater or equal to two but \textbf{otherwise
is arbitrary }Ref\textbf{.[}157\textbf{], }page 121. In the present case
Milnor [26] (and also A'Campo [156]) had shown that the number $\mu $ of
components in the link is given by Eq.(5.33)).; d) the Hosokawa type
Alexander polynomial possess the property: $\hat{\Delta}_{\mathcal{L}}(t)%
\dot{=}\hat{\Delta}_{\mathcal{L}}(\frac{1}{t}).$ Looking at the Alexander
polynomial given by Eq.(5.35) it is straightforward to demonstrate that,
indeed, the equivalence $\hat{\Delta}_{\mathcal{L}}(t)\dot{=}\hat{\Delta}_{%
\mathcal{L}}(\frac{1}{t})$ holds.

\textbf{Remark 5.19}. The above arguments apply, strictly speaking, only to
3d links embedded in $S^{3}$. The Alexander polynomial obtained by A'Campo
[156] describes, however, knots/links embedded \textbf{not} into $S^{2n+1}$
but into \textbf{CP}$^{n}.$This circumstance creates no additional problems
locally (since locally \textbf{CP}$^{n}$ $\supset $ \textbf{C}$^{n+1}\supset 
$ \textbf{R}$^{2n+2}\supset $ S$^{2n+1}).$ Nevertheless, the results related
to the Alexander polynomial derived by A'Campo should be treated with some
caution. We hope, that the Theorems 5.14. and 5.15. provide the necessary
assurance that, indeed, it is possible to extend results for links \ in 3d
to links of dimensions higher than 3.

\textbf{Remark 5.20}. The Alexander polynomial, Eq.(5.32), formally looks
different from that given by Eq.(3.20). This difference is superficial
however since Eq.(5.32) is written for a special case of the Brieskorn-Pham
singularity : when all $a_{i}^{\prime }s=N.$ In this case all roots of unity
factors in Eq.(3.20) originate from the same Eq.(A.3) (with fixed $n=N$).

\textbf{Remark 5.21}. The Corollary 3.7. should be amended now in view of
Eq.(5.32). The mass spectrum is determined actually by two parameters: $n$
and $N$. This option, is only plausible but, in fact, is less convenient
than that discussed below.

\subsubsection{\protect\bigskip Newton's polyhedra and the Hodge spectrum}

Based on the results of Appendix D, we conclude that the above two
parameters control the Hodge spectrum of the B-P singularity.\ As results of
Ref.[36], page 328, indicate, the total number of differential n+1 forms $%
\omega _{b}$ of the type 
\begin{equation}
\omega _{b}=s_{1}^{b_{1}}\cdot \cdot \cdot s_{n+1}^{b_{n+1}}\frac{ds_{1}}{%
s_{1}}\wedge \cdot \cdot \cdot \wedge \frac{ds_{n}}{s_{n}}\wedge \frac{%
ds_{n+1}}{s_{n+1}}  \tag{5.36}
\end{equation}
is given by the Milnor number, Eq.(5.33), (e.g., please, read in addition
the Remark 5.26.below). Hence, this number provides as well the total number
of possible Veneziano amplitudes. This number has a very interesting
geometrical and topological interpretation which we would like to discuss
now. We begin by introducing some definitions.

\textbf{Definition 5.22.} The function \textbf{x}$^{\mathbf{k}%
}=x_{1}^{k_{1}}\cdot \cdot \cdot x_{n}^{k_{n}}$ is called \emph{monomial}
with the (vector) exponent \textbf{k}=($k_{1},...,k_{n})$, $k_{i}\in \mathbf{%
Z}_{+}$ .The monomial is called \emph{Laurent} if $k_{i}\in \mathbf{Z}$
[158].

Let $\mathcal{A}\subset \mathbf{Z}^{\mathbf{k}}$ be a finite set of integer
vectors. Then, this set can be identified with the set of corresponding
monomials. Let \textbf{C}$^{\mathcal{A}}$ be the space of Laurent
polynomials made out of monomials from the set $\mathcal{A}$, i.e. 
\begin{equation}
f(\mathbf{x})=f(x_{1},...,x_{n})=\sum\limits_{\mathbf{k}\in \mathcal{A}}a_{%
\mathbf{k}}x^{\mathbf{k}}.  \tag{5.37}
\end{equation}
Evidently, the B-P variety, Eq.(3.8), is just a special case of the variety
defined by 
\begin{equation}
V(f)=\{\mathbf{z}\in \left( \mathbf{C}^{\ast }\right) ^{\mathbf{k}}\mid f(%
\mathbf{z})=0\}  \tag{5.38}
\end{equation}
where \textbf{C}$^{\ast }=\mathbf{C}\setminus 0.$ Such variety is known in
literature as the \emph{algebraic toric variety} [158].The name ''toric''
will be explained shortly.

\textbf{Definition 5.23.} The \emph{Newton polytope} (polyhedron) $N(f)$
associated with function $f(\mathbf{x})$ defined by Eq.(5.37) is the convex
hull in \textbf{R}$^{n}$ of the set \{$\mathbf{k}:a_{\mathbf{k}}\neq 0\}.$

\textbf{Remark 5.24.} Use of the above definition of $N(f)$ in actual
calculations may or may not be convenient. There are several algorithmically
different (but combinatorially equivalent) ways to define a convex polytope
in \textbf{R}$^{n}[159].$ According to Ref.[159], a convex polytope $%
\mathcal{P}$ is either

a) a convex hull of finite set of points in \textbf{R}$^{n}$ (which is
essentially just the

\ \ \ Definition 5.23) or

b) is the result of an intersection of finitely many half spaces in \textbf{R%
}$^{n}$. In the last

\ \ \ case $\mathcal{P}$ is defined by the set if inequalities: 
\begin{equation}
\mathcal{P}=\{\mathbf{x}\in \mathbf{R}^{n}:(\QTR{sl}{l}_{i},\mathbf{x})\geq
-a_{i}\text{ , }i=1,...,m\}  \tag{5.39}
\end{equation}
The vectors $\QTR{sl}{l}_{i}\in \left( \mathbf{R}^{n}\right) ^{\ast }$ are
from the \emph{dual} space to be defined shortly below and $a_{i}\in \mathbf{%
R}_{i}$, $i=1,...,m$. Let H$_{i}$ be an affine hyperplane in \textbf{R}$^{n}$
then the intersection $\mathcal{P}\cap $H$_{i}$ is called $\emph{face}$ of
the polytope. The boundary $\partial \mathcal{P}$ of $\mathcal{P}$ is the
union of faces.

Let us discuss now the rationale for introducing the dual space. To this
purpose let us recall that the function $f$ is considered to be quasi
homogenous of degree $d$ with exponents $l_{1},...,l_{n}$ if 
\begin{equation}
f(\lambda ^{l_{1}}x_{1},...,\lambda ^{l_{n}}x_{n})=\lambda
^{d}f(x_{1},...,x_{n})  \tag{5.40}
\end{equation}
provided that $\lambda \in \mathbf{C}^{\ast }.$The exponents $l_{i}$
associated with variables $x_{i}$ are called \emph{weights.}

In the light of the above definition, it is clear that each monomial in
Eq.(5.37) is associated with the scalar product of the type: 
\begin{equation}
\sum\limits_{j}\left( l_{j}\right) _{i}k_{j}=d_{i}  \tag{5.41}
\end{equation}
with index $i$ numbering the monomials. Hence, the vector $\mathbf{l}%
=(l_{1},...,l_{n})$ is the dual of the vector \textbf{k}. The name ''toric''
comes from the observation that $\lambda $ can be chosen in such a way that $%
\left| \lambda \right| =1.$ Hence, the usual \emph{topological} torus T$%
^{n}( $which is a product of $n$ circles) defined by 
\begin{equation*}
\text{T}^{n}=\{(e^{2\pi i\varphi _{1}},...,e^{2\pi i\varphi _{n}})\in \left( 
\mathbf{C}^{\ast }\right) ^{n}\},
\end{equation*}
with $\QTR{sl}{\varphi }=(\varphi _{1},...,\varphi _{n})$ running through 
\textbf{R}$^{n}$, is just a subgroup of the \emph{algebraic torus}. The
algebraic torus $\left( \mathbf{C}^{\ast }\right) ^{n}$ is the semigroup
with respect to componentwise multiplication [158], i.e. if $\mathbf{t}\in
\left( \mathbf{C}^{\ast }\right) ^{n}$ then, the action of $\mathbf{t}$ on 
\textbf{x}$^{\mathbf{k}}$ is given by 
\begin{equation*}
\mathbf{t}^{\mathbf{l}}\cdot \mathbf{x}^{\mathbf{k}}=\mathbf{t}^{\mathbf{l}%
\cdot \mathbf{k}}\mathbf{x}^{\mathbf{k}}
\end{equation*}
which is just a special case of Eq.(5.40). Also, Eq.(3.26) is just a special
case of this equation. The scalar product $\mathbf{l}\cdot \mathbf{k}=%
\mathbf{d}$ defines a set $\mathcal{L}$ of hyperplanes. Following Ref.[160]
we call the set of all exponents of monomials that appear in $f(\mathbf{x})$%
, Eq. (5.37), the \emph{support}. It is clear that the members of the
support lie \emph{either inside} of the $N(f)$ or \emph{at} its faces. The
hyperplane set $\mathcal{L}$ \ plays a major role in what follows.

\textbf{Remark 5.25.} It can be shown [161] that the hyperplane set $%
\mathcal{L}$ determines convex polyhedron $\mathcal{P}$ completely and vice
versa (i.e. they are dual to each other). Since the polyhedron $N(f)$
defines(determines) toric variety, Eq.(5.38), this means that the hyperplane
set $\mathcal{L}$ determines the toric variety as well so that \textbf{all
results of} \textbf{this work can be deduced from (encoded in) some
properties of the hyperplane set} $\mathcal{L}$. The theory of these sets is
known in the literature as the theory of $\emph{arrangements.}$Condensed,
physically relevant and pedagogically clear exposition of this theory can be
found in the book by Orlik and Terrao [162]. This book among many other
things discusses work of Schehtman and Varchenko [137] connecting theory of
hyperplane arrangements with the theory of K-Z equations\footnote{%
The systematic exposition of K-Z equations involving Orlik-Solomon algebra
related to hyperplane arrangements can be found in Ref.[78]. Alternative
treatment can be found in recent MIT lecture notes by Varchenko [80].}.

With such background we are ready now to go back to the differential n+1
form, Eq.(5.36), in order to rewrite it in the language of toric varieties.
It should be clear that $\omega _{b}$ in Eq.(5.36) can be presented as
follows 
\begin{equation}
\omega _{b}=\mathbf{x}^{\mathbf{b}}\omega _{0}  \tag{5.42a}
\end{equation}
where \textbf{b}=($b_{1},...,b_{n+1})$ and $\omega _{0}$ is given by 
\begin{equation}
\omega _{0}=\frac{dx_{1}}{x_{1}}\wedge \cdot \cdot \cdot \wedge \frac{%
dx_{n+1}}{x_{n+1}}.  \tag{5.42b}
\end{equation}
Taking into account Eq.s(3.26) and (5.41) and using Eq.(5.42a) we obtain the
hyperplane equation which, in view of Eq.(5.20b), can be written as 
\begin{equation}
-\alpha =\frac{1}{N}\sum\limits_{i}b_{i}=<b>  \tag{5.43}
\end{equation}
and should be compared with Eq.(5.39).

In writing this equation we have taken into account that the eigenvalue $%
\lambda $ of the monodromy operator (Section 3.2) is related to $\alpha $ as 
\begin{equation}
\alpha =-\frac{1}{2\pi i}\ln \lambda  \tag{5.44}
\end{equation}
in accord with notations of the Appendix D.

\textbf{Remark 5.26}. From Eq.(5.20a) it should be clear that the vector 
\textbf{b} is just one of many possible allowed vectors. The totality of
these vectors is given by the Milnor number $\mu ,$ Eq.(5.33). The set of
these vectors is in one-to-one correspondence with the set of eigenvalues $%
\{\lambda \}$ of the monodromy operator, Eq.(3.17).

For the B-P polynomial, Eq.(3.7), it is easy to give useful geometric
description of these vectors. Indeed, consider a hypercube $%
(0,N)^{n+1}=(0,N)\times \cdot \cdot \cdot \times (0,N)$. Then a vector 
\textbf{b} is represented by one of the points \emph{strictly inside} of the
hypercube. If the volume of the unit cube cell is chosen to be unity, then
the number of points inside the cube coincides with the volume $(N-1)^{n+1}$
which is just the Milnor number $\mu ,$Eq.(5.33).

In view of the results of Appendix D, the totality of hyperplanes described
by Eq.(5.43) is in one-to-one correspondence with the totality of the Hodge
numbers associated with the differential forms, Eq.(5.36).We would like to
describe this connection in some detail now following the work of Arnol'd
[163].

From Eq.(D.6) it follows, that -$\alpha $ in Eq.(5.44) should be a positive
rational number. Mathematically, both integer and non integer numbers are
allowed. In the case of integers the corresponding monodromy eigenvalue $%
\lambda =1,$while in the non integer case $\lambda \neq 1.$ In view of
Eq.(3.17a), only non integers should be of physical interest. Physical
relevance of the integer case requires further study. So, let -$\alpha $ be
some integer + a fraction (in Appendix D it is denoted as $\hat{\alpha}).$
The following number-theoretic problem can be posed now: For the set of
integers \textbf{b} lying inside the hypercube $(0,N)^{n+1}$ and for
positive rational number $\hat{\alpha}$ lying in the range 
\begin{equation}
n-p<\hat{\alpha}\leq n-p+1  \tag{5.45}
\end{equation}
(with $p$ being defined either by Eq.(5.9) or in the Appendix D) find how
many vectors \textbf{b} produce -$\alpha $ in Eq.(5.44) provided that $\hat{%
\alpha}$ lies in the above domain.\ 

The Hodge number $h^{pq}$ can be found now if \textbf{in addition} to the
number-theoretic problem just described we would find the number of
solutions of Eq.(5.43) for case when 
\begin{equation}
-\alpha =\hat{\alpha}=n-p+1.  \tag{5.46}
\end{equation}

Denote $h_{\lambda }^{pq}$ the number of solutions for the first problem and 
$h_{1}^{pq}$ the number of solutions for the second problem then, the Hodge
number $h_{{}}^{pq}$ is given by 
\begin{equation}
h_{{}}^{pq}=h_{\lambda }^{pq}+h_{1}^{pq},  \tag{5.47}
\end{equation}
e.g. see Eq.(D.10) of Appendix D.

Geometrically, $h_{\lambda }^{pq}$ is just the number of integer points in
the hypercube lying between the hyperplanes defined by the left and right
sides of the inequality, Eq.(5.45), while $h_{1}^{pq}$ is the number of
integer points lying in the hyperplane defined by Eq.(5.46). According to
the Ref.s[159] and [163], $N(f)$ for the B-P polynomial is combinatorially
equivalent to the hypercube $(0,N)^{n+1}=\Delta $ whose vertices belong to
the integral lattice \textbf{Z}$^{n+1}\subset \mathbf{R}^{n+1}.$ For any
positive integer $k$, let $\mathit{l}(k\Delta )$ be the number of integral
points in $k\Delta \cap $\textbf{Z}$^{n+1}.$ By definition, $\mathit{l}%
(0\Delta )=1.$ In addition, let $\mathit{l}^{\ast }(k\Delta )$ be the number
of integral points in the interior of $k\Delta .$ Based on these
definitions, two generating functions (Poincar$e^{\prime }$ polynomials) can
be constructed now: 
\begin{equation}
P_{\Delta }(t)=\sum\limits_{k\geq 0}\mathit{l}(k\Delta )t^{k}  \tag{5.48a}
\end{equation}
and 
\begin{equation}
Q_{\Delta }(t)=\sum\limits_{k\geq 0}\mathit{l}^{\ast }(k\Delta )t^{k}. 
\tag{5.48b}
\end{equation}
Batyrev [164] proved that, if additionally one defines two related functions
: 
\begin{equation}
\Psi _{\Delta }(t)=(1-t)^{n+2}P_{\Delta }(t)=\sum\limits_{i\geq 0}\psi
_{i}(\Delta )t^{i}  \tag{5.49a}
\end{equation}
and 
\begin{equation}
\Phi _{\Delta }(t)=(1-t)^{n+2}Q_{\Delta }(t)=\sum\limits_{i\geq 0}\varphi
_{i}(\Delta )t^{i},  \tag{5.49b}
\end{equation}
then, 
\begin{equation}
t^{n+2}\Psi _{\Delta }(t^{-1})=\Phi _{\Delta }(t).  \tag{5.50}
\end{equation}
According to Arnol'd, Ref.[163 ], for the B-P variety one obtains 
\begin{equation}
Q_{\Delta }(t)=\left( \frac{t^{\frac{1}{N}}-t}{1-t^{\frac{1}{N}}}\right)
^{n+1}.  \tag{5.51}
\end{equation}
Consider now the limit: $t\rightarrow 1+\varepsilon ,\varepsilon \rightarrow
0^{+},$of the above generating function. An easy calculation produces: 
\begin{equation}
Q_{\Delta }(1)=\mu =(N-1)^{n+1},  \tag{5.52}
\end{equation}
where, again, $\mu $ is Milnor number, Eq.(5.33). According to Eq.(5.33)
this result also allows calculation of the Euler characteristic $\chi
(S_{N}) $ of the complement of the Fermat hypersurface in \textbf{CP}$^{n}$%
and, hence, the Euler characteristic of the Fermat hypersurface as well
[156].

\textbf{Remark 5.27}. The result, Eq.(5.52), is known in the literature as
Koushnirenko theorem [165].

\textbf{Remark 5.28}. Based on the observation that [166] for any K\"{a}hler
manifold $X$ 
\begin{equation}
\chi (X)=\sum\limits_{p,q=0}^{n+1}(-1)^{p+q}h^{pq}(X).  \tag{5.53}
\end{equation}
Danilov and Khovanskii [167] had proposed to look for properties of the
generating function 
\begin{equation}
e(X;x,\bar{x})=\sum\limits_{p,q}e^{pq}(X)x^{p}\bar{x}^{q},  \tag{5.54}
\end{equation}
where $e^{pq}(X)=(-1)^{p+q}h^{pq}(X).$ Because of the connection between $%
N(f)$ and $\chi $ exhibited in Eqs.(5.33) and (5.52), these authors had
demonstrated how to recover the Hodge numbers $h^{pq}$ from generating
functions, Eq.(5.48), of the Newton polyhedron thereby connecting the
generating function $e(X;x,\bar{x})$ with that for the Newton polyhedron.
Useful condensed summary of results of Ref.[167] can be found in the review
paper by Cox [168] .

\textbf{Remark 5.29.} The formula of Arnol'd for $Q_{\Delta }(t)$, Eq.
(5.51), is a special case of more general formula obtained by Steenbrink
[169], Brieskorn (unpublished) and Saito [170]. Their results are
alternative to that obtained by Danilov and Khovanskii and, in our opinion,
more efficient computationally. In the case if $f(\mathbf{x})$, Eq.(5.37),
is linear combination of monomials \textbf{x}$^{\mathbf{k}}$ such that
Eq.(5.41) acquires the form \textbf{l}$\cdot \mathbf{k}=1$ for \textbf{all}
monomials entering $f(\mathbf{x}),$ it is possible to demonstrate that 
\begin{equation}
Q_{\Delta }(t)=\prod\limits_{i=1}^{n+1}(\frac{t^{l_{i}}-t}{1-t^{l_{i}}}) 
\tag{5.55}
\end{equation}
which for the case of B-P polynomial is reduced back to Eq.(5.51).However,
the interpretation of the polynomial $Q_{\Delta }(t)$ by Saito [170] \textbf{%
is different}. According to Saito 
\begin{equation*}
Q_{\Delta }(t)=\sum\limits_{i}t^{\alpha _{i}}
\end{equation*}
with $\alpha _{i}=-\frac{1}{2\pi i}\ln \lambda _{i}$ (e.g. see the Appendix
D and Refs. [135,170]). Hence,$Q_{\Delta }(t)$ is also a generating function
for the Hodge spectrum. Since earlier (and in the Appendix D) we had
mentioned that the spectrum is degenerate, the above generating function
acquires the following form: 
\begin{equation}
Q_{\Delta }(t)=\sum\limits_{\{\alpha \}}n_{\alpha }t^{\alpha }  \tag{5.56}
\end{equation}
with degeneracies $n_{\alpha }=\sum\limits_{l}n_{\alpha ,l}.$ The numbers $%
n_{\alpha ,l}$ participate in yet another generating function 
\begin{equation}
\mathcal{F}=\sum\limits_{\alpha ,l}n_{\alpha ,l}(\alpha ,l)  \tag{5.57}
\end{equation}
with the spectral pair $(\alpha ,l)$ being defined in the Appendix D. Since
both $\alpha $ and $l$ in this pair are related to $p$ and $q$, one can
identify $n_{\alpha ,l}$ with $h_{\lambda }^{pq}$ [135]. Hence, the
generating function $Q_{\Delta }(t)$ allows, in principle, determination of
the Hodge numbers $h_{\lambda }^{pq}.$ Examples are provided in Ref.[135].
The same results (with much less details) can be found in a short paper by
Varchenko and Khovanskii [171].

\bigskip

\section{\protect\bigskip Discussion}

To our knowledge, so far the results from geometry of toric varieties had
been used in the so called mirror symmetry claculations [57\textbf{]}. The
results presented above indicate that actually all results of conformal
field theories and high energy physics could be described from the
standpoint of geometry of toric varieties. This discipline itself is not new
however. Actually, it can be considered as branch of the theory of the
linear algebraic groups to which some results of the theory of Lie groups
and Lie algebras had been adopted. For instance, the result, Eq.(5.56), has
its analogue in the theory of characters in the theory of Lie
groups/algebras and, hence, one can derive the Weil formula for characters,
etc. Under such interpretation, the Milnor number $\mu ,$ Eq.(5.33), plays a
role of the dimension of representation of the Weil/Coxeter reflection
group.With little extra effort one can reobtain Kac-Weil character
formula-cental for developments of conformal field theories. In addition,
obtained results allow the symplectic interpretation enabling us to restore
the underlying physical model producing , for instance, the Veneziano
amplitudes. The results of Atiyah documented in Ref. [40] indicate that this
model is also going to be a string. The combinatorial aspects of the Lie
group theory representations will allow us to restore the Verlinde fusion
rules as a special case of Schubert calculus considered in our earlier work
on Kontsevich-Witten model [64]. All results of this \ earlier work \ and
the work just presented can be reobtained with help of the symplectic
formalism based on the Duistermaat-Heckman formula [40]. If one is
interested in development of p-adic aspects of string theory, the concept of
motives and its latest development culminated in the works by the latest
Field medalist,Vladimir Voevodsky, can be found in the form accessible to
physicists in Refs.[186,187].  Details of the results outlined above are to
be presented in subsequent publications.

\bigskip

\textbf{Acknowledgements}.The author owes this work to Tatyana Zhebentyayeva
(Clemson) whose late night phone call had triggered the chain of thoughts
leading to this work. In addition, the author would like to thank Anatoly
Libgober (UIC), Winnie Li (Penn State U), Michael Rosen (Brown U), Ulrich
Geckeler (U Saarbrucken) for helpful correspondence

\pagebreak

\bigskip

\textbf{Appendix A. Some results from cyclotomic field theory}

\bigskip

Suppose we have the number field $\mathbf{F}_{p}$ of characteristic $p$,
that is the coset \textbf{Z}/$p$\textbf{Z . }Let c$_{i}\in \mathbf{F}_{p}$.
It can take $p$ values. Suppose now that we would like to extend the field $%
\mathbf{F}_{p}.$ This can be accomplished very easy using concepts of
standard linear algebra. To construct the vector space of dimension $n$ we
need to have a set of $n$ linearly independent (basis) vectors \textbf{e}$%
_{i}$ so that any vector \textbf{A} can be decomposed as usual: 
\begin{equation}
\mathbf{A}=\sum\limits_{i=1}^{n}a_{i}\mathbf{e}_{i}.  \tag{A.1}
\end{equation}
Let now the role of $a_{i}$ is to be played by $c_{i\text{ }}$and the role
of \textbf{e}$_{i}$ -by some basis elements $\alpha _{i}$ to be determined
momentarily. Then, by construction, a number 
\begin{equation}
\mathit{\beta }\mathcal{=}\sum\limits_{i=1}^{n}c_{i}\alpha _{i}  \tag{A.2}
\end{equation}
belongs to the extension of the field $\mathbf{F}_{p}.$ Since each $c_{i%
\text{ }}$can have $n$ values, the field $\mathbf{F}_{p}[\alpha ]$ contains $%
p^{n}$ elements and is known in literature as the Galois field \textbf{GF}($%
p^{n})$.The task remains to find the basis set explicitly. The familiar
example of complex numbers helps a lot in this respect. Indeed, the complex
number $i=\sqrt{-1}$ is just solution of the equation $x^{2}+1=0.$ Clearly,
all complex numbers $z$ are given by $z=a1+ib$ with $a$ and $b$ belonging to 
$\mathbf{R,Q}$ or \textbf{Z}. The basis elements are 1 and i. To go beyond
this elementary case it is sufficient to consider all solutions of the
equation 
\begin{equation}
x^{n}-x=0.  \tag{A.3}
\end{equation}
It can be shown [99] that the above equation can be also rewritten as 
\begin{equation}
x^{n}-x=\prod\limits_{i=1}^{n}(x-\alpha _{i})  \tag{A.4}
\end{equation}
with $\alpha _{i}$ being \ a root of Eq.(A.3). In addition, it can be shown
that all roots are distinct and, hence, there are $n$ of them. Naturally,
they can form the basis which we are looking for. Moreover, Eq.(A.3) can be
equivalently rewritten as 
\begin{equation}
x^{n-1}=1  \tag{A.5}
\end{equation}
and, hence, the cyclotomic (that is circular) nature of this equation
becomes clear. Indeed, consider $\zeta _{k}=\exp \{i\alpha \}$ with $\alpha
=k\frac{2\pi }{n-1}.$ If for $k$ we assign values $0,1,2,...,$ $n-2$, we
obtain $n-1$ equidistant points on the circle. It is clear that this
construction can be performed for any $n$ and, hence, if we consider instead
of Eq.(A.5) the equivalent equation 
\begin{equation}
x^{n}=1  \tag{A.5a}
\end{equation}
we shall obtain exactly $n$ values for $\zeta _{k}$. These are independent
of each other and can be used instead of $\alpha _{i}^{\prime }s$ in
Eq.(A.2). Moreover, there actually no need of substituting a\textbf{ll} $%
\zeta _{i}$'s into Eq.(A.2). It is sufficient to use just one of the so
called \textit{primitive} roots of unity in Eq.(A.2). By definition \textbf{%
all} roots of Eq.(A.5a) can be obtained as powers of those which are
considered to be primitive. To obtain primitive roots the following steps
should be taken. For a given $n$ one should look for all $k^{\prime }s$ such
that $k\nmid n$ and $k<n$ . The number of such possibilities is given by the
Euler's function $\varphi (n).$Recall that the coset \textbf{Z}/$n$\textbf{Z}
is a field (and is denoted $\mathbf{F}_{n})$ only if $n$ is some prime
number $p$. An element $c_{i}$ is a \textit{unit} in $\mathbf{F}_{n}$ if ($%
c_{i},n)=1($that is $c_{i}\nmid n)$. Hence, there is one-to-one
correspondence between the units in $\mathbf{F}_{n}$ and the primitive roots
in cyclotomic fields. So if $n=p$ we have $\varphi (p)=p-1$ \ primitive
roots. If we take one of them, say $\zeta ,$ then the desired basis is given
by :$1$, $\zeta ,\zeta ^{2},...,\zeta ^{p-1}.$ Let us renumber these
primitive roots as follows : $\zeta _{1},\zeta _{2},...,\zeta _{\varphi (p)}$%
. Moreover, let us remove for a moment the restriction on $n$ to be a prime,
that is we assume that $n$=$\prod\nolimits_{i=1}^{m}r_{i}$.Then, surely, $%
\varphi (n)=\varphi (r_{1})\cdot \cdot \cdot \varphi (r_{m})$ and for each $%
\varphi (r_{i})$ the cyclotomic polynomial $\Phi _{r_{i}}(x)$ can be defined
now as 
\begin{equation}
(x-\zeta _{1})\cdot \cdot \cdot (x-\zeta _{\varphi (r_{i})})\equiv \Phi
_{r_{i}}(x).  \tag{A.6}
\end{equation}
It is monic irreducible polynomial of degree $\varphi (r_{i})$ by
construction. Using multiplicative property of Euler's function it is clear
that the following identity should hold 
\begin{equation}
x^{n}-1=\prod\limits_{d\mid n}\Phi _{d}(x)  \tag{A.7}
\end{equation}
with $d$ running over all divisors of $n$. For $n=p$ the result can be
considerably simplified by noticing that $x=1$ is the root of Eq.(A.7).This
means, in view of Eq.(A.6), that Eq.(A.7) can be simplified to 
\begin{equation}
x^{p}-1=(x-1)(x^{p-1}+x^{p-2}+\cdot \cdot \cdot +x+1).  \tag{A.8}
\end{equation}
By construction, the second multiplier in the r.h.s of Eq.(A.8) is monic
irreducible (in $\mathbf{R,Q}$ or \textbf{Z) }polynomial. It should be clear
that the roots of the equation 
\begin{equation}
x^{p-1}+x^{p-2}+\cdot \cdot \cdot +x+1=0  \tag{A.9}
\end{equation}
are primitive roots of the cyclotomic Eq.(A.5a). Thus, the connection
between the cyclotomic fields and polynomials is clear. This connection can
be extended further with help of Eq.(A.2). Indeed, let us write 
\begin{equation*}
\mathit{\beta }\mathcal{=}\sum\limits_{i=0}^{n-1}c_{1i}\alpha _{{}}^{i}
\end{equation*}
and also 
\begin{equation*}
\mathit{\beta }^{2}\mathcal{=}\sum\limits_{i=0}^{n-1}c_{2i}\alpha ^{i},
\end{equation*}
etc. and, finally, 
\begin{equation*}
\mathit{\beta }^{n}\mathcal{=}\sum\limits_{i=0}^{n-1}c_{ni}\alpha ^{_{i}}
\end{equation*}
for some $\alpha _{i,j\text{ }}$ in \textbf{F}$_{p}.$ Now we want to look at
equation 
\begin{equation}
\sum\limits_{i=0}^{n}x_{i}\beta ^{i}=0  \tag{A.10}
\end{equation}
with $x_{i}\in $\textbf{F}$_{p}.$ Upon substitution of equations for powers
of $\beta ^{\prime }s$ in Eq.(A.10) the representation of $x_{i}^{\prime }s$
through $\alpha _{i,j\text{ }}^{\prime }s$ can be found and, hence, to
construct equation of order $n$ with $\beta $ being a root. The polynomial
associated with such an equation is irreducible and always can be made monic
[49], page 430. Since for a given prime $p$ there are $p^{n}$ $\beta
^{\prime }s$ according to Eq.(A.2), hence, there are $p^{n}$ monic
irreducible polynomials of degree $n$. If we exclude the trivial case $\beta
=0$, then there are $p^{n}-1$ nontrivial irreducible polynomials.

\textbf{Remark A.1.} \ Let $S_{m}$ be symmetric group acting on complex
space \textbf{C}$^{m}$ by exchanging coordinates. Then 
\begin{equation*}
Y_{m}:=\{(z_{1},...,z_{m})\in \mathbf{C}^{m}\mid z_{i}\neq z_{j}\text{ for }%
i\neq j\}/S_{m}
\end{equation*}
is the set of all unordered m-tuples of distinct complex numbers. It can be
shown (Briesk, malle \&matzat) that the set $Y_{m\text{ }}$is in one-to-one
correspondence with the set of monic irreducible polynomials with roots $%
z_{1},...,z_{m}$ . Define a braid on $m$ strings (and initial point $%
z_{1}^{0},...,z_{m}^{0})$ as homotopy class of a closed paths in $Y_{m}$
with initial and final point at $z_{1}^{0},...,z_{m}^{0}$ . Such constructed
braid is in one-to-one correspondence with the set of monic irreducible
polynomials\footnote{%
Braids are connected directly with the hyperplane arrangements mentioned in
Section 5.3.2. in connection with the K-Z equations.}.

\bigskip

\textbf{Appendix B. Some results from Nielsen-Thurston theory of surface
homeomorphisms}

\bigskip

Let $\mathcal{S}$ be closed orientable Riemann surface of genus $g$.The
first homotopy group, the fundamental group $\pi _{1}(\mathcal{S})$ of
surface $\mathcal{S}$ is made of $2g$ generators $\{x_{i},y_{i}\}$, $i=1-g$
and a single relation so that its presentation is known to be 
\begin{equation}
\pi _{1}(\mathcal{S})=<x_{1},y_{1},...,x_{g},y_{g}\mid \lbrack
x_{1,}y_{1}]\cdot \cdot \cdot \lbrack x_{g},y_{g}]>.  \tag{B.1}
\end{equation}
Nielsen has noted that there is one to one correspondence between
automorphisms of $\pi _{1}(\mathcal{S})$ and surface self-homeomorphisms.
This is summarized in the following proposition

\bigskip

\textbf{Proposition B.1}.(Nielsen [172]) If $g>1$, then every element of $%
Out $($\pi _{1}(\mathcal{S}))$ is represented by a unique isotopy class of
self-homeomorphisms of $\mathcal{S}$.

\bigskip

An important subgroup of $Out$($\pi _{1}(\mathcal{S}))$ is the mapping class
group $\mathcal{M}_{g}$ . Geometrically, this group is finitely generated by
the \textit{Dehn twists} \ in simple closed curves (lamination set) on $%
\mathcal{S}$ whose physical significance was discussed extensively in our
previous work, Refs[4-6]. A simple closed curve $\mathcal{C}$ on the
orientable surface $\mathcal{S}$ has a neighborhood $\mathcal{E}$
homeomorphic to an annulus which is convenient to parametrize by \{[$r$,$%
\theta $]$\mid $1$\leq r<2\}$.The Dehn twist in $\mathcal{C}$ can be
imagined as an automorphism T$_{C}:\mathcal{S}\rightarrow \mathcal{S}$. It
is given by the identity off $\mathcal{E}$ and by [$r$,$\theta $]$%
\rightarrow \lbrack r$,$\theta +2\pi r]$ on $\mathcal{E}$ . We would like to
illustrate these concepts on the simplest example of the punctured torus T$%
^{2}$ using results of our previous work, Ref. [4]. In this case $Out$($\pi
_{1}($T$^{2}))=GL_{2}(\mathbf{Z})$ and $\mathcal{M}_{1,1}=PSL(2,\mathbf{Z}).$
Since any transformation from $PSL(2,\mathbf{Z})$ is obtainable by
projectivisation of $SL(2,\mathbf{Z})$ we discuss everything in terms of $%
SL(2,\mathbf{Z})$ with projectivisation at the end. Any transformation which
belongs to $SL(2,\mathbf{Z})$ is expressible in terms of $2\times 2$ \
matrix \textbf{A} given by 
\begin{equation}
\mathbf{A}=\left( 
\begin{array}{cc}
a & b \\ 
c & d
\end{array}
\right)  \tag{B.2}
\end{equation}
with integer coefficients subject to condition:$\det $\textbf{A}=$ab-cd$=1.
The characteristic polynomial for this matrix is given by 
\begin{equation}
t^{2}-tr\mathbf{A}\text{ }t+\det \mathbf{A}=0.  \tag{B.3}
\end{equation}
This implies that the eigenvalues of \textbf{A} are either :

a) both complex ( when tr\textbf{A}=0,1,-1),

b) both equal to $\pm1$ (when tr\textbf{A}=$\pm2),$

c) distinct and real (when $\left| tr\mathbf{A}\right| $ \TEXTsymbol{>}2). \
\ \ 

If $\frak{F}_{A}$ is toral automorphism then, transformation a) is called 
\textit{periodic} since $\left( \frak{F}_{A}\right) ^{n}=1$ for some $n$
(actually, $n=12$ by the Hamilton-Cayley theorem) , transformation b) is
called \textit{reducible} since it leaves a simple closed curve $\mathcal{C}$
invariant, transformation c) is called (\textit{pseudo) Anosov } if the
(line) vector field on surface (does) does not contain singularities.
Physical significance of this difference is explained and illustrated in our
previous work, Refs.[5,6].

The largest of two eigenvalues is associated with the topological entropy of
the (line)vector flow and is related to the amount of stretching of surface $%
\mathcal{S}$ and, hence, with the dilatation parameter of the
Teichm\"{u}ller theory. The Nielsen-Thurston theory generalizes the above
classification of surface automorphisms to all surfaces of genus $g>1$.
Already Nielsen had realized [172] that for $g>1$ it is more convenient to
study homeomorphisms of surface $\mathcal{S}$ by considering their image on
the universal cover of $\mathcal{S}$ which we choose as Poincare$^{\prime }$
disc model of \textbf{H}$^{2}$, i.e. $int$ \textbf{D} $\cup $ $S_{\infty
}^{1}=$\textbf{H}$^{2}.$ According to Nielsen [172]

\bigskip

\textbf{Proposition B.2}. Any lift \~{h} of \ the surface self-
homeomorphism h: $\mathcal{S}\rightarrow \mathcal{S}$ to the universal cover
of $\mathcal{S}$ extends to a unique self-homeomorphism of the unit disc 
\textbf{D}, i.e. to $int$ \textbf{D }$\cup $ $S_{\infty }^{1}.$

\bigskip

Surface self-homeomorphisms h are associated with the Dehn twists connected
with a set of simple closed nonintersecting curves homotopic to geodesics
(such set is called \emph{geodesic lamination} $\mathcal{L}$). Their lifts
\~{h}($\mathcal{L}$) are associated with some maps of the circle $S_{\infty 
\text{ }}^{1}$extendable (quasi conformally) to the interior of the disc 
\textbf{D} e.g.see Ref.[73]. An image of the closed geodesic on $\mathcal{S}$%
, when lifted to \textbf{H}$^{2},$ is just a segment of a circle whose both
ends lie on $S_{\infty }^{1}$. Since \ geodesics are nonintersecting, circle
segments on $S_{\infty }^{1}$ are also nonintersecting. In this work we are
interested only in the $\mathit{periodic}$ maps of the circle since non
periodic maps can be always systematically approximated by periodic ones as
it is explained in terms of continuous fractions and associated with them
Dehn twists in our earlier work, Ref.[4] or, alternatively, in Milnor's
lecture notes [173] and Ref.[55]. In connection with such circle maps the
following remark is of importance.

\textbf{Remark B.3.} (A variant of Sarkovskii theorem, Ref.[174], page 88).
Let $f$: $S^{1}\rightarrow S^{1}$ be a continuous map of the circle with a
periodic orbit of period 3. If the lift $\tilde{f}$: $\mathbf{R}\rightarrow 
\mathbf{R}$ has also a periodic orbit of period 3 then, $f$ has periodic
orbits of every period. The condition on the lift \ map \~{f} cannot be
dropped. The continuous maps of the circle can be replaced by the piecevise
linear maps while 3 still remains as minimal period.

\textbf{Remark B.4.} As noted by Kontsevich [175], the moduli space problem
makes sense only for Riemann surfaces obeying the following set of
inequalities 
\begin{equation}
g\geq 0,n>0,2-2g-n<0  \tag{B.4}
\end{equation}
with $n$ being the number of distinct marked points (effectively, distinct
boundary components). Boundary components can be eliminated by the \textit{%
Schottky double} construction. This construction can be performed as
follows. If $M$ is a complex manifold with $C_{1},...,C_{n}$ boundary
components, one can consider an exact duplicate of it, say $\hat{M},$ with
the same number of boundary components, say, $\hat{C}_{1},...,\hat{C}_{n}.$\
Evidently, for each point $x\in M$ there is a ''symmetric'' point $\hat{x}%
\in \hat{M}$ . The Schottky double $2M$ is formed as a disjoint union $M\cup 
\hat{M}$ and identifying each point $x\in C_{i}$ with point $\hat{x}\in \hat{%
C}_{i}$ for 1$\leq i\leq n.$ In the simplest case we have initially either
punctured torus, i.e.$g=1,n=1$, or the trice punctured sphere, i.e. $g=0,n=3$%
. In both cases the Schottky double is a double torus. A double torus has 3
geodesics which belong to the geodesic lamination $\mathcal{L}$. The image
of these geodesics lifted to \textbf{H}$^{2}$ produces 3 circular arcs whose
ends lie on $S_{\infty }^{1}$. This is the minimal number of arcs required
for the moduli space problem to make sense. According to Remark B.3. this is
also the minimal period\ for the periodic homeomorphisms of the circle in
view of the Sarkovskii theorem.

In Ref.[176] it is argued that the total number of geodesics on the Schottky
double is $6g-6+3n$. This is the dimension of space of holomorphic quadratic
differentials (real on each of the boundary components). Hence, in
accordance with Teichm\"{u}ller theory [8], it is the dimension of the
Teichm\"{u}ller and, accordingly, the moduli space of such Schottky doubled
surface.

\bigskip

\textbf{Appendix C. Complex multiplication and Hecke operators}

\bigskip

In Section 4b we have introduced both lattice $L^{\prime }$ and sublattice $%
L $ . This caused us to introduce the relationship: $ad-bc=n,n\geq 1.$We
would like now to study physical implications of this relationship. To this
purpose, following Ref.[89], we introduce 3 types of matrices: 
\begin{eqnarray*}
\mathcal{D}_{n} &=&\left\{ \left( 
\begin{array}{cc}
a & b \\ 
c & d
\end{array}
\right) \in M_{2}(\mathbf{Z});\text{ }ad-bc=n\right\} , \\
\mathcal{S}_{n} &=&\left\{ \left( 
\begin{array}{cc}
a & b \\ 
0 & d
\end{array}
\right) \in M_{2}(\mathbf{Z});\text{ }ad=n,a,d>0,0\leq b<d\right\} . \\
\Xi &=&\left\{ \left( 
\begin{array}{cc}
\alpha & \beta \\ 
\gamma & \delta
\end{array}
\right) \in SL_{2}(\mathbf{Z});\text{ }\alpha ,\beta ,\gamma ,\delta \in 
\mathbf{Z,}\alpha \delta -\beta \gamma =1\right\} .
\end{eqnarray*}
Naturally, $\mathcal{S}_{n}\longleftrightarrow SL_{2}(\mathbf{Z})\setminus 
\mathcal{D}_{n}.$ Introduce now the \emph{homothety} operator $R_{\lambda 
\text{ }}$as follows 
\begin{equation}
R_{\lambda \text{ }}L=\lambda L  \tag{C.1}
\end{equation}
where $\lambda \in \mathbf{C}$ and the action of $n^{th}$ Hecke operator $%
T(n)$ on the lattice $L$ is defined by 
\begin{equation}
T(n)L=\sum\limits_{\substack{ L\subset L^{^{\prime }}  \\ \lbrack
L:L^{\prime }]=n}}(L^{\prime }).  \tag{C.2}
\end{equation}
In order to use this equation, it should be rewritten in more explicit form
as 
\begin{equation}
T(n)L=\sum\limits_{\alpha \in \mathcal{S}_{n}}\alpha (L).  \tag{C.3}
\end{equation}
It can be demonstrated [89] that

$a)$ $R_{\lambda }R_{\mu }=R_{\lambda \mu }$ $\forall \lambda ,\mu \in 
\mathbf{C;}$

$b)$ $R_{\lambda }T(n)=T(n)R_{\lambda }$ $\forall \lambda \in \mathbf{C,}%
n\geq 1;$

$c)$ $T(mn)=T(m)T(n)$ $\forall m,n\geq 1$ with $\gcd (m,n)=1;$

$d)$ $T(p^{e})T(p)=T(p^{e+1})+pT(p^{e-1})R_{p}$ for $p$ prime, $e\geq 1$

and $p\nmid n$ , while for $p\mid n$\ \ we get

$d^{\prime })T(p^{e})=\left[ T(p)\right] ^{e}$\ \ .\ \ \ \ \ \ \ \ \ \ \ \ \
\ \ \ \ \ \ \ \ \ \ \ \ \ \ \ \ \ \ 

Since any number $n$ can be decomposed into primes the above rules are
sufficient for any $n,$ prime or not. From the definition of the Hecke
operator it is clear that it is acting on the lattices, more exactly, it
acts on equivalence classes between different lattices. Let now $\tilde{f}%
(L) $ be some function defined on such lattices. We shall assume that this
function is homogenous of degree -$k$. That is 
\begin{equation}
\tilde{f}(\mathbf{Z}\omega _{1}+\mathbf{Z}\omega _{2})=\omega
_{2}^{-k}f(\omega _{2}/\omega _{1}).  \tag{C.4}
\end{equation}
The requirement of homogeneity makes this function invariant with respect to
changes of the basis. Indeed, if \ we have change of the basis given by 
\begin{equation*}
\left( 
\begin{array}{c}
\omega _{2}^{^{\prime }} \\ 
\omega _{1}^{^{\prime }}
\end{array}
\right) =\left( 
\begin{array}{cc}
a & b \\ 
c & d
\end{array}
\right) \left( 
\begin{array}{c}
\omega _{2}^{^{{}}} \\ 
\omega _{1}^{{}}
\end{array}
\right)
\end{equation*}
then, 
\begin{eqnarray}
\tilde{f}(\mathbf{Z}\omega _{1}^{^{\prime }}+\mathbf{Z}\omega _{2}^{\prime
}) &=&\omega _{1}^{\prime -k}f(\omega _{2}^{\prime }/\omega _{1}^{\prime
})=(c\omega _{1}+d\omega _{2})^{-k}(d+c(\omega _{2}/\omega
_{1}))^{k}f(\omega _{2}/\omega _{1})  \notag \\
&=&\omega _{2}^{-k}f(\omega _{2}/\omega _{1}),  \TCItag{C.5}
\end{eqnarray}
provided that $f$ is \emph{modular} function of \emph{weight} $k$, that is 
\begin{equation}
f(\left( 
\begin{array}{cc}
a & b \\ 
c & d
\end{array}
\right) \tau )=(c\tau +d)^{k}f(\tau )\text{ \ for }\left( 
\begin{array}{cc}
a & b \\ 
c & d
\end{array}
\right) \in SL(2,Z).  \tag{C.6}
\end{equation}
As consequence of Eqs.(C.4) and (C.5), function $\tilde{f}(L)$ possess the
homogeneity property 
\begin{equation}
\tilde{f}(\alpha L)=\alpha ^{-k}\tilde{f}(L).  \tag{C.7}
\end{equation}
Consider now action of the Hecke operator on function $\tilde{f}(L).$ In the
case if such a function is a modular form of weight $k$ we obtain 
\begin{equation}
\left( T_{k}(n)f\right) (\tau )=n^{k-1}\sum\limits_{\substack{ L\subset
L^{^{\prime }}  \\ \lbrack L:L^{\prime }]=n}}\tilde{f}(L^{\prime
})=n^{k-1}\sum\limits_{\substack{ ad=n,a\geq 1  \\ 0\leq b<d}}d^{-k}f(\frac{%
a\tau +b}{d}),  \tag{C.8}
\end{equation}
where, following Ref.[89], we defined $\left( T_{k}(n)f\right) (\tau
)=\left( n^{k-1}T(n)\tilde{f}\right) (L).$ The result, Eq.(C.8), was
obtained with help of Eqs.(C.3)and.(C6). In actual calculations one should
take into account that the weight $k$ should be only even number. The
modular forms with odd weight are zero as can be easily shown. The factor $%
n^{k-1}$ looks somewhat artificial at this point. Hence, we would like to
explain its origin now.

The modular function $f(\tau )$ admits the Fourier decomposition $f(\tau
):=\sum c(m)q^{m}$, where $q=\exp (2\pi i\tau )$. The eigenvalue problem 
\begin{equation}
\left( T_{k}(n)f\right) (\tau )=\lambda (n)f(\tau )  \tag{C.9}
\end{equation}
in view of Eq.(C.8) can be written as follows 
\begin{eqnarray}
\lambda (n)f(\tau ) &=&n^{k-1}\sum\limits_{\substack{ ad=n,a\geq 1  \\ 0\leq
b<d}}d^{-k}\sum\limits_{m\in \mathbf{Z}}c(m)\exp (2\pi im(a\tau +b)/d) 
\TCItag{C.10} \\
&=&n^{k-1}\sum\limits_{m\in \mathbf{Z}}\sum\limits_{ad=n,a\geq
1}c(m)d^{-k}e^{2\pi ima\tau /d}\sum\limits_{0\leq b<d}e^{2\pi imb/d}.  \notag
\end{eqnarray}
Since 
\begin{equation*}
\sum\limits_{0\leq b<d}e^{2\pi imb/d}=\left\{ 
\begin{array}{c}
d\text{ if d}\mid m \\ 
0\text{ if d}\nmid m
\end{array}
\right.
\end{equation*}
we have to replace $m$ by $md=mn/a$ thus producing $n/d=a$ resulting in
cancellation of the prefactor $n^{k-1}$. After this, Eq.(C.10) acquires the
following form 
\begin{equation}
\lambda (n)f(\tau )=\sum\limits_{m\in \mathbf{Z}}\sum\limits_{ad=n,a\geq
1}a^{k-1}c(\frac{mn}{a})e^{2\pi ima\tau }.  \tag{C.11}
\end{equation}
The last result can be further rearranged thus producing 
\begin{equation}
\lambda (n)c(l)=\sum\limits_{a\mid \gcd (l,n)}a^{k-1}c(\frac{nl}{a^{2}}). 
\tag{C.12}
\end{equation}
Suppose $\gcd (l,n)=1,$then, surely, $a=1$ which leads us to $\lambda
(n)c(l)=c(nl).$ Let further, $l=1$. This produces :$\lambda (n)c(1)=c(n)$.
From here it follows that $c(1)\neq 0$ and it can actually be taken as 1,
Ref. [89], page 78 and Ref.[100], page 101.If this is so, then we are left
with a very nice result: $\lambda (n)=c(n)$, to be used in the main text.
Moreover, we will also need the following result: 
\begin{equation}
\left( R_{\lambda }\tilde{f}\right) (L)=\tilde{f}(\lambda L)=\lambda ^{-k}%
\tilde{f}(L).  \tag{C.13}
\end{equation}
Using this result in equation 
\begin{equation}
\left( T(p^{e})T(p)\tilde{f}\right) (L)=\left( T(p^{e+1})\tilde{f}\right)
(L)+\left( pT(p^{e-1})R_{p}\tilde{f}\right) (L)  \tag{C.14}
\end{equation}
we obtain 
\begin{equation}
\left( T(p^{e})T(p)\tilde{f}\right) (L)=\left( T(p^{e+1})\tilde{f}\right)
(L)+\left( p^{1-k}T(p^{e-1})\tilde{f}\right) (L).  \tag{C.15}
\end{equation}
Using the definition $\left( T_{k}(n)f\right) (\tau )=\left( n^{k-1}T(n)%
\tilde{f}\right) (L)$ in Eq.(C.15) and multiplying both sides by factor $%
p^{\left( e+1\right) (k-1)}$ produces our final result 
\begin{equation}
T(p^{e})T(p)f=T(p^{e+1})f+p^{k-1}T(p^{e-1})f  \tag{C.16}
\end{equation}
to be discussed from yet another (physical) point of view immediately below.

To this purpose we would like to remind our readers some excerpts from the
graph theory. A graph $G$ consists of finite set of vertices,$V$ and edges $%
E $ along with the following two maps: 
\begin{equation}
E\rightarrow V\times V\text{ \ and }e\rightarrow (o(e),t(e))  \tag{C.17}
\end{equation}
so that $\forall e\in E$ we may have $e\rightarrow \hat{e}$ \ which means
that the edge can change the orientation into opposite so that $o(e)=t(\hat{e%
}).$ The element $o(e)$ is the origin of $e$ and $t(e)$ is the terminus of $%
e $.Two vertices $v_{0}$ and $v_{1}$ are adjacent if there is an edge with $%
o(e)=v_{0}$ and $t(e)=v_{1}.$The order of $G$, denoted as $\left| G\right| ,$%
is determined by the number of vertices in $G$. A walk $W$ in $G$ is a
sequence of oriented edges $e_{1},...,e_{r\text{ }}$such that $%
t(e_{i})=o(e_{i+1})$ for $1\leq i\leq r-1.$ If walk is from $v_{0}=o(e_{1})$
to $v_{r}=t(e_{r})$ it means that the walk is from $v_{0}$ to $v_{r}$ of
length $r$. We shall be interested only in walks \textbf{without backtracking%
} . For such \ walks, let $\left| G\right| =n$ be the order of the graph
then, the matrix $A_{r}$=$\left( a_{i,j}^{\left( r\right) }\right) $ is a
symmetric $n\times n$ matrix with nonnegative integer entries counting the
number of walks of length $r$ from $v_{i}$ to $v_{j}.$ By definition, $A_{1}$
is the \emph{adjacency} matrix [177]. The degree $\deg v_{i\text{ }}$of the
vertex $v_{i}$ is the number of edges emanating at $v_{i}$. Let $D$ be the
diagonal matrix with entries $\deg v_{1},...,\deg v_{n\text{ }}$ along the
diagonal then, matrices $A_{r}$ can be determined recursively from the
following set of equations 
\begin{eqnarray}
A_{1}A_{1} &=&A_{2}+D,  \TCItag{C.18} \\
A_{r}A_{1} &=&A_{r+1}+A_{r-1}(D-I)\text{ for }r\geq 2.  \notag
\end{eqnarray}
The proof of this result can be found in Ref.[178]. In order to use these
recursions we would like to make a few simplifications. First, we would like
to consider only $k$-regular graphs (i.e. such for which degree of all
vertices is the same and equal $k$). Next, let $k=p+1$ for some prime $p$.
And, finally, let $A_{r}=B_{r}-B_{r-2}$ for all $r\geq 1.$ Then, the system
of recursions given by Eq.(C.18) is reduced to 
\begin{equation}
B_{r}B_{1}=B_{r+1}+pB_{r-1}\text{ for }r\geq 1,  \tag{C.19}
\end{equation}
provided that $B_{-1}=0$, $B_{0}=I,$ $B_{1}=A_{1}.$This result should be
compared with Eq.(C.16) (for $k=2$). Evidently, the total agreement is
reached. It should be clear that the edges of \ $G$ represent equivalence
classes between different lattices represented by vertices of \ $G$. The
universal covering of such graph is $p+1$ -valent tree known in the
literature as Bruhat-Tits tree [179]. Should our lattices come from the
lattices whose base is made of more than 2 elements then, instead of trees,
we would have \emph{buildings} [179, 180]. Such necessity indeed occurs for
higher dimensional complex tori (e.g. read Section5) and had been discussed
in connection with geometry of toric varieties (Section 5) by Mumford [181]
already in 1970.

Because of the agreement between Eqs.(C.19) and (C.16), \textbf{it is
sufficient, in principle, to do all relevant physics ''locally'', that is
for the specific prime} $p$, \textbf{in} \textbf{order to restore the
''global'' picture for all primes}. Although this idea is not new, as
discussed in the Introduction, its implementation in the present context may
still require considerable efforts. E.g. to rewrite this paper using the
terminology of motives [38,145] would require entirely new (and much more
advanced) level of presentation. This task is left for the future.

For now we only would like to notice that since $A_{1}$ is the adjacency
matrix, this means that the Hecke operators can be identified with the
adjacency matrix of some graph $G$.This observation makes them already
physical especially because the discrete (combinatorial) Laplacian $\Delta $
for $k$-regular graph $G$ is known to be 
\begin{equation}
\Delta =kI-A_{1}.  \tag{C.20}
\end{equation}
Harmonic analysis for such type of Laplacians is discussed in detail in
Refs.[16,17,182-184]. Higher dimensional analogs of Hecke operators, Hecke
algebras, etc. are discussed in Ref.[185].

\bigskip

\textbf{Appendix D. Veneziano amplitudes from the oscillatory integrals}.

\bigskip

In this appendix we would like to provide the condensed summary of known
results for oscillatory integrals [36,144]. Fortunately, these results are
sufficient for reobtaing the Veneziano amplitudes discussed in the main text.

Following Ref.[36], Ch-r 11, we are interested in study of integrals of the
type 
\begin{equation}
\int\limits_{\lbrack \Gamma ]}e^{\tau f(z)}\omega  \tag{D.1}
\end{equation}
where the function $f$: (\textbf{C}$^{n},0)\rightarrow (\mathbf{C},0)$ is
holomorphic in a neighborhood of its only one critical point located at the
origin, [$\Gamma ]$ is $n-$dimensional homology chain associated with the
cohomological differential $n-$form $\omega .$ The parameter $\tau $ is
expected to be \emph{positive.} At the boundary of the chain [$\Gamma ]$ the 
\emph{real} part of the function $f(z)$ is \emph{negative}. The chain [$%
\Gamma ]$ with such property is called \emph{admissible}. We shall need the
following identity (Lemma 11.2. of Ref.[36]) valid for large $\tau ^{\prime
}s:$%
\begin{equation}
\int\limits_{\lbrack \Gamma ]}e^{\tau f(z)}\omega \approx
\int\limits_{0}^{t_{0}}e^{-\tau t}\left( \int\limits_{\partial _{-t}[\Gamma
]}\omega /df\right) dt  \tag{D.2}
\end{equation}
with $t_{0}$ being a some positive number which eventually can be taken to
the infinity. This result can be easily understood in \ physical \ terms.
Indeed, let us consider an auxiliary integral 
\begin{equation}
I(t)=\int\limits_{[\Gamma ]}\delta (t-f(z))\text{ }\omega  \tag{D.3}
\end{equation}
with $\delta $ being the usual Dirac's delta function. Such type of
integrals formally occur in many calculations, e.g. in many- body quantum
mechanical problems involving the density of states calculations, especially
for the phonon spectra, etc. As it is shown in Ref.[148], Ch-r 4, 
\begin{equation}
I(t)=\int\limits_{\partial _{-t}[\Gamma ]}\omega /df  \tag{D.4}
\end{equation}
so that for $t_{0}\rightarrow \infty $ the r.h.s. of the Eq.(D.2) is just
the Laplace-like transform of the density of states.

Let now $f(z)=z^{N+1}$and $\omega =z^{l-1}dz$ with $1\leq l\leq N.$ Then, $%
I(t)$ is given by (equation (7) on page 305 of Ref.[36]) 
\begin{equation}
I(t)=const\left| t\right| ^{\frac{l}{N+1}-1}  \tag{D.5}
\end{equation}
with $const$=$\frac{\left( \zeta _{j+1}^{l}-\zeta _{j}^{l}\right) }{N+1}.$ \
The phase factor $\zeta _{j}^{l}$ plays the same role here as in Sections
3.2. and 5.2.Using this result in Eq.(D.2) and letting $t_{0}$ to become
infinity produces 
\begin{equation}
\int\limits_{\lbrack \Gamma ]}e^{\tau f(z)}\omega \approx const\Gamma
(-\alpha )\tau ^{\alpha }  \tag{D.6}
\end{equation}
with $\alpha =\frac{-l}{N+1}.$ The result just obtained can be broadly
generalized with help of the Fubini's theorem which can be stated as
follows: 
\begin{equation}
\int\limits_{\lbrack \Gamma _{1}]\times \lbrack \Gamma _{2}]}e^{\tau
(f+g)}\omega \wedge \eta =\int\limits_{[\Gamma _{1}]}e^{\tau f}\omega
\int\limits_{\lbrack \Gamma _{2}]}e^{\tau g}\eta .  \tag{D.7}
\end{equation}
If we combine this formula with the asymptotic result, Eq.(D.2), then, the
Fubini's theorem becomes a statement from the theory of Laplace transforms:
the Laplace transform of the convolution is equal to the product of the
Laplace transforms. Looking at Eq.(D.5) it is easy to write a convolution.
With accuracy up to a constant we obtain 
\begin{equation}
I(\hat{\alpha}+\hat{\beta})=\int\limits_{0}^{t}ds\left| t-s\right| ^{\hat{%
\alpha}}\left| s\right| ^{\hat{\beta}}=t^{\hat{\alpha}+\hat{\beta}+1}B(\hat{%
\alpha}+1,\hat{\beta}+1)  \tag{D.8}
\end{equation}
where $\hat{\alpha}=\frac{l}{N+1}-1$, $\hat{\beta}=\frac{m}{N+1}-1,1\leq
l,m\leq N.$ Evidently, use of this result in the l.h.s. of Fubini's relation
superimposed with Eq.s(D.2-D.5) produces back Eq.(5.24a) of the main text.
Since the results just described and many others can be found in the
comprehensive paper by Varchenko, Ref. [144], there is no need to repeat
here his arguments related to Hodge and mixed Hodge structures associated
with such type of oscillatory integrals. It also should be clear from
reading of the same reference that the result, Eq.(5.32), for the Alexander
polynomials is connected with Fubini's theorem just discussed.

Following Refs.[160,163] we shall call the exponent $\alpha $ in Eq.(D.6) an 
\textbf{order} of the differential form $\omega .$ It is also called the 
\textbf{spectrum} of the critical point of the germ $f(z)=z^{N+1}.$ Let now
\ we have two germs $f$ and $g$ with respective spectrum \{$\alpha _{i}\}$
and \{$\beta _{j}\}$ so that $i=1,...,\mu $ and $j=1,...,\eta .$ Then the
spectrum of the critical point of the germ $f+g$ is obtained $\{\alpha
_{i}+\beta _{j}+1\}.$ For the composition of germs leading to the B-P
variety, e.g.see Eq.s (3.7)-(3.8), the spectrum is made essentially of
numbers which belong to the set $X(S^{1})$ (actually we have to divide these
numbers by $N$) defined by Eq.(5.22a). In view of the Eq.s(3.17a) and (3.26)
each number $\alpha $ is also associated with the eigenvalue $\lambda $ of
the monodromy operator of the critical point, i.e. $\lambda =\exp (\pm 2\pi
i\alpha ).$

The dimensions of spaces of $(p,q)$ differential forms $\omega $ defined by
Eq.(5.9) are given by the Hodge numbers $h^{p,q}.$ These numbers had been
related to the spectrum by Steenbink [169]. We follow, nevertheless, more
recent and \ more clear exposition of this subject presented in Ref.[135] (
Ref.[144] is also very helpful). Let $p+q=m,m\leq n,$ and choose a pair of
numbers $(\hat{\alpha},l)$ where $\hat{\alpha}$ \ is a \emph{spectral}
number $\alpha =-\frac{1}{2\pi i}\ln \lambda $ normalized by the level $p,$%
i.e. $n-p-1<\hat{\alpha}\leq n-p,$ and $l$ is the \emph{weight} number given
by 
\begin{equation}
l=\left\{ 
\begin{array}{c}
p+q\text{ \ \ \ \ if\ \ \ }\lambda \neq 1 \\ 
p+q-1\text{ if }\lambda =1
\end{array}
\right.   \tag{D.9}
\end{equation}
Let $h_{\lambda }^{p,q}$ be the \emph{dimension} (in physics terminology,
the \emph{degeneracy}) of the eigenspace related to the eigenvalue $\lambda .
$ Then, the Hodge numbers $h^{p,q}$ are determined simply by 
\begin{equation}
h^{p,q}=\sum\limits_{\lambda }h_{\lambda }^{p,q}.  \tag{D.10}
\end{equation}
Taking into account the symmetry of the Hodge numbers (and also the numbers $%
h_{\lambda }^{p,q}$ ): 
\begin{equation}
h_{\lambda }^{p,q}=h_{\lambda }^{q,p};h_{\lambda }^{p,q}=h_{\lambda
}^{n-1-q,n-1-p}\text{ for }\lambda \neq 1  \tag{D.11a}
\end{equation}
and 
\begin{equation}
h_{1}^{p,q}=h_{1}^{n-q,n-p}\text{ for }\lambda =1  \tag{D.11b}
\end{equation}
and using this symmetry the following ''sum rule'' is obtained: 
\begin{equation}
\sum_{p,q}h_{\lambda }^{p,q}(p+q+1)+\sum_{p,q}h_{1}^{p,q}(p+q)=n\mu . 
\tag{D.12a}
\end{equation}
where the Milnor number $\mu $ is defined in the main text (for the special
case of B-P variety only!) by \ Eq.(5.54). In addition, there is yet another
sum rule 
\begin{equation}
\sum\limits_{i}\alpha _{i}=\mu (\frac{n}{2}-1).  \tag{D.12b}
\end{equation}
Consider now the simplest example based on analysis of the germ $%
f(z)=z^{N+1}.$ According to Eq.(D.6) we have $\alpha =\frac{-l}{N+1}$ with $%
1\leq l\leq N.$Clearly, $\left| \alpha \right| <1$ so that $\lambda \neq 1,%
\hat{\alpha}=\frac{l}{N+1}.$ In view of the results presented above, we
conclude that $p=0$ and , due to symmetry, also $q=0$ . This produces $%
h_{\lambda }^{0,0}=1$ and $h_{{}}^{0,0}=N=\mu .$ Let us consider less
trivial example of germ of the Fermat-like curve. It is relevant to the
discussion in Section 3.2. In particular, we had obtained the set of
differential forms, Eq.(3.24), distinguished by the numbers $r$ and $s$ (or $%
a=\frac{r}{N}$ and $b=\frac{s}{N}).$ Replacing now $N$ by $N+1$and using
Eq.(D.8) and the discussion which follows this equation, we obtain the
spectrum as $a+b-1$. Clearly, the factor of -1 can be discarded since it
does not carry any additional useful physical information. Hence, we are
left with the spectral set $\frac{m+n}{N+1}$ with $1\leq m,n\leq N.$Using
the arguments presented above the Hodge spectrum can be determined rather
easily on a case by case basis. \textbf{Hence, the particle spectrum is in
one to one correspondence with the Hodge numbers of the spectral set. }The
alternative way of computation of Hodge numbers is presented in Section
5.3.2.

We would like to conclude this appendix by writing down the set of the
Picard -Fuchs equations (e.g. see page 336 of Ref.[36]) for the B-P type of
singularity (Sections 3 and 5). They are given below:\ 
\begin{equation}
\frac{dI^{k}}{dt}=(r_{k}-1)\frac{I^{k}}{t},  \tag{D.13}
\end{equation}
where the subscript $k=(k_{0},...,k_{n})$ numbers different integrals of the
type given by Eq.(D.3) and $r_{k}$ is given by 
\begin{equation*}
r_{k}=\frac{k_{0}+...+k_{n}}{N+1},
\end{equation*}
with $1\leq k_{0},...,k_{n}\leq N.$

\bigskip

\pagebreak

\bigskip

\bigskip

\textbf{References}

\bigskip

[1] G.Veneziano,Construction of a crossing symmetric, Regge-behaved

\ \ \ \ \ amplitude for linearly rising trajectories, Il Nuovo Cimento 57A

\ \ \ \ \ (1968) 190-197.

[2] V. De Alfaro, S. Fubini, G.Furlan, C.Rossetti, Currents in Hadron

\ \ \ \ \ \ Physics, Elsevier, Amsterdam, 1973.

[3] M.Virasoro, Alternative construction of crossing-symmetric amplitudes

\ \ \ \ \ \ with Regge behavior, Phys.Rev. 177 (1969) 2309-2314.

[4] A.Kholodenko, Statistical mechanics of 2+1 gravity from Riemann

\ \ \ \ \ \ zeta function and Alexander polynomial: exact results,
J.Geom.Phys.

\ \ \ \ \ \ 38 (2001) 81-139.

[5] A.Kholodenko, Use of quadratic differentials for description of defects

\ \ \ \ \ \ and textures in liquid crystals and 2+1 gravity, J.Geom.Phys. 33

\ \ \ \ \ \ (2000) 59-102.

[6] A.Kholodenko, Use of meanders and train tracks for description of defects

\ \ \ \ \ and textures in liquid crystals and 2+1 gravity, J.Geom.Phys. 33

\ \ \ \ \ (2000) 23-58.

[7] M.Green, J.Schwarz, E.Witten, Supersting Theory, Vol.1, Cambridge

\ \ \ \ \ University Press, Cambridge, 1987.

[8] Y.Imayoshi, M.Taniguchi, An Introduction to Teichm\"{u}ller Spaces,

\ \ \ \ \ \ Springer, Berlin, 1992.

[9] \ S.Nag, The Complex Analytic Theory of Teichm\"{u}ller Spaces, Wiley

\ \ \ \ \ \ \ Interscience, New York, 1998.

[10] L. Takhtajan, L-P. Teo, Liouville action and Weil-Petersson metric

\ \ \ \ \ \ \ on deformation spaces, global Kleinian reciprocity and
holography,

\ \ \ \ \ \ \ math.CV/0204318.

[11] L.Bekke, P.Freund, p-adic numbers in physics, Phys.Reports

\ \ \ \ \ \ \ \ 233 (1993) 1-66.

[12] E.Artin, Algebraic Numbers and Algebraic Functions, Gordon and

\ \ \ \ \ \ \ \ Breach, London, 1967.

[13] A.Zabrodin, Non-Archimedean strings and Bruhat-Tits trees,

\ \ \ \ \ \ \ \ Comm.Math.Phys.123 (1989) 463-483.

[14] S.Lang, Cyclotomic Fields, Springer-Verlag, Berlin, 1990.

[15] D.Thakur, On Gamma function for function fields,

\ \ \ \ \ \ \ \ in : D.Goss, D.Hayes, M.Rosen (Eds.), The Arithmetic of

\ \ \ \ \ \ \ \ Function Fields, de Gruyter, Berlin, 1992.

[16] A.Terras, Fourier Analysis on Finite Groups and Applications,

\ \ \ \ \ \ \ \ Cambridge University press, Cambridge, 1999.

[17] W. Li, Number Theory With Applications, World Scientific,

\ \ \ \ \ \ \ \ Singapore, 1996.

[18] G.Shimura and Y.Taniyama, Complex Multiplication of Abelian

\ \ \ \ \ \ \ \ Varieties and its Applications to Number Theory,
Publ.Math.Soc.Japan,

\ \ \ \ \ \ \ \ Tokyo, 1961.

[19] \ M.Jacob (Editor), Dual Theory, Elsevier, Amsterdam, 1974.

[20] \ W. Heisenberg, Die beobachtbaren Gr\"{o}ssen in der theorie der

\ \ \ \ \ \ \ \ Elementarteilchen, I, Z.Phys. 120 (1943) 513-538; ibid, II,
673-702.

[21] \ J.Bjorken, S.Drell, Relativistic Quantum Fields, Vol.2, McGraw-Hill,

\ \ \ \ \ \ \ \ \ New York, 1965.

[22] \ S.Novikov, S.Manakov, L.Pitaevskii, V.Zakharov, Theory of Solitons,

\ \ \ \ \ \ \ \ Consultants Bureau, New York, 1984.

[23] \ D.Anosov, S.Aranson, V.Arnol'd, I.Bronstein, V.Grines,

\ \ \ \ \ \ \ \ Yu.Il'yashenko, Ordinary Differential Equations and Smooth

\ \ \ \ \ \ \ \ Dynamical Systems, Springer-Verlag, Berlin, 1997.

[24] L.Landau, On analytic properties of vertex parts in quantum

\ \ \ \ \ \ \ \ field theory, Nucl.Phys.13 (1959) 181-192.

[25] F.Pham, Formules de Picard-Lefschetz generalisees et ramification

\ \ \ \ \ \ \ des integrales, Bull. Soc.Math. France 93 (1965) 333-367.

[26] J.Milnor, Singular Points of Complex Hypersurfaces,

\ \ \ \ \ \ \ \ Princeton University Press, Princeton, N.J., 1968.

[27] E.Brieskorn, Biespiele zur Differentialtopologie von Singularit\"{a}ten,

\ \ \ \ \ \ \ \ Inv.Math.2 (1966) 1-14.

[28] G.Ponzano, T.Regge, E.Speer, M.Westwater, The monodromy rings

\ \ \ \ \ \ \ \ of a class of self-energy graphs, Comm.Math.Phys. 15 (1969)
83-132;

\ \ \ \ \ \ \ \ ibid.18 (1970) 1-64.

[29] J.Edwards, A Treatise on the Integral Calculus, Vol.2,

\ \ \ \ \ \ \ \ Macmillan, London, 1922.

[30] A.Varchenko, Euler's Beta-function, Vandermonde determinant,

\ \ \ \ \ \ \ \ Legendre equation and critical values of linear functions on
hyperplane

\ \ \ \ \ \ \ \ configurations, part I, Sov.Math. Doklady 53 (1989)
1206-1235;

\ \ \ \ \ \ \ \ ibid, part II, 54(1990) 146-155.

[31] S.Chowla, A.Selberg, On Epstein's Zeta-function, J.F\"{u}r die Reine

\ \ \ \ \ \ \ \ und Angew.Math.227 (1967) 86-110.

[32] A.Weil, Sur les periods des integrales Abeliennes, Comm.Pure

\ \ \ \ \ \ \ and Appl.Math.29 (1976) 813-819.

[33] A.Weil, Abelian varieties and the Hodge ring, in: Collected Papers,

\ \ \ \ \ \ \ \ Vol.3, Springer-Verlag, Berlin, 1979.

[34] B.Gross, On the periods of Abelian integrals and formula of Chowla

\ \ \ \ \ \ \ \ and Selberg, Inv.Math.45 (1978) 193-211.

[35] S.Lang, Introduction to Algebraic and Abelian Functions,

\ \ \ \ \ \ \ \ Springer-Verlag, Berlin, 1982.

[36] V.Arnol'd, S.Gussein-Zade, A.Varchenko, Singularities of

\ \ \ \ \ \ \ \ Differentiable Maps, Vol.2, Birkh\"{a}user, Boston, 1988.

[37] P.Deligne, Hodge cycles on Abelian Varieties, in : Lecture Notes

\ \ \ \ \ \ \ \ in Math, Vol.900, Springer-Verlag, Berlin, 1982.

[38] N.N.Schappacher, Periods of Hecke Characters, Lecture Notes

\ \ \ \ \ \ \ in Math. Vol.1301, Springer-Verlag, Belin, 1988.

[39] P.Bertelot, A.Ogus, Notes on Crystalline Cohomology,

\ \ \ \ \ \ \ \ Mathematical Notes, Vol. 21, Princeton University Press,

\ \ \ \ \ \ \ \ Princeton,N.J., 1978.

[40] D.McDuff, D.Salamon, Introduction to Symplectic Topology,

\ \ \ \ \ \ \ \ Clarendon Press, Oxford, 1998.

[41] V.Vladimirov, I.Volovich, E.Zelenov, p-adic Analysis and

\ \ \ \ \ \ \ Mathematical Physics, World Scientific, Singapore, 1994.

[42] D.Goldschmidt, Algebraic Functions and Projective Curves,

\ \ \ \ \ \ \ Springer-Verlag, Berlin, 2003.

[43] M.Rosen, Number Theory in Function Fields,

\ \ \ \ \ \ \ Springer-Verlag, Berlin, 2002.

[44] W.Gilbert, Modern Algebra With Applications,

\ \ \ \ \ \ \ John Wiley\&Sons, New York, 1976

[45] H.Stichtenoth, Algebraic Function Fields and Codes,

\ \ \ \ \ \ \ Springer-Verlag, Berlin,1993.

[46] S.Lang, Undergraduate Algebra, Springer-Verlag,

\ \ \ \ \ \ \ \ Berlin, 1990.

[47] B.Van der Waerden, Algebra, Springer-Verlag, Berlin, 1967.

[48] B.Mazur, On the passage from local to global in number theory,

\ \ \ \ \ \ \ AMS Bulletin 29 (1993) 14-50.

[49] L.Childs, A Concrete Introduction to Higher Algebra,

\ \ \ \ \ \ \ Springer-Verlag, Berlin, 1995.

[50] R.Bott, On the Shape of a Curve,

\ \ \ \ \ \ \ Adv.Math.16 (1975)144-159.

[51] Y.Aurby, M.Perret, A Weil theorem for singular curves,

\ \ \ \ \ \ in : R.Pellikaan, M.Perret, S.Vladut (Eds), Arithmetic,Geometry

\ \ \ \ \ \ and Coding Theory, Walter de Gruyter, Berlin, 1996.

[52] V.Guillemin, A.Pollack, Differential Topology, Prentice

\ \ \ \ \ \ \ \ Hall, New York, 1974.

[53] J.Milnor, Infinite cyclic coverings, in: Collected Papers, Vol.2,

\ \ \ \ \ \ \ Publish or Perish, Inc., Houston, TX, 1995.

[54] A.Katok, B.Hasselblatt, Introduction to the Modern Theory

\ \ \ \ \ \ \ of Dynamical Systems,Cambridge University Press, Cambridge,
1995.

[55] J.Birman, M.Kidwell, Fixed points of pseudo-Anosov

\ \ \ \ \ \ \ diffeomorphisms of surfaces, Adv.Math. 46 (1982) 217-220.

[56] N.Yui, Arithmetic of certain Calabi-Yau varieties and mirror

\ \ \ \ \ \ \ \ symmetry, in : B.Conrad, K.Rubin (Eds), Arithmetic Algebraic

\ \ \ \ \ \ \ \ Geometry, AMS Publications, Providence, RI, 2001.

[57] \ D.Cox, S.Katz, Mirror Symmetry and Algebraic Geometry,

\ \ \ \ \ \ \ \ AMS Publications, Providence, RI, 1999.

[58] \ L.Kauffman, On Knots, Princeton University Press, Princeton, 1987.

[59] \ G.Brude, H.Zieschang, Knots, Walter de Gryter, Berlin, 1985.

[60] \ M.Armstrong, Basic Topology, Springer-Verlag, 1983.

[61] \ D.Rolfsen, Knots and Links, Publish or Perish, Houston, 1990.

[62] \ A.Grothendieck, Modeles de Neron et monodromie, in:

\ \ \ \ \ \ \ \ LNM Vol.288, Springer-Verlag, Berlin, 1972.

[63] \ A.Wiles, Modular elliptic curves and Fermat's last theorem,

\ \ \ \ \ \ \ \ Ann.Math. 141 (1995) 443-551.

[64] \ A.Kholodenko, Kontsevich-Witten model rom 2+1 gravity:

\ \ \ \ \ \ \ \ new exact combinatorial solution, J.Geom.Phys 43 (2002)
45-91.

[65] \ A.Weil. Number of solutions of equations in finite fields,

\ \ \ \ \ \ \ \ AMS Bulletin 55 (1949) 497-508.

[66] \ R.Miranda, Algebraic Curves and Riemann Surfaces,

\ \ \ \ \ \ \ \ AMS Publications, Providence, RI, 1997.

[67] \ D.Rorlich, Points at infinity on the Fermat curves,

\ \ \ \ \ \ \ \ \ Inv. Math. 39 (1977) 95-127.

[68] \ B.Gross, N.Koblitz, Gauss sums and the p-adic $\Gamma $-function,

\ \ \ \ \ \ \ \ Ann.Math. 109 (1979) 569-581.

[69] \ N.Koblitz, p-adic Analysis: A short Course on Recent Work,

\ \ \ \ \ \ \ \ Cambridge University Press, Cambridge, 1980.

[70] \ L.Landau, E.Lifshitz, Mechanics, Nauka, Moscow, 1965.

[71] \ G.Frey, Links between stable elliptic curves and certain

\ \ \ \ \ \ \ \ \ Diophantine equations, Ann.Univ.Saraviensis, Ser.Math.1
(1986) 1-40.

[71] \ H.McKean, V.Moll, Elliptic Curves, Cambidge University Press,

\ \ \ \ \ \ \ \ Cambridge, 1999.

[72] \ M.Mulase, Algebraic theory of the KP equations, in :

\ \ \ \ \ \ \ \ Perspectives in Mathematical Physics, International Press,

\ \ \ \ \ \ \ \ Cambridge, MA, 1994.

[73] \ A.Kholodenko, Boundary conformal field theories,

\ \ \ \ \ \ \ \ \ limit sets of Kleinian groups and holography,

\ \ \ \ \ \ \ \ J.Geom.Phys. 35 (2000) 193-238.

[74] \ S.Chowla, A.Selberg, On Epstein's zeta function,

\ \ \ \ \ \ \ \ \ PNAS 35 (1949) 371-374.

[75] \ K.Ramachandra, Some applications of Kronecker's limit

\ \ \ \ \ \ \ \ formulas, Ann.Math. 80 (1964) 104-148.

[76] \ N.Akhieser, Elements of the Theory of Elliptic

\ \ \ \ \ \ \ \ \ Functions, Nauka, Moscow, 1970.

[77] \ A.Kholodenko, Some thoughts about random walks on figure

\ \ \ \ \ \ \ \ \ eight, Phys.A 289 (2001) 377-408.

[78] \ P.Etingof, I.Frenkel, A.Kirillov, Jr., Lectures on Representation

\ \ \ \ \ \ \ \ \ Theory and Knizhnik-Zamolodchikov Equations, AMS

\ \ \ \ \ \ \ \ \ Publications, Providence, RI, 1998.

[79] \ B.Bakalov, A.Kirillov, Jr., Lectures on Tensor Categories

\ \ \ \ \ \ \ \ \ and Modular Functors, AMS Publications, Providence, R.I.
2001.

[80] \ A.Varchenko, Special Functions, KZ Type Equations and

\ \ \ \ \ \ \ \ \ Representation Theory, Lecture Notes given at MIT in spring

\ \ \ \ \ \ \ \ \ of 2002, Math.QA/0205313.

[81] \ \ C.Itzykson, J-M. Drouffee, Statistical Field Theory, Vol.1.,

\ \ \ \ \ \ \ \ \ Cambridge University Press, Cambridge, 1989.

[82] \ \ P.Di Francesco, P.Mathieu, D.Senechal, Conformal Field Theory,

\ \ \ \ \ \ \ \ \ Springer-Verlag, Berlin, 1997.

[83] \ D.Husem\"{o}ller, Elliptic Curves, Springer-Verlag, Berlin, 1987.

[84] \ G.Hardy, E.Wright, An Introduction to the Theory of Numbers,

\ \ \ \ \ \ \ \ \ Clarendon, Oxford, 1962.

[85] \ \ J.Silverman, The Arithmetic of Elliptic Curves, Springer-Verlag,

\ \ \ \ \ \ \ \ \ Berlin, 1986.

[86] \ Y.Manin, Real multiplication and noncommutative geometry,

\ \ \ \ \ \ \ \ \ math.AG/0202109.

[87] \ Y.Manin, M.Marcolli, Holography principle and arithmetic of

\ \ \ \ \ \ \ \ \ algebraic curves, hep-th/0201036

[88] \ G.Jones, D.Singerman, Complex Functions, Cambridge University

\ \ \ \ \ \ \ \ \ Press, Cambridge, 1987.

[89] \ J.Silverman, Advanced Topics in the Arithmetic of Elliptic Curves,

\ \ \ \ \ \ \ \ \ Springer-Verlag, Berlin, 1994.

[90] \ H.Stark, L-functions at s=1. III. Totally real fields and Hilbert's

\ \ \ \ \ \ \ \ 12th problem, Adv.Math. 22 (1976) 64-84.

[91] \ R.Taylor, A.Wiles, Ring theoretic properties of some Hecke algebras,

\ \ \ \ \ \ \ \ \ Ann.Math. 141 (1995) 553-572.

[92] \ \ K.Ribet, On modular representations of Gal (\textbf{\={Q}}/\textbf{Q%
}) arising from

\ \ \ \ \ \ \ \ \ \ modular forms, Inv.Math. 100 (1990) 431-476.

[93] \ \ A.Knapp, Elliptic Curves, Princeton University Press,

\ \ \ \ \ \ \ \ \ \ Princeton, 1992.

[94] \ \ S.Katok, J.Millson, Eichler-Shimura homology, intersection

\ \ \ \ \ \ \ \ \ \ numbers and rational structures on spaces of modular
forms,

\ \ \ \ \ \ \ \ \ \ AMS Transactions 300 (1987) 737-757.

[95] \ \ R.Sharp, Closed geodesics and periods of automorphic forms,

\ \ \ \ \ \ \ \ \ \ Adv.Math. 160 (2001) 205-216.

[96] \ \ \ B.Birch, H.Swinnerton-Dyer, Notes on elliptic curves. II.

\ \ \ \ \ \ \ \ \ \ Jour. f\"{u}r Reine und Angew.Math. 218 (1965) 79-108.

[97] \ \ Y.Manin, Cyclotomic fields and modular curves,

\ \ \ \ \ \ \ \ \ \ Russ.Math.Surv. 26 (1971) 7-71.

[98] \ \ E.Hecke, Lectures on the Theory of Algebraic Numbers,

\ \ \ \ \ \ \ \ \ Springer-Verlag, Berlin, 1981.

[99] \ \ K.Ireland, M.Rosen, A Classical Introduction to Modern

\ \ \ \ \ \ \ \ \ \ Number Theory, Springer-Verlag, Berlin, 1990.

[100] \ H.Iwanieck, Topics in Classical Automorphic Forms,

\ \ \ \ \ \ \ \ \ \ AMS Publications, Providence, RI, 1997.

[101] \ A.Weil, Jacobi sums as ''Grossencharaktere'',

\ \ \ \ \ \ \ \ \ \ \ AMS Transactions, 73 (1952) 487-495.

[102] \ N. Katz, Travaux de Dwork, Seminaire BOURBAKI,

\ \ \ \ \ \ \ \ \ \ \ \# 409, 1972.

[103] \ J.Cassels, Diophantine equations with special reference to

\ \ \ \ \ \ \ \ \ \ elliptic curves, Jour.Lond.Math.Soc. 41 (1966) 193-291.

[104] \ A.Weil, Adeles and Algebraic Groups, Birkhauser, Boston, 1982.

[105] \ C.Maclachlan, A.Reid, The Arithmetic of Hyperbolic

\ \ \ \ \ \ \ \ \ \ 3-Manifolds, Springer-Verlag, Berlin, 2003.

[106] \ C.Tracy, The emerging role of number theory in exactly

\ \ \ \ \ \ \ \ \ \ solvable models in lattice statistical mechanics,

\ \ \ \ \ \ \ \ \ \ Physica D 25 (1987) 1-19.

[107] Y.Zinoviev, The Ising model and the L-function,

\ \ \ \ \ \ \ \ \ \ Theor.Math.Phys.126 (2001) 66-80.

[108] \ P.Cohen, A C* dynamical system with Dedekind zeta

\ \ \ \ \ \ \ \ \ \ \ partition function and spontaneous symmetry breaking,

\ \ \ \ \ \ \ \ \ \ \ J.Theor.Numbers Bordeaux 11 (1999) 15-30.

[109] \ J-B. Bost, A.Connes, Hecke algebras, type III factors

\ \ \ \ \ \ \ \ \ \ \ and phase transitions with spontaneous symmetry
breaking,

\ \ \ \ \ \ \ \ \ \ \ Sel.Math. 1 (1995) 411-457.

[110] \ A.Weil, Elliptic Functions According to Eisenstein and

\ \ \ \ \ \ \ \ \ \ \ Kronecker, Springer-Verlag, Berlin, 1976.

[111] \ L.Washington, Introduction to Cyclotomic Fields,

\ \ \ \ \ \ \ \ \ \ Springer-Verlag, Berlin, 1997.

[112] \ C.Moreno, The Chowla-Selberg formula, J.Number

\ \ \ \ \ \ \ \ \ \ Theory 17 (1983) 226-245.

[113] \ T.Asai, On a certain function analogous to ln$\left| \eta (z)\right| 
$,

\ \ \ \ \ \ \ \ \ \ Nagoya Math. J. 40 (1970) 193-211.

[114] \ S.Lang, Elliptic Functions, Springer-Verlag, Berlin, 1987.

[115] \ H.Cohn, Advanced Number Theory, Dover

\ \ \ \ \ \ \ \ \ \ Publications, Inc., NY, 1962.

[116] \ H.Tsukada, String Path Integral Realization of Vertex

\ \ \ \ \ \ \ \ \ \ \ Operator Algebras, AMS Memoirs Vol.444, AMS
Publications,

\ \ \ \ \ \ \ \ \ \ \ Providence, RI, 1991.

[117] \ B.Gross, Arithmetic on Elliptic Curves with Complex

\ \ \ \ \ \ \ \ \ \ Multiplication, LNM 776, Springer-Verlag, Berlin, 1980.

[118] \ S.Lang, Complex Multiplication, Springer-Verlag, Berlin, 1983.

[119] \ J.Silverman, The theory of height functions, in :

\ \ \ \ \ \ \ \ \ \ G.Cornell, J.Silverman (Eds), Arithmetic Geometry,

\ \ \ \ \ \ \ \ \ Springer-Verlag, Berlin, 1986.

[120] \ B.Mazur, Arithmetic on curves, AMS Bulletin 14 (1986) 207-259.

[121] \ P.Colmez, Periodes des varietes abeliennes a multiplication complexe,

\ \ \ \ \ \ \ \ \ \ Ann.Math. 138 (1993) 625-683.

[122] \ S.Zhang, Heights of Heegner points on Shimura curves,

\ \ \ \ \ \ \ \ \ \ Ann.Math. 153 (2001) 27-147.

[123] \ J.Polchinski, String Theory, Vol.1,Cambridge University Press,

\ \ \ \ \ \ \ \ \ \ Cambridge, 1998.

[124] \ A.Beilinson, Y.Manin, The Mumford form and the Polyakov measure

\ \ \ \ \ \ \ \ \ \ \ in string theory, Comm.Math.Phys. 107 (1986) 359-376.

[125] \ Y.Manin, The partition function of the string can be expressed in

\ \ \ \ \ \ \ \ \ \ \ terms of theta functions, Phys.Lett. B172 (1986)
184-188.

[126] \ J.Bost,T.Jolicoeur, A holomorphy property and the critical

\ \ \ \ \ \ \ \ \ dimension in string theory from an index theorem,

\ \ \ \ \ \ \ \ \ \ Phys.Lett.B 174 (1986) 273-276.

[127]. J-D.Smit, String theory and algebraic geometry of moduli spaces,

\ \ \ \ \ \ \ \ \ \ \ Comm.Math.Phys. 114 (1988) 645-685.

[128] \ S.Mochizuki, Foundations of p-adic Teichmuller Theory,

\ \ \ \ \ \ \ \ \ \ AMS Publications, Providence, RI , 1999.

[129] A.Ogus, Frobenius and Hodge spectral sequence,

\ \ \ \ \ \ \ \ \ \ Adv. in Math. 162 (2001) 141-172.

[130] \ A.Ogus, Elliptic crystals and modular motives,

\ \ \ \ \ \ \ \ \ \ Adv. in Math. 162 (2001) 173-216.

[131] K.Krasnov, Holography and Riemann surfaces,

\ \ \ \ \ \ \ \ \ Adv. in Theor.Math.Phys. 4 (2000) 929-979.

[132] \ P.Griffiths, On periods of certain rational integrals: I,

\ \ \ \ \ \ \ \ \ \ Ann.Math. 90 (1969) 460-495; ibid II, 496-541.

[133] \ Y.Manin, Algebraic curves over fields with differentiation,

\ \ \ \ \ \ \ \ \ \ AMS Translations 37 (1964) 59-78.

[134] \ Y.Shimizu, K.Ueno, Advances in Moduli Theory,

\ \ \ \ \ \ \ \ \ \ AMS Translations 206 (2002) 1-175.

[135] \ V.Kulikov, Mixed Hodge Structures and Singularities,

\ \ \ \ \ \ \ \ \ \ Cambridge University Press, Cambridge, 1998.

[136] \ S.Lefshetz, Applications of Algebraic Topology:

\ \ \ \ \ \ \ \ \ \ Graphs and Networks, The Picard-Lefshetz Theory

\ \ \ \ \ \ \ \ \ \ and Feynman Integrals, Springer-Verlag, Berlin, 1975.

[137]. V.Schechtman, A.Varchenko, Arrangement of hyperplanes

\ \ \ \ \ \ \ \ \ \ \ and Lie Algebra homology, Inv.Math. 106 (1991) 139-194.

[138] \ E.Brieskorn, H.Knorrer, Plane Algebraic Curves,

\ \ \ \ \ \ \ \ \ \ \ Birkh\"{a}user, Boston, 1986.

[139] \ R.Wells, Differential Analysis on Complex Manifolds,

\ \ \ \ \ \ \ \ \ \ \ Prentice Hall, Inc., NJ, 1973.

[140] \ A.Weil, Introduction a L'etude des Varietes K\"{a}hleriennes,

\ \ \ \ \ \ \ \ \ \ \ Hermann, Paris, 1958.

[141] \ V.Arnol'd, Mathematical Methods of Classical Mechanics,

\ \ \ \ \ \ \ \ \ \ \ Nauka, Moscow, 1974.

[142] \ V.Arnol'd, A.Givental, Symplectic Geometry,

\ \ \ \ \ \ \ \ \ \ \ in: V.Arnol'd, S.Novikov (Eds), Dynamical Systems IV,

\ \ \ \ \ \ \ \ \ \ \ Springer-Verlag, Berlin, 1990.

[143] \ H.Pohlmann, Algebraic Cycles on Abelian varieties of

\ \ \ \ \ \ \ \ \ \ \ complex multiplication type, Ann.Math. 88 (1968)
161-180.

[144] \ A.Varchenko, Asymptotic Hodge structure in the vanishing

\ \ \ \ \ \ \ \ \ \ \ cohomology, Math.USSR, Izvestija 18 (1982) 469-512.

[145] \ U.Jannsen, S.Kleiman, J-P Serre (Eds), Motives,

\ \ \ \ \ \ \ \ \ \ Parts 1and 2, AMS Publications, Providence, RI, 1994.

[146] \ P.Griffiths, Periods of integrals on Algebraic manifolds, I.

\ \ \ \ \ \ \ \ \ \ (Construction and properties of modular varieties),

\ \ \ \ \ \ \ \ \ \ Am.J.of Math. 90 (1968) 568-626.

[147] \ P.Griffiths, Periods of integrals on algebraic manifolds, II.

\ \ \ \ \ \ \ \ \ \ (Local study of the period mapping),

\ \ \ \ \ \ \ \ \ \ Am.J.of Math. 90 (1968) 805-865.

[148] \ R.Hwa, V.Teplitz, Homology and Feynman Integrals,

\ \ \ \ \ \ \ \ \ \ W.A.Benjamin, Inc., NY, 1966.

[149] \ D.Eisenbud,W.Neumann, Three-Dimensional Link Theory

\ \ \ \ \ \ \ \ \ \ and Invariants of Plane Curve Singularities,

\ \ \ \ \ \ \ \ \ \ Princeton University Press, Princeton, 1985.

[150] \ LeTrang , Sur les noeuds algebriques,

\ \ \ \ \ \ \ \ \ \ Compositio Math.25 (1972) 281-321.

[151] \ D.Sumners, J.Woods, The monodromy of reducible plane

\ \ \ \ \ \ \ \ \ \ curves, Inv.Math.40 (1977) 107-141.

[152] \ J.Franks, Homology and Dynamical Systems, Regional Conf.

\ \ \ \ \ \ \ \ \ \ Series in Math.49, AMS Publications, Providence, RI,
1982.

[153] \ M.Sebastiani, R.Thom, Un resultat sur la monodromie,

\ \ \ \ \ \ \ \ \ \ \ Inv.Math.13 (1971) 90-96.

[154] \ M.Oka, On the homotopy types of hypersurfaces defined

\ \ \ \ \ \ \ \ \ \ by weighted homogenous polynomials, Topology 12
(1973)19-32.

[155] \ N.Saveliev, Lectures on the Topology of 3-Manifolds,

\ \ \ \ \ \ \ \ \ \ Walter de Gruyter, Belin, 1999.

[156] \ N.A'Campo, La fonction zeta d'une monodromie,

\ \ \ \ \ \ \ \ \ \ Comm.Math.Helv. 50 (1975) 233-248.

[157] \ K.Murasugi, Knot Theory and its Applications,

\ \ \ \ \ \ \ \ \ \ Birkh\"{a}user, Boston, 1996.

[157] \ A.Kwauchi, A Survey of Knot Theory, Birkh\"{a}user, Boston, 1996.

[158] \ I.Gelfand, M.Kapranov, A.Zelevinski, Discriminants, Resultants and

\ \ \ \ \ \ \ \ \ \ \ Multidimensional Determinants, Birkh\"{a}user, Boston,
1994.

[159] \ V.Buchstaber,T.Panov, Torus Actions and Their Applications

\ \ \ \ \ \ \ \ \ \ \ in Topology and Combinatorics, Univ.Lect.Series Vol.24,

\ \ \ \ \ \ \ \ \ \ \ AMS Publications, Providence, RI, 2002.

[160] \ V.Arnol'd, V.Vaasil'ev,V.Goryunov,O.Lyashko,

\ \ \ \ \ \ \ \ \ \ \ Singularities, local and global theory, in V.Arnold
(Editor),

\ \ \ \ \ \ \ \ \ \ \ Dynamical Systems VI, Springer-Verlag,Berlin, 1993.

[161] \ T.Zaslavsky, Facing up to Arrangements: Face-count

\ \ \ \ \ \ \ \ \ \ \ Formulas for Partitions of Space by Hyperplanes, AMS

\ \ \ \ \ \ \ \ \ \ \ Memoirs, Vol 154, AMS Publications, Providence, RI,
1975.

[162] \ P.Orlik, H.Terao, Arrangements and Hypergeometric Integrals,

\ \ \ \ \ \ \ \ \ \ Math.Sci.Japan Memoirs,Vol.9, Tokyo, 2001.

[163] \ V. Arnol'd, Index of a singularity of the vector field,

\ \ \ \ \ \ \ \ \ \ \ Petrovskii-Oleinik inequalities and mixed Hodge
structures,

\ \ \ \ \ \ \ \ \ \ \ Funct.Analysis and Applications, 12 (1978) 1-14.

[164] \ V.Batyrev, Variations of the mixed Hodge structure of affine

\ \ \ \ \ \ \ \ \ \ \ hypersurfaces in algebraic tori, Duke Math.journ. 69
(1993) 349-409.

[165] \ A.Koushnirenko, Polyedres de Newton et nomres de Milnor,

\ \ \ \ \ \ \ \ \ \ \ Inv.Math.32 (1976) 1-31.

[166] \ \ S.Salamon, On the cohomology of K\"{a}hler and hyper-K\"{a}hler

\ \ \ \ \ \ \ \ \ \ \ manifolds, Topology 35 (1996) 137-155.

[167] \ \ V.Danilov, A.Khovanskii, Newton polyhedra and an algorithm

\ \ \ \ \ \ \ \ \ \ \ for computing Hodge-Deligne numbers,

\ \ \ \ \ \ \ \ \ \ \ Math.USSR-Izv. 29 (1987) 279-298.

[168] \ \ D.Cox, Recent developments in toric geometry, math.AG/9606016.

[169] \ \ J.Steenbrink, Mixed Hodge structure on vanishing cohomology,

\ \ \ \ \ \ \ \ \ \ \ Proc.Nordic Summer Scool, Oslo, 1976.

[170] \ \ M.Saito, On Steenbrink's conjecture, Math.Ann.289 (1991) 703-716.

[171] \ \ A.Varchenko, A.Khovanskii, Sov.Math.Dokl. 32 (1985) 122-127.

[172] \ \ \ J.Nielsen, Untershhungenzur topologie der geslossen zweiseitigen

\ \ \ \ \ \ \ \ \ \ \ \ \ flachen I, Acta Math.50 (1927) 189-358.

[173] \ \ \ J.Milnor, Dynamics in One Complex Variable,

\ \ \ \ \ \ \ \ \ \ \ \ \ Vieweg, Wiesbaden,1999.

[174] \ \ \ W de Mello, S.van Strien, One-Dimensional Dynamics,

\ \ \ \ \ \ \ \ \ \ \ \ \ Springer-Verlag, Berlin, 1993.

[175] \ \ \ \ M.Kontsevich, Intersection theory on the moduli space of

\ \ \ \ \ \ \ \ \ \ \ \ \ \ curves and the matrix Airy function,
Comm.Math.Phys.

\ \ \ \ \ \ \ \ \ \ \ \ \ 147 (1992) 1-23.

[176] \ \ \ F.Tomi, A.Tromba, The Index Theorem for Minimal Surfaces of

\ \ \ \ \ \ \ \ \ \ \ \ \ Higher Genus, AMS Memoirs 117 (1995) 1-78.

[177] \ \ \ \ C.Godsil, G.Royle, Algebraic Graph Theory,

\ \ \ \ \ \ \ \ \ \ \ \ \ \ Springer-Verlag, Berlin, 2001.

[178] \ \ \ \ A.Pitzer, Ramanujan Graphs, Studies in Adv.Math.

\ \ \ \ \ \ \ \ \ \ \ \ \ 7 (1998) 159-178.

[179] \ \ \ \ J-P. Serre, Trees, Springer-Verlag, Berlin, 1980.

[180] \ \ \ \ K.Brown, Buildings, Springer-Verlag,Berlin,1991.

[181] \ \ \ \ D.Mumford, Further applications, LNM 339 (1973) 165-190.

[182] \ \ \ \ A.Figa-Talamaca, C.Nebbia, Harmonic Analysis and

\ \ \ \ \ \ \ \ \ \ \ \ \ \ Representation Theory for Groups Acting on
Homogenous

\ \ \ \ \ \ \ \ \ \ \ \ \ \ Trees, Cambridge University Press, Cambridge,
1991.

[183] \ \ \ \ A.Lubotsky, Discrete Groups, Expanding Graphs and

\ \ \ \ \ \ \ \ \ \ \ \ \ \ Invariant Measures, Birkh\"{a}user, Boston, 1994.

[184] \ \ \ \ \ P.Diaconis, Group Representations in Probability and

\ \ \ \ \ \ \ \ \ \ \ \ \ \ Statistics, Inst.of Math.Statistics Lecture
Notes Series,

\ \ \ \ \ \ \ \ \ \ \ \ \ \ Vol.11, Hayward, CA, 1988.

[185] \ \ \ \ \ A.Kreig, Hecke Algebras, AMS Memoirs, Vol.435,

\ \ \ \ \ \ \ \ \ \ \ \ \ \ AMS Publications, Providence, RI, 1990.

[186] \ \ \ \ \ M.Kontsevich, D.Zagier, Periods, in: B.Engquist, 

\ \ \ \ \ \ \ \ \ \ \ \ \ \ W.Schid (Eds), Mathematics Unlimited,
Springer-Verlag,

\ \ \ \ \ \ \ \ \ \ \ \ \ \ \ Berlin, 2001.

[187] \ \ \ \ \ \ E.Frielander, A.Suslin, The work of Vladimir Voevodsky,

\ \ \ \ \ \ \ \ \ \ \ \ \ \ \ AMS Notices 50 (2003) 214-217.

\ \ \ \ \ \ \ \ \ \ \ \ \ \ 

\ \ \ \ \ \ \ \ \ 

\end{document}